%% file: thesis.tex
\documentstyle[epsfig,amssymb,pthes12]{pthesis}

\def\dalemb#1#2{{\vbox{\hrule height .#2pt
        \hbox{\vrule width.#2pt height#1pt \kern#1pt
                \vrule width.#2pt}
        \hrule height.#2pt}}}

\def\0{{\sst{(0)}}}
\def\1{{\sst{(1)}}}
\def\2{{\sst{(2)}}}
\def\3{{\sst{(3)}}}
\def\4{{\sst{(4)}}}
\def\5{{\sst{(5)}}}
\def\6{{\sst{(6)}}}
\def\7{{\sst{(7)}}}
\def\8{{\sst{(8)}}}

\def\Z{\rlap{\sf Z}\mkern3mu{\sf Z}}
\def\R{\rlap{\rm I}\mkern3mu{\rm R}}

\def\td{\tilde}
\def\wtd{\widetilde}

\let\a=\alpha \let\b=\beta

  \let\D=\Delta  
\let\X=\Xi

\def\nn{\nonumber} \def\bd{\begin{document}} \def\ed{\end{document}}
\def\ds{\documentstyle} \let\fr=\frac \let\bl=\bigl \let\br=\bigr
\let\Br=\Bigr \let\Bl=\Bigl
\let\bm=\bibitem
\let\na=\nabla
\let\pa=\partial \let\ov=\overline
\newcommand{\be}{\begin{equation}}
\newcommand{\ee}{\end{equation}}
\def\ba{\begin{array}}
\def\ea{\end{array}}
\def\ft#1#2{{\textstyle{{\scriptstyle #1}\over {\scriptstyle #2}}}}
\def\fft#1#2{{#1 \over #2}}
\def\del{\partial}
\def\sst#1{{\scriptscriptstyle #1}}
\def\oneone{\rlap 1\mkern4mu{\rm l}}
\def\ie{{\it i.e.\ }}
\def\via{{\it via}}
\def\semi{{\ltimes}}
\def\str{{\rm str}}
\def\jm{{\rm j}}
\def\im{{\rm i}}
\def\bOmega{{{\bar\Omega}}}
\def\Qn{{{Q_{\sst{\rm N}}}}}

\def\mapright#1{\smash{\mathop{-\!\!\!-\!\!\!-\!\!\!-\!\!\!-\!\!\!
             \longrightarrow}\limits^{#1}}}
\def\maprightt#1#2{\smash{\mathop{-\!\!\!-\!\!\!-\!\!\!-\!\!\!-\!\!\!
             \longrightarrow}\limits^{#1}_{#2}}}

\newcommand{\ho}[1]{$\, ^{#1}$}
\newcommand{\hoch}[1]{$\, ^{#1}$}
\newcommand{\bea}{\begin{eqnarray}}
\newcommand{\eea}{\end{eqnarray}}
\newcommand{\ra}{\rightarrow}
\newcommand{\lra}{\longrightarrow}
\newcommand{\Lra}{\Leftrightarrow}
\newcommand{\ap}{\alpha^\prime}
\newcommand{\bp}{\tilde \beta^\prime}
\newcommand{\tr}{{\rm tr} }
\newcommand{\Tr}{{\rm Tr} }

\begin{document}
\thispagestyle{empty}

        \begin{flushright}
        UPR/ 962-T\\
        \hfill{\bf hep-th/0110084}\\
        \end{flushright}

\begin{center}

{\large {\bf Branes and Brane Worlds in M-Theory}}

\vspace{20pt}

J.F. V\'{a}zquez-Poritz

\vspace{10pt}

{\it Department of Physics and Astronomy,\\ University of Pennsylvania,
Philadelphia, PA 19104}

\footnote{\footnotesize This work is supported in part by the DOE
grant DOE-FG02-95ER40893}

\end{center}

\newpage

\begin{frontmatter}

        \title{Branes and Brane Worlds in M-Theory}
	\author{Justin Federico V\'{a}zquez-Poritz} 
	\graduategroup{Physics and Astronomy}
        \supervisor{Mirjam Cveti\v{c}}
        \gradchair{Randall Kamien}  
        \maketitle
        \legalname{Justin Federico V\'{a}zquez-Poritz}
%        \makecopyright
        \begin{dedication} 
		\centerline{For my parents, Dorothy Poritz and Paul Gauvin}
 	\end{dedication}

	\begin{acknowledgements} 

I would like to thank my advisor, Mirjam Cveti\v{c}, for sharing her
ideas, time and advice. I am also grateful to Hong L\"{u} for the ideas he
shared with me and the research we did together.

I would like to thank Chris Pope for the research we did together.

I am grateful to Vijay Balasubramanian and Matt Strassler for their advice
and for participating in my dissertation committee, as well as Larry
Gladney and Charles Kane.

I have also benefitted from discussions with Gary Shiu and Asad
Naqvi. Discussions with Finn Larsen played a crucial role in initially
starting off my research.

I am deeply grateful to my wife Anee for all her support and love.

I would like to thank my family for a lifetime of love and encouragement.

	\end{acknowledgements}

	\begin{abstract} 

The search for a theory which unifies all fundamental physics has
culminated in M-theory, whose solitonic p-brane solutions offer a wealth
of non-perturbative phenomena. In a particular regime of M-theory, there
is a duality between gauge theories and the near-horizon region of certain
p-branes, a concrete  example of which is the AdS/CFT correspondence. 

I find a new class of warped Anti-de Sitter solutions which arise as the
near-horizon region of various semi-localized brane intersections. This
provides an example of five-dimensional AdS originating in eleven-dimensional
supergravity, as well as four and six-dimensional AdS in Type IIB string theory,
cases which do not arise from direct products of spaces. This enables us
to study four-dimensional gauge theories which are dual to
eleven-dimensional supergravity solutions. The dual gauge theories of AdS
in warped spacetimes have reduced supersymmetry, which is pertinent to the
study of viable supersymmetric extensions of the Standard Model. 

In addition, I probe various supergravity solutions via the absorption of
various fields. In particular, I calculate exact absorption probabilities
which provide finite-energy probes of supergravity solutions away from the
conformal limit. For the case of the D3-brane, it is shown that particles
belonging to the same supermultiplet on the gauge theory side exhibit
identical absorption probabilities on the supergravity side. In
this manner, D3-branes are also probed away from the extremal limit.

Lastly, I discuss brane world scenarios, and how they may
arise from the near-horizon region of various p-brane
configurations. Motivated by the dual non- \\ commutative gauge theory, I
demonstrate how a background B field mimics a negative four-dimensional
cosmological constant, such that in both cases there is a massive
four-dimensional graviton.

	\end{abstract}

	\tableofcontents

	\listoffigures

%	\listoftables

\end{frontmatter}

\include{thesisintro}

\include{thesiswarped}

\include{thesisabs}

\include{thesisworld}

\include{thesiscon}

\appendix

\include{appendixA}

\include{appendixB}

\include{appendixC}

\pagebreak

\end{document}

%% file: thesisintro.tex
\chapter{Introduction}

\section{Historical Perspective on Unification Physics}

\subsection{General Relativity and the Standard Model}

Our current understanding of nature at the most fundamental level
is contained within two theories, Einstein's theory of General
Relativity and the Standard Model of particle physics. General
Relativity describes gravity as a consequence of spacetime
curvature. This theory correctly predicts physics on large scales,
such as the bending of light by the Sun's gravitational field, the
orbital motion of Mercury and the large-scale structure and
dynamics of the entire universe.

On the other hand, the Standard Model agrees with experiments down
to scales 100,000,000,000,000 smaller than are resolvable by the
human eye. This theory unifies the electromagnetic,
strong nuclear and weak nuclear forces. A quantum field theoretic
description is employed, in which force-carrier particles exist in
quantized packets of matter.

Gravity is normally insignificant for small-distance physics, the
domain in which the Standard Model comes into play. However,
nature does not completely decouple its phenomena into two
separate regimes. In a strong gravitational field, such as in the
vicinity of a black hole central singularity or during an early
epoch of the universe, aspects of both curved spacetime and particle
physics play a crucial role. In certain limits, the semi-classical
approach of quantum field theory in curved spacetime may suffice,
which treats spacetime as a classical manifold. For example, for
field perturbations near a black hole, one may neglect the
back-reaction that the mass-energy of the fields have on the
background spacetime. This enables one to calculate thermodynamic
properties of the black hole, such as temperature. However, close
to the spacetime singularity of the black hole, it is believed
that a quantum theory of gravity is required. One might be tempted
to argue that the region near a spacetime singularity is cloaked
by the surface of an event horizon. That is, any information which
is only deciphered by a quantum theory of gravitation might be
locked away within the black hole, never making contact with the
outside universe. However, as briefly mentioned above, even at the
semi-classical level it is predicted that black holes radiate
energy. Whether or not this energy carries information has long
been a subject of debate. Regardless of the outcome, as a black
hole evaporates, the surface of the event horizon gets ever closer
to the central singularity, until a full quantum theory of gravity
is needed in order to describe processes taking place outside of
the event horizon as well. Thus, if we take General Relativistic 
solutions seriously, then we are forced to consider
regimes which incorporate both gravity and particle physics. Many
physicists now believe that the quantization of the gravitational
field is linked to its unification with the fields of the Standard
Model.

\subsection{Early unification}

The idea of unification has practically always been a background
theme in physics. Newtonian gravitation explained Earth's
attraction and celestial mechanics as two results of the same
force. During the late $19^{th}$ century, the Maxwellians combined
electric and magnetic phenomena into a unified field description.
It was found that the speed of electromagnetic waves was the same
as that of light, leading to the brilliant insight that light is
actually oscillations of the electromagnetic field.

These preceding examples serve to demonstrate the overall idea of
unification, that of describing natural phenomena with as few free
parameters as possible. In light of this (no pun intended) it may
be rather disappointing, even without considerations of gravity,
if the Standard Model is the end of the road. This is because it
contains about twenty free parameters, whose values are specified
only by experiment. One would hope that a fundamental theory of
nature could be completely specified by self-consistency, and would only
need experiments to boast its predictive power. 

It can still be argued that the Standard Model has incredible predictive
power once values to its twenty parameters have been assigned. In fact,
it agrees with practically all of our observations of the physical
world. This brings into question the purpose of theoretical physics. If
our job as theorists is simply to create theories that predict what we
observe and may lead to practical applications, then our job has been
finished for quite some time now. In fact, most physicists are at a loss
to find practical applications to Special or General
Relativity\footnote{The Gravitational Positioning System (GPS) does use
General Relativity.}.

It is our search for the underlying truth that drives us to go
further. A working model does not necessarily reflect fundamental
truth. For example, Newton gave us equations with which to make
predictions but he explicitly states that it is up to the reader to
think of a mechanism for gravity. Albert Einstein found that mechanism
to be the fabric of spacetime itself.

However, his theory of General Relativity is mutually exclusive with our
present-day theories of particle physics, since the former considers
spacetime as a classical manifold whereas the latter relies on quantum
field theories. This incompatibility makes the idea of unification all 
the more tantalizing.

\subsection{Relativity}

At the beginning of the $20^{th}$ century, Einstein unified
electrodynamics and kinematics via Special Relativity.
Geometrically, this amounts to melding space and time into a
manifold known as 'spacetime.' A generalized Pythagorean Theorem
measures the coordinate-invariant spacetime distance:
%%%%%%
\be ds^2=-dt^2+dx_i^2 \equiv dx_{\mu}^2. \ee
%%%%%%
The negative sign divides spacetime into three space-like dimensions and
one time-like dimension ($3+1$ dimensions). Effects of this division
include the length contraction and time dilation of measurements on
co-moving frames. Dynamical equations for particles come about by
extremizing the length of their "worldline", the distance traveled
through spacetime. 

Differential geometry extends this notion of invariant distance
to curved manifolds by allowing the coefficients of the coordinate
quadratics to vary as functions of spacetime:
%%%%%%
\be ds^2=g_{\mu \nu}(x^{\mu})dx^{\mu}dx^{\nu}, \ee
%%%%%%
where $g_{\mu \nu}(x^{\mu})$ is known as the metric and
parameterizes the geometry of spacetime. General Relativity
makes the conceptual leap of equating the geometry of spacetime
with gravitation. Field equations describe the interplay between
the dynamical geometry of spacetime and the motion of matter from
an action principle. That is, particles minimize their path
through the geometry of spacetime, while the geometry changes as a
result of mass-energy content and motion.

\subsection{Higher-dimensional General Relativity}

Many mathematical generalizations of General Relativity have been
explored in the hopes of including non-gravitational forces in a
purely geometrical formulation. Most early attempts, including
those made by Einstein himself, were somewhat of a 'top-down
approach', in which it was hoped that macroscopic physics could
shed light on microscopic phenomena, via the dynamics of a
classical geometric manifold. This is not surprising, considering
that quantum mechanics was still new and somewhat ad hoc at the
time. A strong motivation for Einstein's search of a unified field
theory was, in fact, to rid our fundamental description of nature of
quantum indeterminism. There is a touch of irony, then, that
quantum gravity is currently seen by many as a guiding light
towards unification.

One of the first attempts at classically extending General
Relativity was made by Kaluza a mere four years after the
publication of Einstein's theory. Classical theory offers no constraints
on the dimensionality of spacetime. While it is with absolute certainty
that we observe $3+1$ dimensions, there may be additional dimensions
which have yet to be detected, existing on compact submanifolds of a
larger dimensional spacetime. Kaluza originally considered a fifth
dimension that is "compactified" on a circle of radius $L$, where $L$ is
below the current observable length scales. The corresponding
five-dimensional metric can be written in the form
%%%%%%%%%%%
\be ds_5^2=e^{2\alpha \phi}ds_4^2+e^{2\beta \phi}(dz+{\cal A})^2,
\ee
%%%%%%%%%%%
where $\alpha$ and $\beta$ are non-zero constants which can be
chosen for convenience.

The geometric degrees of freedom within $ds_4^2$, the observable
portion of spacetime, describe four-dimensional gravity independently
of the fifth dimension. In addition, ${\cal A}=A_{\mu}dx^{\mu}$ is an
additional degree of freedom called the Kaluza-Klein vector
potential, which corresponds to the electromagnetic potential in
the original setup and has a field strength given by ${\cal F}=d 
{\cal A}$.  The warping factor is given by the scalar field $\phi$
\footnote{In the original Kaluza-Klein setup, $\phi$ was regarded as a
somewhat embarrassing extra degree of freedom and was set to
zero. However, in order for the higher-dimensional equations of
motion to be satisfied provided that those of the lower dimension
are satisfied, $\phi$ must be retained. Current higher-dimensional
theories refer to $\phi$ as the 'dilaton', which carries
information on the conformal structure of spacetime.}.

While this extension of Relativity does succeed in classically 
describing gravity and electromagnetism as the result of the curvature
of a five-dimensional spacetime, on its own it was not believed to offer
any predictions, which is why Einstein waited for two years before
publicly supporting it\footnote{Five-dimensional Relativity does
predict gravito-electromagnetic waves, which oscillate between
gravitational and electromagnetic modes.}.

During the next half-century, there was much progress in quantum
physics, and the "top-down" approach to unification was abandoned by
most physicists. It is rumored, however, that even on his deathbed
Albert Einstein asked for a pencil and paper, in the hope that with his
remaining last minutes he may stumble upon a glimmer of a unified field
theory. 

\subsection{Quantum divergences}

Quantum field theories had fantastic predictive power
and successfully described electromagnetism, strong and weak
nuclear forces in a unified framework known as the Standard Model.
The idea of force-carrying particles was a key constituent of
fundamental physics. 'Bottom-up' approaches were attempted to
incorporate gravity into the Standard Model, with force-carrying
particles called 'gravitons'. However, a quantum field theoretic
approach to gravity has not been successful because the
gravitational coupling parameter is too large for gravitational
interactions to be renormalizable. That is, interactions cannot be
dissected into a series of virtual processes that lead to finite
results. Thus, practically every attempt to formulate a unified
field theory which includes the covariance of General Relativity,
or general coordinate transformation invariance, and the quantum
mechanics of the Standard Model have led to inconsistencies.

\subsection{Supersymmetry}

At this point, it is relevant to note that almost all known
fundamental theories have underlying symmetry principles.
Classical mechanics has space and time invariances and space
rotational invariance. Special Relativity adds space-time
rotational invariance to these symmetries. General Relativity
localizes the above invariances, which means that the degree of
translation and rotation may vary as functions of space and time. On
the other hand, electromagnetism is based on abelian gauge
invariance, which is an arbitrariness in the electromagnetic field
potentials corresponding to local invariance of the action with respect
to wave function rotations in a complex plane. The Standard Model
is based on non-abelian gauge invariance, in which the order of
field transformations matters.

For quite a number of decades, quantum mechanics did not seem to be
based on any symmetry principle, which may partially explain the
difficulties in unifying the covariance of General Relativity with
quantum physics. However, quantum mechanics introduces a new concept for
particles: bosons (integer-spin) and fermions (half-integer spin).
This leads to the possibility of a new symmetry, called
'supersymmetry' (SUSY), which is the invariance of the
theory under interchange of bosonic and fermionic particles. The
SUSY transformation acting twice over turns a particle back into
itself, but it can be located at a different point in spacetime.
Thus, one would expect that the corresponding superalgebra is
closely related to the Poincar\'{e} (spacetime rotations plus
translations) algebra. In fact, the anti-commutators of the
supercharges $Q$ which generate SUSY transformations have the
structure
%%%%%%%%
\be \{ Q,Q\} =\Gamma^{\mu}P_{\mu}, \ee
%%%%%%%%%
where the presence of the four-momentum operator $P_{\mu}$ implies
Poincar\'{e} invariance. Local supersymmetry, for which the
supersymmetric transformations depend on 
\\ spacetime, is known as 'supergravity', and automatically includes
general coordinate invariance. In other words, supergravity requires
Einstein's General Relativity; even though quantum mechanics is
incompatible with gravity, it predicts the existence of gravity!

\subsection{Classical string theory}

A revolutionary idea, initially manifesting itself as an unsuccessful
attempt to understand hadronic interactions, is that the basic
constituents of matter are $1+1$-dimensional objects called
"strings". Analogous to particles, strings obey an action
principle. That is, the area of their worldsheet, the
two-dimensional surface swept out by the string as it moves
through spacetime, is extremized.

The vast number of particles seen in nature are hypothesized to
correspond to various excitations of strings. In much the same way
as a musical string on a violoncello or guitar produces notes at
different frequencies, the discrete vibrations of a string
correspond to a whole spectrum of particles. Thus, interactions
between particles, which on the level of quantum field theory is
described by the exchange of virtual particles, are described in
string theory by the splitting and joining of strings. This path
to unification certainly exemplifies the "bottom-up" approach, in
that a drastic change to the sub-microscopic nature of matter
potentially results in the unification of all forces. It seems
logical that, given only one fundamental building block of matter,
there can only be one type of interaction. 

While the Standard Model has about twenty free parameters, which are
assigned values through experiments rather than theoretical
consistency, string theory has none, except perhaps for the string
tension. However, we have not yet described how the physics of
gravitation and quantum field theory arise from this rather simple
notion of strings.

\subsection{Quantizing the string}

The first step towards endeavor of deriving General Relativity and
particle physics from a common fundamental source lies within the 
quantization of the classical string action. At a given momentum,
quantized strings exist only at discrete energy levels, each level
containing a finite number of string states, or particle
types. There are huge energy gaps between each level, which means that
the directly observable particles belong to a small subset of string
vibrations. In principle, a string has harmonic frequency modes ad
infinitum. However, the masses of the corresponding particles get
larger, and decay to lighter particles all the quicker \cite{polch}.

Most importantly, the ground energy state of the string contains a
massless, spin-two particle. There are no higher spin particles,
which is fortunate since their presence would ruin the consistency
of the theory. The presence of a massless spin-two particle is
undesirable if string theory has the limited goal of explaining
hadronic interactions. However, previous attempts at a quantum
field theoretic description of gravity had shown that the
force-carrier of gravity, known as the graviton, had to be a
massless, spin-two particle. Thus, in string theory's comeback as a
potential "theory of everything", a curse turns into a blessing.

Once again, as with the case of supersymmetry and supergravity, we have
the astonishing result that quantum considerations require the
existence of gravity! From this vantage point, right from the start the
quantum divergences of gravity are swept away by the extended string. 
Rather than being mutually exclusive, as it seems
at first sight, quantum physics and gravitation have a symbiotic
relationship. This reinforces the idea that quantum gravity is a
mandatory step towards the unification of all forces.

\subsection{Supersymmetry makes a second entrance}

Unfortunately, the ground state energy level also includes
negative-mass particles, known as tachyons. Such particles have
light speed as their limiting minimum speed, thus violating
causality. Tachyonic particles generally suggest an instability,
or possibly even an inconsistency, in a theory. Since tachyons
have negative mass, an interaction involving finite input energy
could result in particles of arbitrarily high energies together
with arbitrarily many tachyons. There is no limit to the number of
such processes, thus preventing a perturbative understanding of
the theory.

An additional problem is that the string states only include
bosonic particles. However, it is known that nature certainly
contains fermions, such as electrons and quarks. Since
supersymmetry is the invariance of a theory under the interchange
of bosons and fermions, it may come as no surprise, post priori,
that this is the key to resolving the second issue. As it turns
out, the bosonic sector of the theory corresponds to the spacetime
coordinates of a string, from the point of view of the conformal
field theory living on the string worldvolume. This means that
the additional fields are fermionic, so that the particle spectrum can
potentially include all observable particles. In addition, the
lowest energy level of a supersymmetric string is naturally
massless, which eliminates the unwanted tachyons from the theory
\cite{polch}.

The inclusion of supersymmetry has some additional bonuses.
Firstly, supersymmetry enforces the cancellation of zero-point
energies between the bosonic and fermionic sectors. Since gravity
couples to all energy, if these zero-point energies were not
canceled, as in the case of non-supersymmetric particle physics,
then they would have an enormous contribution to the cosmological
constant. This would disagree with the observed cosmological
constant being essentially zero relative to the energy scales of
particle physics.

Also, the weak, strong and electromagnetic couplings of the
Standard Model differ by several orders of magnitude at low
energies. However, at high energies, the couplings take on almost
the same value-- almost but not quite. It turns out that a
supersymmetric extension of the Standard Model appears to render
the values of the couplings identical at approximately $10^{16}$
GeV. This may be the manifestation of the fundamental unity of
forces.

It would appear that the "bottom-up" approach to unification is
winning. That is, gravitation arises from the quantization of
strings. To put it another way, supergravity is the low-energy
limit of string theory, and has General Relativity as its own
low-energy limit.

\subsection{The dimension of spacetime}

String theory not only predicts the particle spectrum and
interactions of nature, but the dimension of spacetime itself! In
the process of quantizing the superstring, certain symmetries of
the action are lost unless the number of spacetime dimensions is
ten. Initially, since this is grossly inconsistent with every-day
observations, this seems to indicate that we should look elsewhere
for a "theory of everything." However, almost a century ago,
Kaluza and Klein had explored the possibility of extra small
dimensions. At the time, this idea seemed to be an unnecessary
extension of General Relativity with no additional predictions.

Thus, the "top-down" approach to unification that once motivated
higher dimensional extensions of General Relativity enters the arena of 
fundamental physics once again, in which the gauge fields of
the Standard Model are the result of ripples in a
higher-dimensional spacetime.

In the original Kaluza-Klein Relativity, there was only one extra
compact dimension. This didn't leave much topological freedom
\footnote{A slightly different possibility that was initially
explored by Horava and Witten is that the one extra dimension can
be compactified on an $S^1/Z_2$ orbifold, which leads an embedding
of ten-dimensional $E_8 \times E_8$ heterotic string theory in
eleven-dimensional M-theory \cite{hor}.}, since the extra dimension can
only be
curled up into a circle $S^1$. However, in this modern version of
Kaluza-Klein Relativity, six extra compact dimensions are
required. Greater number of compact dimensions brings the
possibility of more complicated topologies. For example, with two
extra dimensions, the compact part of the space could have the
topology of a sphere, a torus, or any higher genus surface.

Higher dimensions are not simply unwanted additions that are
required for theoretical consistency. They do have phenomenological
implications. For instance, compactifying on a three-dimensional complex
space known as Calabi-Yau manifold leads, in the low-energy limit, to a
four-dimensional ${\cal N}=1$ supersymmetric theory, the effective
field theory that underlies many supersymmetric theories of particle
phenomenology. Moreover, the geometrical and topological properties of
the extra dimensions determine the number of particle generations, the
particle species in each generation and the low-energy Lagrangian. Thus,
determining the shape of the extra dimensions is crucial for
understanding the low-energy predictions of string theory.

Unfortunately, there are thousands of Calabi-Yau manifolds to choose
from. It is hoped that the dynamics of string theory constrain the
possible shapes of the compact dimensions.

\subsection{"Who ordered the extra theories of everything?"}

Another rather unsettling issue is that there appeared
to be not one but five seemingly different but mathematically consistent
superstring theories: the $E_8 \times E_8$ heterotic string, the $SO(32)$
heterotic string, the $SO(32)$ Type I string, and Types IIA and IIB
strings. Each of these theories corresponded to a different way in
which fermionic degrees of freedom could be added to the string
worldsheet. 

Even before looking at the low-energy physics predicted by a particular
compactification of string theory, it is crucial to choose {\it
which} of the five string theories one is discussing!

\section{M-Theory and Non-Perturbative Solitons}

\subsection{M-Theory}

Superstrings provided a perturbatively finite theory of gravity which,
after compactification down to $3+1$ dimensions, seemed potentially
capable of explaining the strong, weak and electromagnetic forces of the
Standard Model, including the required chiral representations of quarks
and leptons. However, in addition to there being five different
superstring theories, many important questions seemed incapable of being
answered within the framework of the weak-coupling perturbation
expansion for which these theories were probed: 

How do strings break supersymmetry? 

How do strings choose the correct vacuum state? 

How do strings explain the smallness of the cosmological constant? 

How do strings supply a microscopic description of black holes?

Also, supersymmetry constrains the upper limit on the number of
spacetime dimensions to be eleven. Why, then, do superstring theories
stop at ten? In fact, before the "first string revolution" of the
mid-1980's, many physicists sought super-unification in 
eleven-dimensional supergravity. Solutions to this most primitive
supergravity theory include the elementary supermembrane and its dual
partner, the solitonic superfivebrane. These are supersymmetric objects
extended over two and five spatial dimensions, respectively. This brings
to mind another question: why do superstring theories generalize
zero-dimensional point particles only to one-dimensional strings, rather
than $p$-dimensional objects?

During the "second superstring revolution" of the mid-nineties it was
found that, in addition to the $1+1$-dimensional string solutions,
string theory contains soliton-like Dirichlet branes \cite{polch}. These
D$p$-branes
have $p+1$-dimensional worldvolumes, which are hyperplanes in
$9+1$-dimensional spacetime on which strings are allowed to end. If a
closed string collides with a D-brane, it can turn into an open string
whose ends move along the D-brane. The end points of such an open string
satisfy conventional free (Neumann) boundary conditions along the
worldvolume of the D-brane, and fixed (Dirichlet) boundary conditions
are obeyed in the $9-p$ dimensions transverse to the D-brane \cite{polch}.

D-branes make it possible to probe string theories
non-perturbatively, i.e., when the interactions are no longer
assumed to be weak. This more complete picture makes it evident
that the different string theories are actually related via a
network of "dualities." $T$-dualities relate two different string
theories by interchanging winding modes and Kaluza-Klein states, via $R
\rightarrow \alpha^{\prime}/R$. For example, Type IIA string theory
compactified on a circle of radius $R$ is equivalent to Type IIB string
theory compactified on a circle of radius $1/R$. We have a similar
relation between $E_8 \times E_8$ and $SO(32)$ heterotic string theories.
While T-dualities remain manifest at weak-coupling, S-dualities are less
well-established strong/weak coupling relationships. For example, the
$SO(32)$ heterotic string is believed to be S-dual to the $SO(32)$ Type
I string, while the Type IIB string is self-S-dual\footnote{There is a
duality of dualities, in which the T-dual of one theory is the S-dual of
another.} \footnote{Compactification on various manifolds often leads to
dualities. The heterotic string compactified on a six-dimensional torus
$T^6$ is believed to be self-S-dual. Also, the heterotic string on $T^4$
is dual to the type II string on four-dimensional K3. The heterotic
string on $T^6$ is dual to the Type II string on a Calabi-Yau
manifold. The Type IIA string on a Calabi-Yau manifold is dual to the
Type IIB string on the mirror Calabi-Yau manifold.} \cite{polch}.  

This led to the discovery that all five string theories are actually
different sectors of an eleven-dimensional non-perturbative theory, known
as M-theory. When M-theory is compactified on a circle $S^1$ of radius
$R_{11}$, it leads to the Type IIA string, with string coupling constant
$g_s=R_{11}^{3/2}$. Thus, the illusion that this string theory is
ten-dimensional is a remnant of weak-coupling perturbative
methods. Similarly, if M-theory is compactified on a line segment
$S^1/Z_2$, then the $E_8 \times E_8$ heterotic string is recovered. As
previously mentioned, a string (no pun intended) of dualities relates
each of these string theories to all of the rest.

Just as a given string theory has a corresponding supergravity in its
low-energy limit, eleven-dimensional supergravity is the low-energy limit
of M-theory. Since we do not yet know what the full M-theory actually is,
many different names have been attributed to the "M", including Magical,
Mystery, Matrix, and Membrane! In this thesis, whenever we refer to
"M-theory", we mean the theory which subsumes all five string theories and
whose low-energy limit is eleven-dimensional supergravity.

We now have an adequate framework with which to understand a wealth
of non-perturbative phenomena. For example, electric-magnetic duality in
$D=4$ is a consequence of string-string duality in $D=6$, which in turn is
the result of membrane-fivebrane duality in $D=11$. Furthermore, the exact 
electric-magnetic duality has been extended to an effective duality of
non-conformal $\cal N$ $=2$ Seiberg-Witten theory\footnote{Seiberg-Witten 
theory has led to insights on quark confinement. However, this relies on
an, at present, unphysical supersymmetric QCD.}, which can be derived
from M-theory. In fact, it seems that all supersymmetric quantum field
theories with any gauge group could have a geometrical interpretation
through M-theory, as worldvolume fields propagating on a common
intersection of stacks of $p$-branes wrapped around various cycles of
compactified manifolds.

In addition, while perturbative string theory has vacuum degeneracy
problems due to the billions of Calabi-Yau vacua, the non-perturbative
effects of M-theory lead to smooth transitions from one Calabi-Yau
manifold to another\footnote{Now the question to ask is not why do we
live in one topology but rather why do we live in a particular corner of
the unique topology. M-theory might offer a dynamical explanation of 
this.}.

While supersymmetry ensures that the high-energy values of the Standard
Model coupling constants to meet at a common value, which is consistent
with the idea of grand unification, the gravitational coupling constant
just misses this meeting point. However, a particular compactification
of M-theory envisioned by Horava and Witten, in which the Standard Model
fields live on a four-dimensional spacetime while gravity propagates in
five dimensions, allows the size of the fifth dimension to be chosen so
that the gravitational coupling constant meets the other three at high
energy. In fact, this may occur at much less energy than the
originally-thought $10^{19}$ GeV, which leads to various interesting 
cosmological effects. 

In fact, M-theory may resolve long-standing cosmological and quantum
gravitational problems. For example, M-theory accounts for a microscopic 
description of black holes by supplying the necessary non-perturbative
components, namely $p$-branes. As Chapter 1 will show in more detail,
this solves a the problem of counting black hole entropy by internal
degrees of freedom.

\subsection{Black hole thermodynamics}

Since particle accelerators cannot currently probe the energy
scale of string theory \footnote{Supersymmetry, which is an
essential component for consistent string theories, may be tested
as soon as 2009, with the search for supersymmetric partners of
known particles.}, we hope that cosmology and astrophysics will
provide a testing ground. In particular, black holes provide a
theoretical test that low-energy string theory agrees with the
predictions of quantum field theory in curved spacetime. During
the 1970's, it was found that the entropy of black holes is
given by
%%%%%%%%%
\be S=\frac{A}{4G}, \ee
%%%%%%%%
where $A$ is the surface area of the event horizon and $G$ is Newton's
gravitational constant. However, the microscopic meaning of black hole
entropy, in terms of counting the degrees of freedom, was far from
clear; it was believed that a black hole did not possess any degrees of
freedom other than its energy, charges and angular momenta.

Understanding of the microscopics of black holes came with the
proposition that certain types of black holes are actually made up of
collections of D-branes. With this model, a detailed microscopic
derivation of black hole entropy was provided by counting the ways in
which branes could be configured to form a black hole.

\subsection{The Holographic Principle}

As shown above, black hole entropy scales like the surface area of 
the event horizon rather than the volume within. This contradicts our
naive intuition about the extensivity of thermodynamic entropy which we
have gained from quantum field theories. This motivated 't Hooft and
Susskind to conjecture that the degrees of freedom describing the system
are characterized by a quantum field theory with one fewer space
dimensions.

We have previously discussed how supergravity, in which all fields are the
result of fluctuations in the spacetime geometry, represents a
"top-down" approach to unification. On the other hand, the excitations of
the string yields the particle spectrum of quantum field theory, which is 
a "bottom-up" approach. The duality of these descriptions is embodied
within the Holography Principle 

In 1997, Maldacena found a specific and precise example of this
idea for the case of Anti-de Sitter spacetimes, now known as the
"AdS/CFT correspondence" \cite{malda}. The notion that certain
gravitational descriptions are dual to quantum field theories may provide
understanding of how these descriptions can be unified.

In particular, AdS/CFT implies that the weakly-coupled gravity
theory is dual to the strongly-coupled super-conformal field
theory. This allows us to probe the quantum field theory side
non-perturbatively via supergravity. Anti-de Sitter spacetimes
arise as the near-horizon regions of certain $p$-branes, which 
indicates a dual CFT on the brane's worldvolume.

\section{Summary of this thesis}

While AdS$_d$ for dimensionality $d \le 3$ is fairly common as
near-horizon solutions in both string theories and M-theory, 
higher-dimensional AdS spaces are not. For example, AdS$_5 \times S^5$
arises as the near-horizon region of D3-branes in type IIB string
theory. The dual quantum field theory is ${\cal N}=4$
supersymmetric four-dimensional U(N) gauge theory.

However, for some time, $AdS_5$ did not have a similar known
origin in eleven-dimensional supergravity. If M-theory is the
long-sought after ultimate theory of nature, then it is crucial to
understand how our observable world is embedded within its
low-energy limit. Thus, $AdS_5$ arising as a near-horizon region of a
brane configuration within eleven-dimensional supergravity would present
us with an ideal case for which to explore four-dimensional dual field
theories.

I describe how examples of AdS in warped spacetimes can arise in the
near-horizon geometry of semi-localized, multi-intersections of
$p$-branes, including the case of warped AdS$_5$ in eleven-dimensional
supergravity.

One of the initial motivations for the AdS/CFT correspondence came
from absorption cross-section calculations of fields on the
background curved spacetimes of $p$-branes. That is, the dual
field theory provides a precise mechanism for the absorption of
fields into black holes composed of $p$-branes, in agreement with
classical results. The absorption cross-sections in supergravity
backgrounds yields information on correlation functions of corresponding
operators of the strongly-coupled dual gauge theory. However,
absorption calculations have mostly been done for very low energy.
Higher-energy calculations yield information on the dual gauge
theory away from the conformal fixed point, in a regime for which
conformal symmetry is broken.

I calculate exact absorption calculations for various
fields on $p$-brane spacetimes. In addition, absorption
calculations are performed for near-extremal $p$-branes, thereby
probing the worldvolume field theory at finite temperature. In
the midst of AdS/CFT, however, it should not be forgotten that
absorption calculations shed light on the nature of supergravity
solutions themselves. In particular, it is found that the
absorption cross-section has a remarkably simple oscillatory
pattern with respect to the energy of the scattered field.

I discuss the possibility that extra dimensions may be unseen and yet
infinite in extent, as opposed to the Kaluza-Klein idea of compact
dimensions. The observable dimensions of our universe may be on a
$p$-brane embedded in a higher-dimensional space. While open strings
corresponding to gauge fields are constrained to move along the brane,
for a long time  it was believed that all closed strings corresponding
to gravitons were free to move within the dimensions transverse to
the brane, which disagrees with the observed four-dimensional law of
gravity. Randall and Sundrum recently found that, in the case of
warped higher-dimensional spaces such as AdS, there is a bound
massless graviton. This means that, even for an extra dimension of
infinite extent, four-dimensional gravity is reproduced in the
low-energy limit. 

I explore the localization of gravity in the case of a brane in a
background B field, which is of interest since the dual gauge theory is
on noncommutative space. It is found that the localized graviton is
massive. Randall and Karch have recently found that, for the case of
negative cosmological constant, there is a bound massive graviton.
Thus, it would appear that, in the context of localized gravity
within Randall-Sundrum brane worlds, a non-commutative gauge
theory mimics the effect of a non-vanishing cosmological constant!

%% file: thesiswarped.tex
\chapter{Warped AdS}

\section{Introduction}

\subsection{"QCD strings"}

String theory began from attempts to formulate a theory of
hadronic interactions. After quantum chromodynamics (QCD) entered the
theoretical arena, string theory was abandoned, only to be upgraded
shortly afterwards as a candidate for a theory of all fundamental
interactions of the universe \cite{wsg}. 

Recent developments demonstrate that at least some strongly-coupled
gauge theories have a dual string description. That is, a 
dual string theory provides exact information about certain gauge
theories at strong coupling, a regime which was intractable by
previously known perturbative methods. The most established examples of
this duality are super-conformal gauge theories which are dual to
supergravity theories on Anti-de Sitter spacetime \cite{malda}.

The idea of dual string theories has been around for quite some time. 
During the 1970's, 't Hooft proposed to generalize the $SU(3)$ gauge
group of QCD to $SU(N)$ and take the large $N$ limit while keeping
$g_{YM}^2N$ fixed. In this limit, the sum over Feynman graphs of a given
topology can be regarded as the sum over world sheets of a hypothetical
"QCD string." The closed string coupling constant goes as $N^{-1}$, so
that in the large $N$ limit we have a weakly-coupled string theory. The
spectrum of these free closed strings is the same as that of glueballs in
large $N$ QCD, while open strings can describe mesons. If a method
is developed to calculate this spectra and they are found to be discrete,
then this would be an elegant explanation of quark confinement.

After much work done in search of an exact gauge field/string duality, it
was speculated that such strings may actually live in five
dimensions. In 1997, Maldacena made the AdS/CFT conjecture, in which conformal
field theories are dual to supergravities with Anti-de Sitter spacetimes,
which have constant negative curvature. In particular, "QCD strings" might
actually be Type IIB superstrings living in five non-compact
($AdS_5$) and five compact ($X^5$) dimensions, where $X^5$ is a
positively curved Einstein space \cite{klebTASI}.

We will now give a brief description of the intriguing discoveries which
led up to the Holographic Principle, a particular example of
which is the celebrated AdS/CFT conjecture.

\subsection{Dp-branes and Black Holes}

As previously mentioned, black hole entropy scales as the surface area
of the event horizon, rather than the volume within. This motivated 't
Hooft and Susskind to conjecture that the degrees of freedom describing
the system are characterized by a quantum field theory of one fewer
spatial dimensions \cite{peet}.

It had been believed that the state of a black hole was completely
specified by its energy, charges and angular momenta. This left no
internal degrees of freedom to be counted by the entropy, thus
making the meaning of this thermodynamic quantity rather illusive.

The resolution of this puzzle came about through the study of Dp-branes,
soliton-like solutions of Types IIA and IIB supergravity which
carry Ramond-Ramond charge and are specified by the metric
%%%%%%
\be 
ds^2=H^{-1/2}(r)\big( -f(r)dt^2+dx_1^2+..+dx_p^2\big)
+H^{1/2}(r)(f(r)^{-1}dr^2+r^2d\Omega_{8-p}^2),
\ee
%%%%%%
and a dilaton
%%%%%%%
\be 
e^{\phi}=H^{(3-p)/4}(r), 
\ee
%%%%%%%%
where the harmonic function is given by
%%%%%%
\be 
H(r)=1+\frac{R^{7-p}}{r^{7-p}}. 
\ee
%%%%%%%%
Also,
%%%%%%%%
\be 
f(r)=1-\frac{r_o^{7-p}}{r^{7-p}}, 
\ee
%%%%%%%%
where $r_o$ is the nonextremality parameter. In the extremal case,
when $r_o$ vanishes, the mass saturates the lower (BPS) bound for
a given choice of charges. The coordinates $t,x_i$ describe the brane
worldvolume while $r$ and the $8-p$-sphere coordinates describe the space
transverse to the brane.

Polchinski illuminated the importance of these solutions with his
discovery that Dp-branes are the fundamental objects in string
theory which carry RR charges \cite{polch}. These p-brane supergravity
solutions were then identified with the long-range background
fields produced by stacks of parallel Dp-branes, for which the
constant $R$ is given by
%%%%%%%%%
\be R^{7-p}=\alpha^{\prime (7-p)/2}g_s N (4\pi
)^{(5-p)/2}\Gamma\big( \frac{7-p}{2}\big) , \ee
%%%%%%%%
where $N$ is the number of coincident Dp-branes.

More complicated systems were considered, in which D-branes of
different dimensionalities were intersecting and possibly wrapped
on compact manifolds. It was found that such configurations could
be identified with black holes and black p-brane solutions of
supergravity with the appropriate charges. This provides the
microscopic description of black holes that was needed to count
the degrees of freedom that is parametrized by entropy.

Strominger and Vafa were the first to build this type of
correspondence between black hole solutions and D-branes \cite{sv}. In
particular, they began with an intersection of D1 and D5-branes in
ten-dimensional type IIB theory. After compactification on a
five-dimensional manifold, this brane configuration corresponds
to a five-dimensional black hole carrying two separate $U(1)$
charges. This is a generalization of the four-dimensional
Reissner-Nordstrom black hole.

It was now possible to understand black hole entropy in terms of
the degrees of freedom living on the D-branes. Strominger and Vafa
calculated the Bekenstein-Hawking entropy as a function of the
charges and successfully reproduced the result provided by
macroscopic entropy \cite{sv} \footnote{The extremal Strominger-Vafa black
hole preserves 1/8 of the supersymmetries present in the vacuum,
which ensures that the number of states does not change as the
coupling is increased to the regime in which the D-brane
configuration corresponds with a black hole. This correspondence
was quickly generalized to near-extremal solutions.}.

As mentioned in the introduction of this thesis, the association
of black hole entropy with area would imply that black holes have
a temperature. In the classical regime this is nonsensical, since
nothing can escape from inside the event horizon of a black hole.
Thus, Hawking's proposal that black holes radiate energy was
initially met with ridicule. However, Hawking used a
semi-classical calculation, in which quantum processes occur
within a fixed, classical backdrop of spacetime, allowing energy
to quantum tunnel through the event horizon. Since such processes
involve the annihilation of the the original particles by their
anti-particle partners, this brings about the issue as to whether
information is lost. It is hoped that the D-brane description of
black holes can be used to resolve this question.

Black hole radiation implies that scattering processes of black
holes have grey-body factors. This is reflected in the absorption
of particles, which can be pictured in two different ways. In the
semi-classical approach, a particle is attracted by long-range
gravity of the black hole, tunnels through an effective potential
barrier, and is absorbed by the event horizon. In the D-brane
picture, the incident particle travels through flat space and
decays into a number of new particles which are constrained to
live on the brane intersection.

The thermodynamics and scattering processes of black holes seem to
indicate that there is an underlying Holographic Principle at
work-- a deep equivalence between gauge theories and gravity.
Since it would appear that the roots of gravity and quantum field
theory are incompatible on the level of particle physics, it is
quite surprising that they are, in some ways, equivalent on the
deeper level of string theory. However, a set of precise examples
of this notion was not found until 1997.

\subsection{The AdS/CFT Correspondence}

The celebrated AdS/CFT conjecture of Maldacena states, in its most
familiar form, that string theory in the near-horizon geometry of
a large number of $N$ coincident D3-branes ($AdS_5 \times S^5$) is
completely equivalent to the low-energy $U(N)$ $\cal N$ $=4$
Super-Yang-Mills (SYM) gauge theory in four dimensions, which describes
the excitations on the brane \cite{malda}. 

To be more specific, consider a stack of N parallel D3-branes in the
"zero slope limit," for which $\alpha^{\prime} \rightarrow 0$ and the
masses of all massive string modes go to infinity. In this limit, the
gravitational coupling $\kappa \sim g_s \alpha^{\prime} \rightarrow 0$,
so that the bulk closed string modes decouple from the massless open
string modes on the brane. In addition, all higher derivative terms of the
worldvolume fields vanish in the action. Thus, dynamics on the brane is
completely described by the low-energy theory of the massless open string
modes, the $U(N)$ $\cal N$ $=4$ Super-Yang-Mills theory in $3+1$
dimensions. 

In this limit, the supergravity metric becomes that of flat
spacetime. That is, the open string modes on the brane decouple from
the bulk closed string modes; from the point of view of a distant
observer, the brane disappears from the geometry. Closed strings propogate
in the decoupled flat spacetime of the bulk while decoupled open strings
obey  a non-trivial field theory on the brane. 

On the field theory side, the masses of the lowest energy level
of strings stretched between two D-branes separated by a distance $r$ are
given by $u R^2=r/\alpha^{\prime}$. In the decoupling limit, these are
important degrees of freedom. In order to keep their energies finite, we
keep $u$ fixed as we take the limit $\alpha^{\prime} \rightarrow 0$. 

Now we consider what happens to the supergravity metric in this limit.
The metric for a stack of extremal D3-branes is
%%%%%%%%
\be
ds^2=H^{-1/2} (-dt^2+dx_1^2+dx_2^2+dx_3^2)+H^{1/2}(dr^2+r^2d\Omega_5^2).
\ee
%%%%%%%%
The decoupling limit $\alpha^{\prime} \rightarrow 0$ with fixed $u$
corresponds to the "near-horizon" limit $r \ll R$, for which the constant
"$1$" in the harmonic function $H$ can be neglected. The metric becomes
%%%%%%%
\be
ds^2=\alpha^{\prime} R^2 \Big(
u^2(-dt^2+dx_1^2+dx_2^2+dx_3^2)+\frac{du^2}{u^2}+d\Omega _5^2
\Big) ,
\ee
%%%%%%%%
which describes the space $AdS_5 \times S^5$. Note that $\alpha^{\prime}$
becomes a constant overall factor in this limit. By the change of
coordinates $z=1/u$, this metric can be expressed as
%%%%%%%%
\be
ds^2=\alpha^{\prime} R^2 \Big(
\frac{1}{z^2}(-dt^2+dx_1^2+dx_2^2+dx_3^2+dz^2)+d\Omega_5^2 \Big),
\ee
%%%%%%%%
which corresponds to the conformally-flat frame in the five-dimensional
supergravity after a dimensional-reduction over $S^5$. This property of
$AdS$ spaces is also reflected in the constancy of the dilaton
field. Thus, branes whose spacetimes approach $AdS$ in the near-horizon
region are non-dilatonic, such as the D3-brane of type IIB string
theory.

The AdS/CFT conjecture makes the claim that the full string theory on the
near-horizon geometry of $AdS_5 \times S^5$ is dual to the
four-dimensional $\cal N$ $=4$ Super-Yang-Mills gauge theory on the
brane. Note that the string coupling constant is related to the Yang-Mills
coupling constant by $g_s=g_{YM}^2$. This can be understood qualitatively
as two open strings, each of which has a corresponding factor of
$g_{YM}$, coming together to form a closed string \cite{wsg}.

The radius of curvature of the space in string units is $\sqrt{4\pi
g_s N}$, which means that the supergravity solution can only be trusted
for $1 \ll g_s N$. This corresponds to strong 't Hooft coupling in the
gauge theory. In the limit $g_s N \rightarrow \infty$ and $g_s \rightarrow
0$, both string loops and stringy $\alpha^{\prime}$ effects may be
neglected. In order to have these limits simultaneously, we require that
$N \rightarrow \infty$. Thus, classical supergravity on $AdS_5 \times S^5$
should be dual to the large $N$ limit of $U(N)$, $\cal N$ $=4$ SYM theory
in the limit of strong 't Hooft coupling \cite{klebTASI}.

A stronger form of the AdS/CFT conjecture asserts that for $N \rightarrow
\infty$ and $1 \ll g_s N$ but finite, so that $g_s \rightarrow 0$
still, $1/(g_s N)$ corrections on the field theory side correspond to
classical string theoretic corrections to supergravity. In the strongest
version of the AdS/CFT conjecture, we may take $N$ to be finite, and
$1/N$ corrections in the field theory correspond to string loop effects.

The AdS/CFT Conjecture is extremely powerful, even in its weakest form.
It provides a way to probe gauge theory at the previously inaccessible
regime of strong 't Hooft coupling via classical supergravity
calculations. 

Since the AdS/CFT correspondence has to do with a gauge theory at
strong coupling, calculations tend to be difficult on one side of
the duality and relatively simple on the other side. Thus, this
correspondence is both incredibly useful and rather difficult to check
directly. However, the symmetries of the theory should be independent of
the parameters and may be compared directly.

On the supergravity side, the bosonic symmetry group includes
SO(4,2), the isometry group of $AdS_5$, as well as the global
symmetry group SO(6), the rotational symmetry of $S^5$. These groups match
with the bosonic global symmetry group on the field theory side, where
SO(4,2) is the conformal group in $3+1$ dimensions and $SO(6)$ is locally
SU(4), which is the R-symmetry group of ${\cal N}=4$ SYM gauge theory. In
addition, both theories have 32 fermionic global symmetry generators,
which combine with the bosonic symmetries to yield a supergroup
SU(2,2$|$4) on both sides \cite{klebTASI}.

In addition, there should be an exact correspondence between operators in
the field theory and particle states in $AdS_5 \times S^5$. As we shall
discuss in the next chapter on absorption by branes, explicit calculations
of correlation functions show complete agreement between the supergravity
and field theory sides of the AdS/CFT correspondence.

\subsection{Other Examples of the Holographic Principle}

As already mentioned, the AdS/CFT correspondence is one particular example
of the Holographic principle. Another example which is discussed in
Chapter 4 of this thesis is that of branes in a background B field. It is
believed that, in a particular decoupling limit, the near-horizon physics
of a brane in a background B field corresponds to a gauge theory in
non-commutative geometry \cite{itzhaki,maldacena}. Also, it is believed
that a non-dilatonic brane
excited above extremality corresponds to a gauge theory in finite
temperature \cite{malda}. Also, separation of branes in bulk space
corresponds to the
Coloumb Branch of the dual non-conformal gauge theory, in which
certain scalar fields have nonzero expectation values \cite{costa}.

The duality between type IIB strings on $AdS_5 \times S^5$ and the $\cal
N$ $=4$ SYM gauge theory is naturally generalized to dualities between
backgrounds of the form $AdS_5 \times X^5$ and other conformal gauge
theories. The five-dimensional compact space $X^5$ is required to be an
Einstein manifold of positive curvature, i.e., one for which $R_{\mu
\nu}=\Lambda g_{\mu \nu}$ with $\Lambda > 0$. The number of isometries on
$X^5$ determine the number of supersymmetries in the dual gauge theory
\cite{klebTASI}.

A simple example of $X^5$ which is locally equivalent to $S^5$ is the
orbifold-type $S^5/\Gamma$, where $\Gamma$ is a discrete subgroup of 
$SO(6)$. The dual gauge theory is the IR limit of the world-volume theory
on a stack of $N$ D3-branes placed at the conifold singularity of 
$R^6/\Gamma$ \cite{klebTASI}.

An example for which $X^5$ is not locally equivalent to $S^5$ is the
Romans compactification on $X^5=T^{1,1}=\big( SU(2) \times
SU(2)\big) /U(1)$. The dual gauge theory is the conformal limit of the
world-volume theory on a stack of $N$ D3-branes placed at the singularity
of a certain Calabi-Yau manifold known as the conifold \cite{klebTASI}
\footnote{Recently, Klebanov and Strassler have constructed an exact
non-singular supergravity solution of a D3-brane on a conifold that is
deformed by an additional three-form flux. This is an example of the
resolution of a naked singularity at the level of supergravity, rather
than relying on stringy effects. The infrared phenomena of the dual
non-conformal $\cal N$ $=1$ supersymmetric gauge theory include
confinement, chiral symmetry breaking, a glueball spectrum, baryons and
domain walls separating inequivalent vacua \cite{klebstrass}. Another
mechanism for resolving the naked singularity involves going away from
extremality. This corresponds to a finite-temperature gauge theory in
which chiral symmetry is restored above a critical temperature. Exploring
this transition to deconfinement is of particular current interest
\cite{klebnon}.}.

We now turn to the possibility of warped AdS spaces, that are of the form
$f(\alpha) (AdS_m \times S^n)$, where $\alpha$ is an angular parameter of
the $S^n$ portion. The warping factor $f(\alpha)$ reduces the isometries
of the geometry, which implies that the dual gauge theory has reduced 
supersymmetries. Thus, such warped AdS solutions are of interest as
gravity duals of strongly-coupled field theories with less than maximal
supersymmetry. In addition, $AdS_d$ solutions arise in various theories
for particular values of $d$ only in warped spacetimes. A particular
example of this is warped $AdS_5$ in eleven-dimensional supergravity
\cite{oz,warped}. 

\subsection{Warped AdS}

       Anti-de Sitter (AdS) spacetimes naturally arise as the
near-horizon geometries of non-dilatonic $p$-branes in supergravity
theories.  The metric for such a solution is usually the direct sum of
AdS and an internal sphere.  These geometries are of particular
interest because of the conjecture that supergravity on such a
background is dual to a conformal field theory on the boundary of the
AdS \cite{malda,gkp,wit}.  Examples include all the anti-de Sitter
spacetimes AdS$_d$ with $2\le d\le 7$, with the exception of $d=6$.
The origin of AdS$_6$ is a little more involved, and it was first
suggested in \cite{fkpz} that it was related to the ten-dimensional
massive type IIA theory.  Recently, it was shown that the massive type
IIA theory admits a warped-product solution of AdS$_6$ with $S^4$
\cite{bo}, which turns out to be the near-horizon geometry of a
semi-localised D4/D8 brane intersection \cite{youm}.  It is important
that the warp factors depend only on the internal $S^4$ coordinates,
since this implies that the reduced theory in $D=6$ has AdS spacetime
as its vacuum solution.  The consistent embedding of $D=6$, $N=1$
gauged supergravity in massive type IIA supergravity was obtained in
\cite{d6gauge}.  Ellipsoidal distributions of the D4/D8 system were
also obtained, giving rise to AdS domain walls in $D=6$, supported by
a scalar potential involving 3 scalars \cite{dist}.

   In fact, configurations with AdS in a warped spacetime are not rare
occurrences.  In \cite{oz}, a semi-localised M5/M5 system \cite{youm}
was studied, and it was shown that the near-horizon geometry turns out
to be a warped product of AdS$_5$ with an internal 6-space.  This
makes it possible to study AdS$_5$/CFT$_4$ from the point of view of
M-theory.  In this paper, we shall consider AdS with a warped
spacetime in a more general context and obtain such geometries for
all the AdS$_d$, as the near-horizon limits of semi-localised multiple
intersections in both type IIA and type IIB theories.

   The possibility of this construction is based on the following
observations.  As is well known, a non-dilatonic $p$-brane has the
near-horizon geometry AdS$_d \times S^n$. The internal $n$-sphere can
be described geometrically as a foliation of $S^p\times S^q$ surfaces
with $n=p+q+1$, and so, in particular, if $n\ge 4$
the $n$-sphere can be viewed in terms of a foliation with $S^3\times
S^{n-4}$ surfaces, {\it viz.}
%%%%%
\be
d\Omega_n^2 = d\a^2 + \cos^2\a\, d\Omega_3^2 + \sin^2\a\,
d\Omega_{n-4}^2\,.\label{folia}
\ee
%%%%%
When a non-dilatonic $p$-brane with an
$n$-sphere in the transverse space intersects with a Kaluza-Klein
monopole (a Taub-NUT with charge $\Qn$) in a semi-localised manner,
the net result turns out to be effectively a coordinate transformation
of a solution with a distribution of pure $p$-branes with no NUT
present.  The round $S^3$ in (\ref{folia}) becomes the cyclic lens
space $S^3/Z_{\Qn}$ with metric
%%%%%
\be
d\bOmega_3^2 = \ft14 d\Omega_2^2 + \ft14 (\fft{dy}{\Qn} + \omega)^2\,,
\label{lens}
\ee
%%%%%
where $d\omega=\Omega_2$ is the volume form of the unit 2-sphere.
This metric retains the same local structure as the standard round
3-sphere, and it has the same curvature tensor, but the $y$ coordinate
on the $U(1)$ fibres is now identified with a period which is $1/\Qn$
of the period for $S^3$ itself.  We can now perform a dimensional
reduction, or a T-duality transformation, on the fibre coordinate $y$,
and thereby obtain AdS in a warped spacetime.  The warp factor depends
only on the internal ``latitude'' coordinate $\a$, but is independent
of the lower-dimensional spacetime coordinates.  In fact, the M5/M5
system with AdS$_5$ found in \cite{oz} can be obtained in precisely
such a manner from the D3-brane by using type IIA/IIB T-duality.  Note
that an isotropic $p$-brane can be viewed as carrying a single unit of
NUT charge.  Although this semi-localised way of introducing a
Taub-NUT seems trivial, in that it amounts to a coordinate
transformation, performing Kaluza-Klein reduction on the fibre
coordinate does create a non-trivial intersecting component, since the
Kaluza-Klein 2-form field strength now carries a non-trivial
flux. This fact was used in \cite{create} to construct multi-charge
$p$-branes starting from flat spacetime.

    An analogous procedure can instead be applied to the anti-de
Sitter spacetime, rather than the sphere, in the near-horizon limit
AdS$_d\times S^n$ of a non-dilatonic $p$-brane. AdS$_d$ can be described in 
terms of a foliation of AdS$_p\times S^q$ surfaces with $d=p+q+1$ and so,
in particular, for $d\ge 4$ it can be expressed as a foliation of
AdS$_3\times S^{d-3}$:
%%%%%
\be
ds_{\rm{AdS}_d}^2 = d\rho^2 + \cosh^2\rho\, ds_{\rm{AdS}_3}^2 +
\sinh^2\rho\, d\Omega_{d-4}^2\,.
\ee
%%%%%%
In the presence of a pp-wave that is semi-localised on the
world-volume of the $p$-brane, the AdS$_3$ turns out to have the
form of a $U(1)$ bundle over AdS$_2$ \cite{s3twist},
%%%%%
\be
ds_{\rm{AdS_3}}^2 = -r^2\,W^{-1}\, dt^2 + \fft{dr^2}{r^2} + r^2\,
W\, (dy + (W^{-1}-1)) dt)^2\,,\label{ads3a}
\ee
%%%%
where $W=1 + Q_w/r^2$, and $Q_w$ is the momentum carried by the
pp-wave.  This is precisely the structure of the extremal BTZ black
hole \cite{btz}.  We can now perform a Kaluza-Klein reduction, or
T-duality transformation, on the fibre coordinate $y$.  In the
near-horizon limit where the ``1'' in $W$ can be dropped, we obtain
AdS$_2$ in a warped spacetime with a warp factor that depends only on
the foliation coordinate, $\rho$.

    A T-duality transformation on such a fibre coordinate of AdS$_3$
or $S^3$ has been called Hopf T-duality \cite{s5twist}.  It has the
effect of (un)twisting the AdS$_3$ or $S^3$.  The effect of this
procedure on the six-dimensional dyonic string, whose near-horizon
limit is AdS$_3\times S^3$, was extensively studied in \cite{s3twist}.
We apply the same technique to AdS$_3$ or $S^3$ geometries that are
themselves factors in the foliation surfaces of certain larger-dimensional
AdS spacetimes or spheres.

We consider the semi-localised D3/NUT system and
show that the effect of turning on the NUT charge $\Qn$ in the
intersection is merely to convert the internal 5-sphere, viewed
as a foliation of $S^1\times S^3$, into a foliation of $S^1\times
(S^3/Z_{\Qn})$, where $S^3/Z_{\Qn}$ is the cyclic lens space of order
$\Qn$.  We can then perform a T-duality transformation on the Hopf
fibre coordinate of the lens space and thereby obtain an AdS$_5$ in a
warped spacetime as a solution in M theory, as the near-horizon
geometry of a semi-localised M5/M5 system.

We consider a semi-localised D3/pp-wave system,
for which the AdS$_5$ becomes a foliation of a circle with the
extremal BTZ black hole, which is locally AdS$_3$ and can be viewed as
a $U(1)$ bundle over AdS$_2$.  We then perform a Hopf T-duality
transformation on the fibre coordinate to obtain a solution with
AdS$_2$ in a warped spacetime in M-theory, as the near-horizon
geometry of a semi-localised M2/M2 system.

We apply the same analysis to the M2/NUT
and M2/pp-wave systems, and the M5/NUT and M5/pp-wave systems,
respectively; we obtain various configurations of AdS in warped
spacetimes by performing Kaluza-Klein reductions and Hopf T-duality
transformations on the fibre coordinates.

We consider the D4/D8 system, which has the
near-horizon geometry of a warped product of AdS$_6$ and $S^4$. We
perform a Hopf T-duality transformation on the fibre coordinate of
the foliating lens space of $S^4$, and thereby embed AdS$_6$ in a
warped spacetime solution of type IIB theory.

\section{D3/NUT systems and AdS$_5$ in M-theory from T-duality}

      AdS$_5$ spacetime arises naturally from type IIB theory as the
near-horizon geometry of the D3-brane.  Its origin in M-theory is more
obscure.  One way to embed the AdS$_5$ in M-theory is to note that
$S^5$ can be viewed as a $U(1)$ bundle over $CP^2$, and hence we can
perform a Hopf T-duality transformation on the $U(1)$ fibre
coordinate.  The resulting M-theory solution becomes AdS$_5\times
CP^2\times T^2$ \cite{s5twist}.  However, this solution is not
supersymmetric at the level of supergravity, since $CP^2$ does not
admit a spin structure.  {\it Charged} spinors exist but, after making
the T-duality transformation, the relevant electromagnetic field is
described by the winding-mode vector and it is only in the full
string theory that states charged with respect to this field arise. It
was therefore argued in \cite{s5twist} that the lack of supersymmetry
(and indeed of any fermions at all) is a supergravity artifact and
that, when the full string theory is considered, the geometry is
supersymmetric.  Such a phenomenon was referred as ``supersymmetry
without supersymmetry'' in \cite{s7twist}.

        Recently, AdS$_5$ in warped eleven-dimensional spacetime was
constructed in \cite{oz}.  It arises as the near-horizon limit of the
semi-localised M5/M5 intersecting system. After performing a T-duality
transformation, the warped spacetime of the near-horizon limit becomes
AdS$_5\times (S^5/Z_{\Qn})$. In this section, we shall review this
example in detail and show that the M5/M5 system originates from a
semi-localised D3/NUT intersection in type IIB supergravity.

\subsection{D3/NUT system}

        Any $p$-brane with a transverse space of sufficiently high
dimension can intersect with a NUT.  The D3/NUT solution of type IIB
supergravity is given by
%%%%%%%%
\bea
ds_{10\rm{IIB}}^2 &=& H^{-1/2}(-dt^2 + dw_1^2 + \cdots + dw_3^2) +
H^{1/2}\Big(dx_1^2 + dx_2^2\nn\\
&&\phantom{xxxxxx} K(dz^2+ z^2\, d\Omega_2^2) + K^{-1}(dy +
\Qn\, \omega)^2 \Big)\,,\label{d3nut}\\
F_5&=&dt\wedge d^3w\wedge dH^{-1} + {*(dt\wedge d^3w\wedge dH^{-1})}
\,,\nn
\eea
%%%%%%
where $z^2 = z_1^2 + z_2^2 + z_3^2$, and $\omega$ is a 1-form
satisfying $d\omega = \Omega_2$.  The solution can be best
illustrated by the following diagram:

\bigskip\bigskip
\centerline{
\begin{tabular}{c|cccccccccccc}
&$t$ & $w_1$ & $w_2$ & $w_3$ & $x_1$ & $x_2$ & $z_1$ & $z_2$ &
$z_3$ & $y$ & \\ \hline
D3&$\times$ & $\times$ & $\times$ & $\times$ &$-$ &$-$ &$-$ &$-$ &$-$
&$-$ & $H$ \\
NUT&$\times$ & $\times$ & $\times$ & $\times$ & $\times$ & $\times$
&$-$ &$-$ &$-$ &$*$ & $K$ \\
\end{tabular}}
\bigskip

\centerline{Diagram 1.  The D3/NUT brane intersection.  Here
$\times$ and $-$ denote the}
\centerline{$\phantom{Diagram 1. }$
worldvolume and transverse space coordinates respectively,}
\centerline{$\phantom{xxxx}$ and $*$ denotes the fibre coordinate
of the Taub-NUT.}
\bigskip\bigskip

        The function $K$ is associated with the NUT component of the
intersection; it is a harmonic function in the overall transverse
Euclidean 3-space coordinatised by $z_i$.  The function $H$ is
associated with the D3-brane component.  It satisfies the equation
%%%%%
\be
\del_{\vec z}^2 H + K\, \del_{\vec x}^2 H=0\,.
\ee
%%%%%
Equations of this type were also studied in
\cite{cvet1,cvet2,tseyt1,horo,tseyt2,lpint,ity,02128,02210,03038}.
In the absence of NUT charge, {\it i.e.} $K=1$, the function $H$ is
harmonic in the the transverse 6-space of the D3-brane.  When the NUT
charge $\Qn$ is non-zero, $K$ is instead given by
%%%%%
\be
K=1 + \fft{\Qn}{z}\,,\label{knut}
\ee
%%%%%
and the function $H$ cannot be solved analytically, but only in terms
of a Fourier expansion in $\vec x$ coordinates.  The usual way to
solve for the solution is to consider the zero-modes in the Fourier
expansion.  In other words, one assumes that $H$ is independent of
$\vec x$.  The consequence of this assumption is that the resulting
metric no longer has an AdS structure in its near-horizon region.  In
\cite{youm}, it was observed that an explicit closed-form solution for
$H$ can be obtained in the case where the ``1'' in function K is
dropped.  This solution is given by \cite{youm}
%%%%%%
\be
K = \fft{\Qn}{z}\,,\qquad H = 1 + \sum_k\fft{Q_k}{(|\vec x - \vec
x_{0k}|^2 + 4\Qn\, z)^2} \,.
\ee
%%%%%
In this paper, we shall consider the case where the D3-brane is
located at the origin of the $\vec x$ space and so we have
%%%%%%%
\be
H = 1 + \fft{Q}{(x^2 + 4 \Qn\, z)^2}\,,
\ee
%%%%%%
where $x^2 = x^i \,x^i$.  Thus, the D3-brane is also localised in the
space of the $\vec x$ as well.  Let us now make a coordinate
transformation
%%%%
\be
x_1=r\, \cos\a\, \cos\theta\,,\quad
x_2=r\, \cos\a\, \sin\theta\,,
\quad z = \ft14 \Qn^{-1}\, r^2\, \sin^2\a\,.
\label{d2nutvar}
\ee
%%%%
In terms of the new coordinates, the metric for the solution becomes
%%%%%%
\bea
ds_{10\rm{IIB}}^2 &=& H^{-1/2}(-dt^2 + dw_1^2 + dw_2^2 + dw_3^2) +
H^{1/2}(dr^2 + r^2\, dM_5^2)\,,\nn\\
H&=&1 + \fft{Q}{r^4}\,.
\eea
%%%%%%
where
%%%
\be
dM_5^2 = d\a^2 + c^2\,d\theta^2 + \ft14s^2\Big(d\Omega_2^2
+(\fft{dy}{\Qn} + \omega)^2\Big)\,,
\ee
%%%%
and $s=\sin\a$, $c=\cos\a$.  Thus, we see that $dM_5^2$ describes a
foliation of $S^1$ times the lens space $S^3/Z_{\Qn}$.  For a unit NUT
charge, $\Qn=1$, the metric $dM_5^2$ describes the round 5-sphere and
the solution becomes an isotropic D3-brane.  It is interesting to note
that the regular D3-brane can be viewed as a semi-localised D3-brane
intersecting with a NUT with unit charge.\footnote{An analogous
observation was also made in \cite{create}, where multi-charge
solutions were obtained from flat space by making use of the fact that
$S^3$ can be viewed as a $U(1)$ bundle over $S^2$.  In other words,
flat space can be viewed as a NUT, with unit charge, located on the
$U(1)$ coordinate.}  In the near-horizon limit $r\rightarrow 0$, where
the constant 1 in the function $H$ can be dropped, the metric becomes
AdS$_5\times M_5$:
%%%%%
\bea
ds_{10\rm{IIB}}^2 &=& Q^{-1/2}\, r^2\, (-dt^2 + dw^i dw^i) +
Q^{1/2}\fft{dr^2}{r^2} +\\ & & Q^{1/2}\Big(d\a^2 + c^2\,d\theta^2 +
\ft14s^2(d\Omega_2^2
+(\fft{dy}{\Qn} + \omega)^2)\Big) \label{d3nuthorizon}
\eea
%%%%%

\subsection{M5/M5 system and AdS$_5$ in M-theory}

      Since the near-horizon limit of a semi-localised D3-brane/NUT is
a direct product of AdS$_5$ and an internal 5-sphere that is a
foliation of a circle times a lens space, it follows that if we
perform a T-duality transformation on the $U(1)$ fibre coordinate $y$,
we shall obtain AdS$_5$ in a warped spacetime as a solution of the
type IIA theory.  The warp factor is associated with the scale factor
$s^2$ of $dy^2$ in (\ref{d3nuthorizon}). This type of Hopf T-duality
has the effect of untwisting a 3-sphere into $S^2\times S^1$
\cite{s3twist}.  If one performs the T-duality transformation on the
original full solution (\ref{d3nut}), rather than concentrating on its
near-horizon limit, then one obtains a semi-localised NS5/D4 system of
the type IIA theory, which can be further lifted back to $D=11$ to
become a semi-localised M5/M5 system, obtained in \cite{youm}.  In
\cite{oz}, the near-horizon structures of these semi-localised branes
of M-theory were analysed, and AdS$_5$ was obtained as a warped
spacetime solution.  We refer the readers to Ref.~\cite{oz} and shall
not discuss this solution further, but only mention that, from the
above analysis, it can be obtained by implementing the T-duality
transformation on the coordinate $y$ in (\ref{d3nuthorizon}).

\section{D3/pp-wave system}

\subsection{D3/pp-wave system and extremal BTZ black hole}

       In this section, we study the semi-localised pp-wave
intersecting with a D3-brane.  The solution is given by
%%%%%%
\bea
ds_{10\rm{IIB}}^2 &=& H^{-1/2}\Big(-W^{-1}\, dt^2 +
W\,(dy + (W^{-1}-1)dt)^2 +dx_1^2 + dx_2^2\Big)\nn\\
&&+ H^{1/2} (dz_1^2 + \cdots dz_6^2)\,,\label{d3wave}\\
F_\5&=&dt\wedge dy\wedge dx_1\wedge dx_2\wedge dH^{-1} +
{*(dt\wedge dy\wedge dx_1\wedge dx_2\wedge dH^{-1})}\,,\nn
\eea
%%%%%%%%
The solution can be illustrated by the following diagram

\bigskip\bigskip
\centerline{
\begin{tabular}{c|cccccccccccc}
&$t$ & $y$ & $x_1$ & $x_2$ & $z_1$ & $z_2$ & $z_3$ & $z_4$ &
$z_5$ & $z_6$ & \\ \hline
D3&$\times$ & $\times$ & $\times$ & $\times$ &$-$ &$-$ &$-$ &$-$ &$-$
&$-$ & $H$ \\
wave&$\times$ & $\sim$ & $-$ & $-$ & $-$ & $-$
&$-$ &$-$ &$-$ &$-$ & $W$ \\
\end{tabular}}
\bigskip

\centerline{Diagram 2.  The D3/pp-wave brane intersection.  Here
$\sim$ denotes the wave coordinate.}

\bigskip\bigskip

    In the usual construction of such an intersection, the harmonic
functions $H$ and $W$ depend only on the overall transverse space
coordinates $\vec z$.  The near-horizon limit of the solution then
becomes K$_5\times S^5$, where K$_5$ is the generalised Kaigorodov
metric in $D=5$, and the geometry is dual to a conformal field theory
in the infinite momentum frame \cite{kaig}.  On the other hand, the
semi-localised solution is given by \cite{youm}
%%%%%%%%
\be
H=\fft{Q}{|\vec z|^4}\,,\qquad
W= 1 + Q_w (|\vec x|^2 + \fft{Q}{|\vec z|^2})\,.
\ee
%%%%%
We now let
%%%%
\be
x_1 = \fft{1}{r}\,\cos\a\, \cos\theta\,,\quad
x_2 = \fft{1}{r}\,\cos\a\, \sin\theta\,,\qquad
z_i = \fft{r\, Q^{1/2}}{\sin\a}\, \nu_i\,,\label{d3wavenewcord}
\ee
%%%%%%
where $\nu_i$ coordinates, satisfying $\nu_i\,\nu_i=1$, define a
5-sphere with the unit sphere metric $d\Omega_5^2 = d\nu_i\, d\nu_i$.
Using these coordinates, the metric of the semi-localised D3/wave
system becomes
%%%%%%%%%
\be
ds_{10\rm{IIB}}^2 = Q^{1/2}\, s^{-2}\, (ds_{\rm{AdS}_3}^2 + d\a^2 +
c^2\, d\theta^2 +s^2\, d\Omega_5^2)\,,\label{d3wavehorizon}
\ee
%%%%%
where $ds_{\rm{AdS}_3}^2$ is given by
%%%%%
\bea
ds_{\rm{AdS}_3}^2 &=& -r^2\, W^{-1}\, dt^2 + r^2\,W\, (dy + (W^{-1}
-1)dt)^2 + \fft{dr^2}{r^2}\,,\nn\\
W&=& 1+ \fft{Q_w}{r^2}\,.\label{ads3}
\eea
%%%%%
Note that the above metric is exactly the extremal BTZ black hole
\cite{btz}, and hence it is locally AdS$_3$.  Thus we have
demonstrated that the semi-localised D3/pp-wave system is in fact a
warped product of AdS$_3$ (the extremal BTZ black hole) with a
7-sphere, where $S^7$ is described as a foliation of $S^1\times S^5$
surfaces.\footnote{A D3-brane with an $S^3\times \R$ worlvolume was
obtained in \cite{papadop}.  In that solution, which was rather
different from ours, the dilaton was not constant.}  Note that the
metric (\ref{d3wavehorizon}) can also be expressed as a direct product
of AdS$_5\times S^5$, with the AdS$_5$ metric written in the following
form:
%%%%%
\be
ds_5^2 = s^{-2}(ds_{\rm{AdS}_3}^2 + d\a^2 + c^2\, d\theta^2)\,.
\label{ads5a}
\ee
%%%%%
Making a coordinate transformation $\tan(\a/2) = e^\rho$, the metric
becomes
%%%%%
\be
ds_5^2 = d\rho^2 + \sinh^2\rho\, d\theta^2 +
\cosh^2\rho\, ds_{\rm{AdS}_3}^2\,,
\ee
which is precisely the AdS$_5$ metric written as a foliation of a
circle times AdS$_3$\footnote{Since this metric can be expressed as
warped $AdS_3$ or $AdS_5$, at first sight it appears that there may
be two dual gauge theories corresponding to this geometry. However, this
may involve a topological change in the boundary.}.

    The extremal BTZ black hole occurs \cite{ss} as the near-horizon
geometry of the boosted dyonic string in six-dimensions, which can be
viewed as an intersection of a string and a 5-brane in $D=10$.  The
boosted D1/D5 system was used to obtain the first stringy
interpretation \cite{sv} of the microscopic entropy of the
Reissner-Nordstr\"om black hole in $D=5$.  The boosted dyonic string
has three parameters, namely the electric and magnetic charges $Q_e$,
$Q_m$, and the boost momentum parameter $Q_w$.  On the other hand, the
extremal BTZ black hole itself has only two parameters: the
cosmological constant, proportional to $\sqrt{Q_e\, Q_m}$, and the
mass (which is equal to the angular momentum in the extremal limit),
which is related to $Q_w$.  (Analogous discussion applies to $D=4$
\cite{cl}.)  In our construction of the BTZ black hole in warped
spacetime, the original configuration also has only two parameters,
namely the D3-brane charge $Q$, related to the cosmological constant
of the BTZ black hole, and the pp-wave charge, associated with the
mass.

\subsection{NS1/D2 and M2/M2 systems and AdS$_2$}

          We can perform a T-duality transformation on the coordinate
$y$ in the previous solution.  The D3-brane is T-dual to the D2-brane,
and the wave is T-dual to the NS-NS string.  Thus the D3/pp-wave
system of the type IIB theory becomes an NS1/D2 system in the type IIA
theory, given by
%%%%%%
\bea
ds_{10\rm{IIA}}^2 &=& W^{1/4}H^{3/8} \Big[-(WH)^{-1}\, dt^2 +
H^{-1}\,(dx_1^2+ dx_2^2) + W^{-1}\,dy_1^2\,,\nn\\
&&\phantom{xxxxxxxxxxx} + dz_1^2 + \cdots dz_6^2\Big]\,,\nn\\
e^{\phi} &=& W^{-1/2}\, H^{1/4}\,,\label{ns1d2}\\
F_\4 &=& dt\wedge dx_1\wedge dx_2\wedge dH^{-1}\,,\qquad
F_\3 = dt\wedge dy_1\wedge dW^{-1}\,.\nn
\eea
%%%%%
This solution can be represented diagrammatically as follows:

\bigskip\bigskip
\centerline{
\begin{tabular}{c|ccccccccccc}
&$t$ & $x_1$ & $x_2$ & $y_1$ & $z_1$ & $z_2$ & $z_3$ & $z_4$ &
$z_5$ & $z_6$ & \\ \hline
D2&$\times$ & $\times$ & $\times$ & $-$ &$-$ &$-$ &$-$ &$-$ &$-$
&$-$ & $H$ \\
NS1&$\times$ & $-$ & $-$ & $\times$ & $-$ &$-$ &$-$ &$-$ &$-$
&$-$ & $W$ \\
\end{tabular}}
\bigskip

\centerline{Diagram 3.  The NS1/D2 brane intersection.}
\bigskip\bigskip

          In the near-horizon limit where the 1 in $W$ is
dropped, the metric of the NS1/D2 system (\ref{ns1d2}), in terms of
the new coordinates (\ref{d3wavenewcord}), becomes
%%%%%
\be
ds_{10}^2 =Q_w^{1/4}Q^{5/8}\,s^{-5/2}\, \Big(ds_{\rm{AdS}_2}^2 +
d\a^2 + c^2\, d\theta^2 + s^2\, d\Omega_5^2
+ (Q_w\,Q)^{-1}\,s^4\, dy_1^2\,\Big)\,,
\ee
%%%%
where
%%%%%%
\be
ds_{\rm{AdS}_2}^2 =-\fft{r^4\, dt^2}{Q_w} +\fft{dr^2}{r^2}\,.
\label{ads2}
\ee
%%%%%%%
Thus we see that the near-horizon limit of the NS1/D2 system is a
warped product of AdS$_2$ with a certain internal 8-space, which is a
warped product of a 7-sphere with a circle.

            We can further lift the solution back to $D=11$, where it
becomes a semi-localised M2/M2 system,
%%%%%
\bea
ds_{11}^2 &=& (WH)^{1/3}\Big[-(WH)^{-1}\, dt^2 +
H^{-1}\,(dx_1^2 + dx_2^2) + W^{-1}\,(dy_1^2 + dy_2^2)\,,\nn\\
&&\phantom{xxxxxxxxxxx} + dz_1^2 + \cdots +dz_6^2\Big]\,,\nn\\
F_\4 &=& dt\wedge dx_1\wedge dx_2\wedge dH^{-1} +
       dt\wedge dy_1\wedge dy_2\wedge dW^{-1}\,.\label{m2m2}
\eea
%%%%%
The configuration for this solution can be summarised in the following
diagram:

\bigskip\bigskip
\centerline{
\begin{tabular}{c|cccccccccccc}
&$t$ & $x_1$ & $x_2$ & $y_1$ & $y_2$ & $z_1$ & $z_2$ & $z_3$ & $z_4$ &
$z_5$ & $z_6$ & \\ \hline
M2&$\times$ & $\times$ & $\times$ & $-$ &$-$ &$-$ &$-$ &$-$ &$-$ &$-$
&$-$ & $H$ \\
M2&$\times$ & $-$ & $-$ & $\times$ & $\times$ & $-$ &$-$ &$-$ &$-$ &$-$
&$-$ & $W$ \\
\end{tabular}}
\bigskip

\centerline{Diagram 4.  The M2-M2 brane intersection.}
\bigskip\bigskip

   It is straightforward to verify that the near-horizon geometry of
this system is a warped product of AdS$_2$ with a certain 9-space,
namely
%%%%%
\be
ds_{11}^2 = Q_w^{1/3}Q^{2/3}\,s^{-8/3}\,
(ds_{\rm{AdS}_2}^2 + d\a^2 + c^2\, d\theta^2 + s^2\, d\Omega_5^2
+ (Q_w\,Q)^{-1}\,s^4\, (dy_1^2 + dy_2^2))\,,
\ee
%%%%
where $ds_{\rm{AdS}_2}^2$ is an AdS$_2$ metric given by (\ref{ads2}),
and the internal 9-space is a warped product of a 7-sphere and a
2-torus.

\subsection{Further possibilities}

       Note that in the above examples, we can replace the round
sphere $d\Omega_5^2$ by a lens space of the following form:
%%%%%
\be
d\Omega_5^2 = d\td \a^2 + \td c^2\, d\td \theta^2 + \td s^2
\, (d\wtd \Omega_2^2 + (\fft{d\td y}{\wtd Q_{\sst{\rm N}}} +
\td \omega)^2)\,,
\ee
%%%%%
where $\td c\equiv \cos\td \a$, $\td s\equiv \sin\td \a$ and $d\td
\omega = \wtd \Omega_2$.  This can
be viewed as an additional NUT with charge $\wtd Q_{\sst{\rm N}}$
intersecting with the system.  We can now perform a Kaluza-Klein
reduction or T-duality transformation on the fibre coordinate $\td y$,
leading to many further examples of warped products of AdS$_2$ or
AdS$_3$ with certain internal spaces.  The warp factors again depend
only on the coordinates of the internal space.  These geometries can
be viewed as the near-horizon limits of three intersecting branes,
with charges $Q$, $\Qn$ and $\wtd Q_{\sst{\rm N}}$.  Of course, this
system can equally well be obtained by replacing the horospherical
AdS$_5$ in (\ref{d3nuthorizon}) with (\ref{ads5a}).

       For example, let us consider the M2/M2 system with an
additional NUT component.  The solution of this semi-localised
intersecting system is given by
%%%%%
\bea
ds_{11}^2 &=& (WH)^{1/3}\Big[-(WH)^{-1}\, dt^2 +
H^{-1}\,(dx_1^2 + dx_2^2) + W^{-1}\,(dy_1^2 + dy_2^2)\,,\nn\\
&&\phantom{xxxxxxxxxxx} + K(dz^2 + z^2\, d\Omega_2^2) +
K^{-1}(dy + \Qn\,\omega)^2 + du_1^2 + du_2^2\Big]\,,\nn\\
F_\4 &=& dt\wedge dx_1\wedge dx_2\wedge dH^{-1} +
       dt\wedge dy_1\wedge dy_2\wedge dW^{-1}\,.\label{m2m2nut}
\eea
%%%%%
where the functions $H$, $W$ and $K$ are given by
%%%%%%
\be
H=\fft{Q}{(|\vec u|^2 + 4\Qn\, z)^2}\,,\quad
W=1 + Q_w(|\vec x|^2 + \fft{Q}{|\vec u|^2 + 4\Qn\, z})\,,\quad
K=\fft{\Qn}{z}\,.
\ee
%%%%%%
We illustrate this solution in the following diagram:

\bigskip\bigskip
\centerline{
\begin{tabular}{c|cccccccccccc}
&$t$ & $x_1$ & $x_2$ & $y_1$ & $y_2$ & $z_1$ & $z_2$ & $z_3$ & $y$ &
$u_1$ & $u_2$ & \\ \hline
M2&$\times$ & $\times$ & $\times$ & $-$ &$-$ &$-$ &$-$ &$-$ &$-$ &$-$
&$-$ & $H$ \\
M2&$\times$ & $-$ & $-$ & $\times$ & $\times$ & $-$ &$-$ &$-$ &$-$ &$-$
&$-$ & $W$ \\
NUT& $\times$ & $\times$ & $\times$ & $\times$ & $\times$ & $-$
& $-$ & $-$ & $*$ &$\times$ & $\times$ & $K$
\end{tabular}}
\bigskip

\centerline{Diagram 5.  The M2/M2/NUT brane intersection.}
\bigskip\bigskip

          The near-horizon structure of this solution is basically the
same as that of the M2/M2 system with the round $S^3$ in the foliation
replaced by the lens space $S^3/Z_{\Qn}$.  We can now perform
Kaluza-Klein reduction on the fibre coordinate $y$ and the solution
becomes the semi-localised D2/D2/D6 brane intersection, given by
%%%%%
\bea
ds_{10\rm{IIA}}^2 &=& (WH)^{3/8}K^{-1/8}\,\Big[-(WH)^{-1}\, dt^2 +
H^{-1}\,(dx_1^2 + dx_2^2) + W^{-1}\,(dy_1^2 + dy_2^2)\,,\nn\\
&&\phantom{xxxxxxxxxxx} + K(dz^2 + z^2\, d\Omega_2^2) +
du_1^2 + du_2^2\Big]\,,\nn\\
F_\4 &=& dt\wedge dx_1\wedge dx_2\wedge dH^{-1} +
       dt\wedge dy_1\wedge dy_2\wedge dW^{-1}\,.\label{d2d2d6}\\
e^{\phi}&=& (W\, H)^{1/4} K^{-3/4}\,,\qquad
F_\2 = \Qn\,\Omega_2\,.
\eea
%%%%%
The solution can be illustrated by the following diagram:

\centerline{
\begin{tabular}{c|ccccccccccc}
&$t$ & $x_1$ & $x_2$ & $y_1$ & $y_2$ & $z_1$ & $z_2$ & $z_3$ &
$u_1$ & $u_2$ & \\ \hline
D2&$\times$ & $\times$ & $\times$ & $-$ &$-$ &$-$ &$-$ &$-$ &$-$
&$-$ & $H$ \\
D2&$\times$ & $-$ & $-$ & $\times$ & $\times$ & $-$ &$-$ &$-$ &$-$
&$-$ & $W$ \\
D6& $\times$ & $\times$ & $\times$ & $\times$ & $\times$ & $-$
& $-$ & $-$ &$\times$ & $\times$ & $K$
\end{tabular}}
\bigskip

\centerline{Diagram 6.  The D2/D2/D6 brane intersection.}
\bigskip\bigskip

\section{M2/NUT and AdS$_4$ in type IIB from T-duality}

       In this section, we apply an analogous analysis to the
M2-brane.  We show that the semi-localised M2-brane intersecting
with a NUT is in fact an isotropic M2-brane with the internal
7-sphere itself being described as a foliation of a regular $S^3$
and lens space $S^3/Z_{\Qn}$, where $\Qn$ is the NUT charge.
Reducing the system to $D=10$, we obtain a semi-localised D2/D6
system whose near-horizon geometry is a warped product of AdS$_4$
with an internal 6-space.  We also show that a semi-localised
pp-wave intersecting with the M2-brane is in fact a warped product
of AdS$_3$ (the BTZ black hole) and an 8-space.  The system can be
reduced to $D=10$ to become a semi-localised D0/NS1 intersection.

\subsection{M2-brane/NUT system}

         The solution for the intersection of an M2-brane and a NUT is
given by
%%%%%%
\bea
ds_{11}^2 &=& H^{-2/3}\,(-dt^2 + dw_1^2 + dw_2^2) + H^{1/3}\Big(dx_1^2 +
\cdots + dx_4^2\nn\\
&&\phantom{xxxx}+ K(dz^2 + z^2 d\Omega_2^2) + K^{-1} (dy + \Qn\, \omega)^2
\Big)\,,\nn\\
F_\4 &=& dt\wedge dw_1\wedge dw_2\wedge dH^{-1}\,,\label{m2nut}
\eea
%%%%%%
where $z^2 = z_1^2 + z_2^2 + z_3^2$ and $d\omega = \Omega_2$.
The solution can be illustrated by the following diagram:

\bigskip\bigskip
\centerline{
\begin{tabular}{c|cccccccccccc}
&$t$ & $w_1$ & $w_2$ & $x_1$ & $x_2$ & $x_3$ & $x_4$ & $z_1$ & $z_2$ &
$z_3$ & $y$ & \\ \hline
M2&$\times$ & $\times$ & $\times$ & $-$ &$-$ &$-$ &$-$ &$-$ &$-$ &$-$
&$-$ & $H$ \\
NUT&$\times$ & $\times$ & $\times$ & $\times$ & $\times$ & $\times$
&$\times$ &$-$ &$-$ &$-$ &$*$ & $K$ \\
\end{tabular}}
\bigskip

\centerline{Diagram 7.  The M2/NUT brane intersection.}
\bigskip\bigskip

        If the function $K$ associated with the NUT components of the
intersection takes the form $K=\Qn/z$, then the function $H$
associated with the M2-brane component can be solved in the
semi-localised form
%%%%%%%
\be
H = 1 + \fft{Q}{(|\vec x|^2 + 4 \Qn\, z)^3}\,.
\ee
%%%%%%
Thus, the solution is also localised on the space of the $\vec x$
coordinates.  Let us now make a coordinate transformation
%%%%%
\be x_i=r\, \cos\a\, \mu_i,,\qquad z = \ft14 \Qn^{-1}\,
r^2\, \sin^2\a\,,
\label{m2nutvar}
\ee
%%%%
where $\mu_i\,\mu_i =1$, defining a 3-sphere, with the unit 3-sphere
metric given by $d\Omega_3^2 = d\mu_i\, d\mu_i$.  In terms of the new
coordinates, the metric for the solution becomes
%%%%%%%
\bea
ds_{11}^2 &=& H^{-2/3} (-dt^2 + dw_1^2 + dw_2^2) + H^{1/3}(dr^2 +
r^2\, dM_7^2)\,,\nn\\
H&=& 1 + \fft{Q}{r^6}\,,
\eea
%%%%%
where
%%%%%
\be
dM_7^2 = d\a^2 + c^2\, d\Omega_3^2 + \ft14 s^2\,
\Big(d\Omega_2^2 + (\fft{dy}{\Qn} + \omega)^2\Big)\,.
\ee
%%%%%
Thus we see that $dM_7^2$ is a foliation of a regular 3-sphere,
together with a lens space $S^3/Z_{\Qn}$.  When $\Qn=1$ the metric
$dM_7^2$ describes a round 7-sphere and the solution becomes an
isotropic M2-brane.  Interestingly, the regular M2-brane can be viewed
as an intersecting semi-localised M2-brane with a NUT of unit
charge. In the near-horizon limit $r\rightarrow 0$, where the 1 in the
function $H$ can be dropped, the metric becomes AdS$_4\times M_7$.

\subsection{D2-D6 system}

      In the M2-brane and NUT intersection (\ref{m2nut}), we can
perform a Kaluza-Klein reduction on the $y$ coordinate.  This gives
rise to a semi-localised intersection of D2-branes and D6-branes:
%%%%%
\bea
ds_{10\rm{IIA}}^2 &=& H^{-5/8} K^{-1/8}\, (-dt^2 + dw_1^2 + dw_2^2) +
H^{3/8} K^{-1/8}\, (dx_1^2 + \cdots + dx_4^2)\nn\\
&&\phantom{xxxxxx} H^{3/8} K^{7/8}\, (dz_1^2 + dz_2^2 + dz_3^2)
\,,\nn\\
e^{\phi} &=& H^{1/4} K^{-3/4}\,,\label{d2d6}\\
F_\4&=&dt\wedge d^2w\wedge dH^{-1}\,,\qquad
F_2=e^{-3/2\phi} {*(dt\wedge d^2w\wedge d^4x\wedge dK^{-1})}\,.\nn
\eea
%%%%%%
The solution can be illustrated by the following diagram

\centerline{
\begin{tabular}{c|ccccccccccc}
&$t$ & $w_1$ & $w_2$ & $x_1$ & $x_2$ & $x_3$ & $x_4$ & $z_1$ & $z_2$ &
$z_3$ &\\ \hline
D2&$\times$ & $\times$ & $\times$ & $-$ &$-$ &$-$ &$-$ &$-$ &$-$
&$-$ & $H$ \\
D6&$\times$ & $\times$ & $\times$ & $\times$ & $\times$ & $\times$
&$\times$ &$-$ &$-$ &$-$ & $K$ \\
\end{tabular}}
\bigskip

\centerline{Diagram 8. The D2/D6 brane intersection.}
\bigskip\bigskip

   Again, in the usual construction of a D2-D6 system, the harmonic
functions $H$ and $K$ are taken to depend only on the overall
transverse space coordinates $\vec z$. In the semi-localized
construction, the function $H$ depends on $\vec x$ as well.  In terms
of the new coordinates defined in (\ref{m2nutvar}), the metric becomes
%%%%%%%%
\be
ds_{10\rm{IIA}}^2 = (\fft{r\,s}{2\Qn})^{1/4}\Big[H^{-5/8}(-dt^2 + dw_1^2 +
dw_2^2) + H^{3/8}(dr^2 + r^2(d\a^2 + c^2\, d\Omega_3^2 + \ft14 s^2\,
d\Omega_2^2)\Big]\,.
\ee
%%%%%
Thus, in the near-horizon limit where the 1 in $H$ can be dropped, the
solution becomes a warped product of AdS$_4$ with an internal 6-space:
%%%%%
\be
ds_{10\rm{IIA}}^2 =(2\Qn)^{-1/4}Q^{3/8}\,s^{1/4}\,
(ds_{\rm{AdS}_4}^2 + d\a^2 + c^2\, d\Omega_3^2 +
\ft14 s^2\, d\Omega_2^2)\,,
\ee
%%%%%
where $ds_4^2$ is the metric on AdS$_4$, given by
%%%%%%
\be
ds_{\rm{AdS}_4}^2 = \fft{r^4}{Q}(-dt^2 + dw_1^2 +
dw_2^2) + \fft{dr^2}{r^2}\,.
\ee
%%%%%%
The internal 6-space is a warped product of a 4-sphere with a 2-sphere.

\subsection{AdS$_4$ in type IIB from T-duality}

         In the above discussion, we found that our starting point is
effectively to replace the round 7-sphere of the M2-brane by the
foliation of a round 3-sphere together with a lens space
$S^3/Z_{\Qn}$.  We can also replace the round 3-sphere by another lens
space $S^3/Z_{\wtd Q_{\sst{\rm N}}}$, given by
%%%%
\be
d\bOmega_3^2 = \ft14 \Big(d\wtd\Omega_2^2 +
(\fft{d\td y}{\wtd Q_{\sst{\rm N}}} + \omega)^2\Big)\,.
\ee
%%%%
As discussed in the appendix, the lens space arises from introducing a
NUT around the fibre coordinate $\td y$, with NUT charge $\wtd
Q_{\sst{\rm N}}$.  The system can then be viewed as the near-horizon
limit of three intersecting branes, with charges $Q$, $\Qn$ and $\wtd
Q_{\sst{\rm N}}$.  For example, with this replacement the D2/D6 system
becomes a D2/D6/NUT system. Performing a T-duality transformation on
the fibre coordinate $\td y$, the $S^3$ untwists to become $S^2\times
S^1$.  The resulting type IIB metric is given by
%%%%%%
\be
ds_{10\rm{IIB}}^2 =
\Big(\fft{Q\, s\, c}{4\Qn\,\wtd Q_{\sst{\rm N}}}\Big)^{1/2}\,
\Big(ds_{\rm{AdS}_4}^2 + d\a^2 + \ft14 c^2\, d\wtd\Omega_2^2 +
\ft14 s^2\,d\Omega_2^2 +
\fft{(4\Qn\,\wtd Q_{\sst{\rm N}})^2}{Q\, s^2\, c^2}\, d\td
y^2\Big)\,.\label{d3d5ns5horizon}
\ee
%%%%
This metric can be viewed as describing the near-horizon geometry of
a semi-localised D3/D5/NS5 system in the type IIB theory.  This metric
(\ref{d3d5ns5horizon}) provides a background for consistent reduction
of type IIB supergravity to give rise to four-dimensional gauged
supergravity with AdS background.

         In order to construct the semi-localised D3/D5/NS5
intersecting system in the type IIB theory, we start with the
D2/D6/NUT system, given by
%%%%%
\bea
ds_{10\rm{IIA}}^2 &=& H^{-5/8} K^{-1/8}\, (-dt^2 + dw_1^2 + dw_2^2) +
H^{3/8} K^{7/8}\, (dz_1^2 + dz_2^2 + dz_3^2)\nn\\
&&+H^{3/8} K^{-1/8}\, ( \wtd K\,(dx^2 + x^2\, d\wtd \Omega_2^2) +
\wtd K^{-1} (dy + \wtd Q_{\sst{\rm N}}\, \wtd \omega)^2)\,,\nn\\
e^{\phi} &=& H^{1/4} K^{-3/4}\,,\label{d2d6nut}\\
F_\4&=&dt\wedge d^2w\wedge dH^{-1}\,,\qquad
F_2=e^{-3/2\phi} {*(dt\wedge d^2w\wedge d^4x\wedge dK^{-1})}\,.\nn
\eea
%%%%%%
where $x^2 = x_1^2+ x_2^2 + x_3^2$ and the functions $H$, $K$ and
$\wtd K$ are given by
%%%%%
\be
H=1 + \fft{Q}{(4\wtd Q_{\sst{\rm N}}\, x + 4 \Qn\, z)^3}\,,\quad
K=\fft{\Qn}{z}\,,\quad \wtd K= \fft{\wtd Q_{\sst{\rm N}}}{x}\,.
\ee
%%%%%%
It is instructive to illustrate the solution in the following diagram:

\bigskip\bigskip
\centerline{
\begin{tabular}{c|ccccccccccc}
&$t$ & $w_1$ & $w_2$ & $x_1$ & $x_2$ & $x_3$ & $y$ & $z_1$ & $z_2$ &
$z_3$ &\\ \hline
D2&$\times$ & $\times$ & $\times$ & $-$ &$-$ &$-$ &$-$ &$-$ &$-$
&$-$ & $H$ \\
D6&$\times$ & $\times$ & $\times$ & $\times$ & $\times$ & $\times$
&$\times$ &$-$ &$-$ &$-$ & $K$ \\
NUT& $\times$ & $\times$ & $\times$ & $-$ & $-$ & $-$ & $*$ &
$\times$ & $\times$ & $\times$ & $\wtd K$
\end{tabular}}
\bigskip

\centerline{Diagram 9. The D2/D6/NUT system}
\bigskip\bigskip

We can now perform the T-duality on the coordinate $y$, and obtain the
semi-localised D3/D5/NS5 intersection of the type IIB theory, given by
%%%%%%%
\bea
ds_{10\rm{IIB}}^2 &=& H^{-1/2}(K\,\wtd K)^{-1/4}\, \Big[
-dt^2 + dw_1^2 + dw_2^2\nn\\
&& H\, \wtd K\, (dx_1^2 + dx_2^2 + dx_3^2) +
K\, \wtd K\, dy^2 + H\, K\, (dz_1^2 + dz_2^2 + dz_3^2)\Big]\,.
\eea
%%%%%%
It is straightforward to verify that the near-horizon structure of the above
D3/D5/NS5 system is of the form (\ref{d3d5ns5horizon}).   The solution
can be illustrated by the following diagram:

\centerline{
\begin{tabular}{c|ccccccccccc}
&$t$ & $w_1$ & $w_2$ & $x_1$ & $x_2$ & $x_3$ & $y$ & $z_1$ & $z_2$ &
$z_3$ &\\ \hline
D3&$\times$ & $\times$ & $\times$ & $-$ &$-$ &$-$ &$\times$ &$-$ &$-$
&$-$ & $H$ \\
D5&$\times$ & $\times$ & $\times$ & $\times$ & $\times$ & $\times$
&$-$ &$-$ &$-$ &$-$ & $K$ \\
NS5& $\times$ & $\times$ & $\times$ & $-$ & $-$ & $-$ & $-$ &
$\times$ & $\times$ & $\times$ & $\wtd K$
\end{tabular}}
\bigskip

\centerline{Diagram 10. The D3/D5/NS5 system}
\bigskip\bigskip

\section{M2/pp-wave system}

\subsection{M2/pp-wave system}

          The M2/pp-wave solution is given by
%%%%%%
\bea
ds_{11}^2 &=& H^{-2/3}(-W^{-1}\, dt + W\, (dy + (W^{-1}-1)dt)^2 +
dx^2) + H^{1/3}(dz^2 + z^2\, d\Omega_7^2)\,,\nn\\
F_\4 &=& dt\wedge dy\wedge dx\wedge dH^{-1}\,.\label{m2wave}
\eea
%%%%%
The solution can be illustrated by the following diagram:

\bigskip\bigskip
\centerline{
\begin{tabular}{c|cccccccccccc}
&$t$ & $y_1$ & $x_1$ & $z_1$ & $z_2$ & $z_3$ & $z_4$ & $z_5$ & $z_6$ &
$z_7$ & $z_8$ & \\ \hline
M2&$\times$ & $\times$ & $\times$ & $-$ &$-$ &$-$ &$-$ &$-$ &$-$
&$-$ &$-$ & $H$ \\
wave&$\times$ & $\sim$ & $-$ & $-$ & $-$ & $-$
&$-$ &$-$ &$-$ &$-$ & $-$ & $W$ \\
\end{tabular}}
\bigskip

\centerline{Diagram 11. The M2/pp-wave brane intersection.}
\bigskip\bigskip

   When both functions $H$ and $W$ are harmonic on the overall
transverse space of the $z^i$ coordinates, the metric becomes a direct
product of the Kaigorodov metric with a 7-sphere in the near-horizon
limit.  Here, we instead consider a semi-localised solution, with $H$
and $K$ given by
%%%%%%
\be
H=\fft{Q}{z^6}\,,\qquad W= 1 + Q_w\, (x^2 + \fft{Q/4}{z^4})\,.
\ee
%%%%%
Making the coordinate transformation
%%%%%
\be
x=\fft{\cos\a}{r}\,,\qquad z^2 = \fft{r\, Q^{1/2}}{2\sin\a}\,,
\ee
%%%%%
the metric becomes AdS$_4\times S^7$, with
%%%%%%
\be
ds_{11}^2 = \fft{Q^{1/3}}{4s^2}(ds_{\rm{AdS}_3}^2 + d\a^2)+
 Q^{1/3}\, d\Omega_7^2\,.\label{m2wavehorizon}
\ee
%%%%%
Here $ds_{\rm{AdS}_3}^2$ is the metric of AdS$_3$ (the BTZ black
hole), given by (\ref{ads3}).  Thus, we have demonstrated that the
semi-localised M2/pp-wave system is a warped product of AdS$_3$ and an
8-space.  Making the coordinate transformation $\tan(\a/2) = e^\rho$,
the first part of (\ref{m2wavehorizon}) can be expressed as
%%%
\be
ds_4^2 = d\rho^2 + \cosh^2\rho\, ds_{\rm{AdS}_3}^2\,.
\ee
%%%%%
This is AdS$_4$ expressed as a foliation of AdS$_3$.

\subsection{The NS1/D0 system}

         Reducing the above solution on the coordinate $y_1$, it
becomes an intersecting NS1/D0 system, with
%%%%%%
\bea
ds_{10\rm{IIA}} &=& H^{-3/4}W^{-7/8}\Big(-dt^2 + W\, dx^2 + W\,H\,(dz_1^2 +
\cdots + dz_8^2)\Big)\,,\nn\\
F_\3&=& dt\wedge dx\wedge dH^{-1}\,,\qquad
F_\2= dt\wedge dW^{-1}\,,\nn\\
e^{\phi} &=& H^{-1/2}\, W^{3/4}\,.\label{ns1d0}
\eea
%%%%%%
The metric of the near-horizon region describes a warped product
of AdS$_2$ with an 8-space:
%%%%%
\be
ds_{10\rm{IIA}}^2 = 8^{-3/4}Q^{3/8}Q_w^{1/8}\, s^{-9/4}\,
(ds_{\rm{AdS}_2}^2 + d\a^2 +4s^2 d\Omega_7^2)\,,
\ee
%%%%
where $ds_{\rm{AdS}_2}^2$ is the metric of AdS$_2$, given by
(\ref{ads2}).  The NS1/D0 system can be illustrated by the following
diagram:

\bigskip\bigskip
\centerline{
\begin{tabular}{c|cccccccccccc}
&$t$ & $x_1$ & $z_1$ & $z_2$ & $z_3$ & $z_4$ & $z_5$ & $z_6$ &
$z_7$ & $z_8$ & \\ \hline
NS1&$\times$ & $\times$ & $-$ &$-$ &$-$ &$-$ &$-$ &$-$
&$-$ &$-$ & $H$ \\
D0&$\times$ & $-$ & $-$ & $-$ & $-$
&$-$ &$-$ &$-$ &$-$ & $-$ & $W$ \\
\end{tabular}}
\bigskip

\centerline{Diagram 12.  The NS1/D0 brane intersection.}
\bigskip\bigskip

    In the M2/pp-wave and NS1/D0 systems, the internal space has a
round 7-sphere.  We can replace it by foliating of two lens spaces
$S^3/Z_{\Qn}$ and $S^3/Z_{\wtd Q_{\sst{\rm N}}}$. This can be achieved by
introducing two NUTs in the intersecting system.  We can then perform
Kaluza-Klein reductions or
T-duality transformations on the two associated fibre coordinates of
the lens spaces.  The resulting configurations can then be viewed as
the near-horizon geometries of four intersecting $p$-branes, with
charges $Q$, $Q_w$, $\Qn$ and $\wtd Q_{\sst{\rm N}}$

\section{M5/NUT and M5/pp-wave systems}

\subsection{M5/NUT and NS5/D6 systems}

      The solution of an M5-brane intersecting with a NUT is given by
%%%%%
\bea
ds_{11}^2\!\!\! &=&\!\!\! H^{-1/3}(-dt^2 + dw_1^2 + \cdots +dw_5^2) +
H^{2/3}(dx_1^2 + K\, (dz^2 + z^2 d\Omega_2^2) +
K^{-1}(dy + \omega)^2)\,,\nn\\
F_\4 &=& {*(dt\wedge d^5w\wedge dH^{-1})}\,.\label{m5nut}
\eea
%%%%%
The solution can be illustrated by the following diagram:

\bigskip\bigskip
\centerline{
\begin{tabular}{c|cccccccccccc}
&$t$ & $w_1$ & $w_2$ & $w_3$ & $w_4$ & $w_5$ & $x_1$ & $z_1$ & $z_2$ &
$z_3$ & $y$ & \\ \hline
M5&$\times$ & $\times$ & $\times$ & $\times$ &$\times$ &$\times$
&$-$ &$-$ &$-$ &$-$ &$-$ & $H$ \\
NUT&$\times$ & $\times$ & $\times$ & $\times$ & $\times$ & $\times$
&$\times$ &$-$ &$-$ &$-$ &$*$ & $K$ \\
\end{tabular}}
\bigskip

\centerline{Diagram 13.  The M5/NUT brane intersection.}
\bigskip\bigskip

In the usual construction where the harmonic functions $H$ and $K$
depend only the $z$ coordinate, the metric does not have an AdS
structure in the near-horizon region.  Here, we instead consider a
semi-localised solution, given by
%%%%%%
\be
H=1 + \fft{Q}{(x^2 + 4\Qn\, z)^{3/2}}\,,\qquad
K = \fft{\Qn}{z}\,.
\ee
%%%%
After an analogous coordinate transformation, we find that the metric
can be expressed as
%%%%%%
\bea
ds_{11}^2 &=& H^{-1/3}(-dt^2 + dw_i\, dw_i) + H^{2/3}(dr^2+ r^2\,
dM_4^2)\,,\nn\\
dM_4^2 &=& d\a^2 + \ft14 s^2\, (d\Omega_2^2 + (\fft{dy}{\Qn} +
\omega)^2)\,.
\eea
%%%%
Thus, in the near-horizon limit, the metric is AdS$_7\times M_4$, where
$M_4$ is a foliation of a lens space $S^3/Z_{\Qn}$.

          We can dimensionally reduce the solution (\ref{m5nut}) on
the fibre coordinate $y$. The resulting solution is the NS-NS 5-brane
intersecting with a D6-brane:

\bigskip\bigskip
\centerline{
\begin{tabular}{c|ccccccccccc}
&$t$ & $w_1$ & $w_2$ & $w_3$ & $w_4$ & $w_5$ & $x_1$ & $z_1$ & $z_2$ &
$z_3$ & \\ \hline
NS5&$\times$ & $\times$ & $\times$ & $\times$ &$\times$ &$\times$
&$-$ &$-$ &$-$ &$-$ & $H$ \\
D6&$\times$ & $\times$ & $\times$ & $\times$ & $\times$ & $\times$
&$\times$ &$-$ &$-$ &$-$ & $K$ \\
\end{tabular}}
\bigskip

\centerline{Diagram 14. The NS5/D6 brane intersection.}
\bigskip\bigskip

The solution is given by
%%%%%
\bea
ds_{10\rm{IIA}}^2 &=& H^{-1/4}K^{-1/8}\, (-dt^2 + dw_i\, dw_i) + H^{3/4}
K^{-1/8}\, dx^2 + H^{3/4} K^{7/8}\, dz_i\, dz_i\,,\nn\\
e^{\phi} &=& H^{1/2} K^{-3/4}\,,\qquad
F_\3 = e^{\phi/2} {*(dt\wedge d^5w\wedge dH^{-1})}\,,\nn\\
F_\2 &=& e^{-3\phi/2} {*(dt\wedge d^5w\wedge dx\wedge dK^{-1})}\,,
\eea
%%%%%%
In the near-horizon limit, the metric becomes a warped product of
AdS$_7$ with a 3-space
%%%%%
\be
ds_{10\rm{IIA}}^2 = \fft{Q^{3/4}}{(2\Qn)^{1/4}}\,s^{1/4}\,
(\fft{r}{Q}(-dt^2 + dw_i\, dw_i) + \fft{dr^2}{r^2} + d\a^2
+ \ft14 s^2\, d\Omega_2^2)\,.
\ee
%%%%%%%%

\subsection{M5/pp-wave and D0/D4 system}

       The solution of an M5-brane with a pp-wave is given by
%%%%%%%
\bea
ds_{11}^2 &=& H^{-1/3}(-W^{-1}\, dt^2 + W\, (dy_1 + (W^{-1} -1)dt)^2 +
dx_1^2 + \cdots + dx_4^2)\nn\\
&&\phantom{xxxxxxxxxxxxx} + H^{2/3}\,
(dz_1^2 + \cdots + dz_5^2)\,,\nn\\
F_4&=&{*(dt\wedge dy_1\wedge d^4x\wedge dH^{-1})}\,.\label{m5wave}
\eea
%%%%%%
The solution can be illustrated by the following diagram:

\bigskip\bigskip
\centerline{
\begin{tabular}{c|cccccccccccc}
&$t$ & $y_1$ & $x_1$ & $x_2$ & $x_3$ & $x_4$ & $z_1$ & $z_2$ & $z_3$ &
$z_4$ & $z_5$ & \\ \hline
M5&$\times$ & $\times$ & $\times$ & $\times$ &$\times$ &$\times$
&$-$ &$-$ &$-$&$-$ &$-$ & $H$ \\
wave&$\times$ & $\sim$ & $-$ & $-$ & $-$ & $-$
&$-$ &$-$ &$-$ &$-$ & $-$ & $W$ \\
\end{tabular}}
\bigskip

\centerline{Diagram 15. The M5/pp-wave brane intersection.}
\bigskip\bigskip

  We shall consider semi-localised solutions, with the functions $H$
and $W$ given by
%%%
\be
H=\fft{Q}{z^3}\,,\qquad W=1 + Q_w\, (x^2 + \fft{4Q}{z})\,.
\ee
%%%%%
Using analogous coordinate transformations, we find that the metric
of the semi-localised M5/pp-wave system becomes
%%%%%%
\be
ds_{11}^2 = 4Q^{2/3}\, s^{-2} (ds_{\rm AdS_3}^2 + d\a^2 + c^2\,
d\Omega_3^2) + Q^{2/3}\, d\Omega_4^2\,,\label{m5wavehorizon}
\ee
%%%%
where $ds_{\rm{AdS}_3}^2$, given by (\ref{ads3}), is precisely the
extremal BTZ black hole and hence is is locally AdS$_3$. After making
the coordinate transformation $\tan(\a/2) = e^\rho$, the first part of
the metric (\ref{m5wavehorizon}) can be expressed as
%%%%%
\be
ds_7^2 = d\rho^2 + \sinh^2\rho\, d\Omega_3^2 +
\cosh^2\rho\, ds_3^2\,.
\ee
%%%%%%
This is AdS$_7$ written as a foliation of AdS$_3$ and $S^3$.

         Performing a dimensional reduction of the solution
(\ref{m5wave}) on the coordinate $y_1$, we obtain a D0/D4 intersecting
system, given by
%%%%%%
\bea
ds_{10\rm{IIA}}^2 &=& H^{-3/8} W^{-7/8}(-dt^2 + W\, dx_i\, dx_i +
H\, W\, dz_i\, dz_i)\,,\nn\\
e^{\phi} &=& H^{-1/4}\, W^{3/4}\,,\qquad
F_\2=dt\wedge dW^{-1}\,,\nn\\
F_4&=& e^{-\phi/2} {*(dt\wedge d^4x\wedge dH^{-1})}\,.\label{d0d4}
\eea
%%%%%
The near-horizon limit of the semi-localised D0/D4 system is a
warped product of AdS$_2$ with an 8-space:
%%%%%
\be
ds_{10\rm{IIA}}^2= 2^{9/4}Q^{3/4}\, Q_w^{1/8}\, s^{-9/4}
\, (ds_2^2 + d\a^2 + c^2\, d\Omega_3^2 + \ft14 s^2\, d\Omega_4^2)\,,
\ee
where $ds_2^2$ is given by (\ref{ads2}).  We illustrate this
intersecting system with the following diagram

\bigskip\bigskip
\centerline{
\begin{tabular}{c|ccccccccccc}
&$t$ & $x_1$ & $x_2$ & $x_3$ & $x_4$ & $z_1$ & $z_2$ & $z_3$ &
$z_4$ & $z_5$ & \\ \hline
D4&$\times$ & $\times$ & $\times$ &$\times$ &$\times$
&$-$ &$-$ &$-$&$-$ &$-$ & $H$ \\
D0&$\times$ & $-$ & $-$ & $-$ & $-$
&$-$ &$-$ &$-$ &$-$ & $-$ & $W$ \\
\end{tabular}}
\bigskip

\centerline{Diagram 16. The D0/D4 brane intersection.}
\bigskip\bigskip

        In this example in the internal space the round $S^3$ and
$S^4$ can be replaced by a lens space $S^3/Z_{\Qn}$ and the foliation of
a lens space $S^3/Z_{\wtd Q_{\sst{\rm N}}}$, respectively.  We can
then perform Kaluza-Klein reductions or T-duality transformations on
the fibre coordinates of the lens spaces, leading to four-component
intersections with charges $Q$, $Q_w$, $\Qn$ and $\wtd Q_{\sst{\rm
N}}$.

\section{AdS$_6$ in type IIB from T-duality}

       So far in this paper we have two examples of intersecting
D$p$/D$(p+4)$ systems in the type IIA theory that give rise to warped
products of AdS$_{p+2}$ with certain internal spaces, namely for $p=0$
and $p=2$.  It was observed \cite{bo} also that the D4/D8 system,
arising from massive type IIA supergravity, gives rise to the warped
product of AdS$_6$ with a 4-sphere in the near-horizon limit:
%%%%%%
\be
ds_{10\rm{IIA}}^2 = s^{1/12}\, (ds_{\rm AdS_6}^2 + g^{-2}(d\a^2 + c^2\,
d\Omega_3^2)) \,.\label{ads6type2a}
\ee
%%%%%%
Note that the D4/D8 system is less trivial than the previous examples,
in the sense that it cannot be mapped by T-duality to a non-dilatonic
$p$-brane intersecting with a NUT or a wave.

       We can now introduce a NUT in the intersecting system which
has the effect, in the near-horizon limit, of replacing the round
3-sphere by a lens space, given in (\ref{lens}).  We can then perform
a Hopf T-duality transformation and obtain an embedding of AdS$_6$ in
type IIB theory:
%%%%%%
\be
ds_{10}^2 = c^{1/2}\, \Big[ds_{\rm{AdS}_6}^2 +
g^{-2}(d\a^2 + \ft14 c^2\, d\Omega_2^2) +  s^{2/3}\, c^{-2}\,
dy^2\Big]\,.\label{ads6type2b}
\ee
%%%%%
This solution can be viewed as the near-horizon geometry of an
intersecting\\ D5/D7/NS5 system.  It provides a background for the exact
embedding of six-dimensional gauged supergravity in type IIB theory.

         The D5/D7/NS5 semi-localised solution can be obtained by
performing the T-duality on the D4/D8/NUT system.  The solution is
given by
%%%%%%%
\bea
ds_{10\rm{IIB}}^2 &=& (H_1\, K)^{-1/4}\Big(-dt^2 + dw_1^2 + \cdots +
dw_4^2 + H_1\, K\, (dx_1^2 + dx_2^2 + dx_3^2)\nn\\
&&\phantom{xxxxxx} + H_2\, K\, dy^2 + H_1\, H_2\, dz^2\Big)\,.
\eea
%%%%%
The functions $H_1$, $H_2$ and $K$ are given by
%%%%%
\be
H_1 =1+ \fft{Q_1}{(4\Qn\, |\vec x| + \ft{4Q_2}{9}\, z^3)^{5/3}}\,,\quad
H_2=Q_2\, z\,,\quad K = \fft{\Qn}{|\vec x|}\,.
\ee
%%%%%
It is straightforward to verify that the near-horizon structure of
this system is of the form (\ref{ads6type2b}). The solution can be
illustrated by the following:

\bigskip\bigskip
\centerline{
\begin{tabular}{c|ccccccccccc}
&$t$ & $w_1$ & $w_2$ & $w_3$ & $w_4$ & $x_1$ & $x_2$ & $x_3$ &
$y$ & $z$ & \\ \hline
D5&$\times$ & $\times$ & $\times$ &$\times$ &$\times$
&$-$ &$-$ &$-$&$-$ &$-$ & $H_1$ \\
D7&$\times$ & $\times$ & $\times$ & $\times$ & $\times$
&$\times$ &$\times$ &$\times$ &$-$ & $-$ & $H_2$ \\
NS5& $\times$ & $\times$ & $\times$ & $\times$ & $\times$ &
$-$ & $-$ & $-$ & $-$ & $\times$ & $K$ \\
\end{tabular}}
\bigskip

\centerline{Diagram 17. The D5/D7/NS5 brane intersection.}
\bigskip\bigskip

\section{Conclusion}

We have obtained various AdS spacetimes warped with
certain internal spaces in eleven-dimensional and type IIA/IIB
supergravities.  These solutions arise as the near-horizon geometries
of more general semi-localised multi-intersections of M-branes in
$D=11$ or NS-NS branes or D-branes in $D=10$.  We achieve this by
noting that any bigger sphere (AdS spacetime) can be viewed as a
foliation involving $S^3$ (AdS$_3$).  Then the $S^3$ (AdS$_3$) can be
replaced by a three-dimensional lens space (BTZ black hole), which
arise naturally from the introduction of a NUT (pp-wave). We can
then perform a Kaluza-Klein reduction or Hopf T-duality transformation
on the fibre coordinate of the lens space (BTZ black hole).

          It is important to note that the warp factor depends only on
the internal foliation coordinate but not on the lower-dimensional
spacetime coordinates.  This implies the possibility of finding a
larger class of consistent dimensional reduction of eleven-dimensional
or type IIA/IIB supergravity on the internal space, giving rise to
gauged supergravities in lower dimensions with AdS vacuum solutions.
The first such example was obtained in \cite{d6gauge}. We obtain further
examples for possible consistent embeddings of lower-dimensional gauged
supergravity in $D=11$ and $D=10$.  For example, we obtain the vacuum
solutions for the embedding of the six and four-dimensional gauged AdS
supergravities in type IIB theory and for the embedding of the
seven-dimensional gauged AdS supergravity in type IIA theory.

%% file: thesisabs.tex
\chapter{Absorption by Branes}

\section{Introduction}

\subsection{Low-energy Absorption Cross-sections and AdS/CFT}

The absorption cross-sections of $p$-branes, in particular the D1-D5
brane intersection and the D3-brane, offered an early hint of an exact
duality between string theory and supersymmetric gauge theory. The
low-energy dynamics of a stack of D3-branes is described by $\cal N$ $=4$
supersymmetric Yang-Mills theory \cite{malda}. This gauge theory is
conformally invariant, which corresponds to a constant dilaton
background. This implies that the dilaton fluctuation satisfies the
minimally-coupled scalar equation
%%%%%%%%%
\be
\partial_{\mu}(\sqrt{-g} g^{\mu \nu} \partial_{\nu} \phi)=0.
\ee
%%%%%%%%%
The extremal D3-brane metric is
%%%%%%%%
\be
ds^2=H^{-1/2}(-dt^2+dx_1^2+dx_2^2+dx_3^2)+H^{1/2}(dr^2+r^2d\Omega_5^2),
\ee
%%%%%%%
where the harmonic function $H=1+R^4/r^4$. The wave equation for the
$\ell$th partial wave of a minimally-coupled (massless) scalar is
%%%%%%%
\be
\Big[ \frac{1}{\rho^5} \partial_{\rho}\rho^5
\partial_{\rho}+1+\frac{\omega^4 R^4}{\rho^4}-\frac{\ell (\ell+4)}{\rho^2}
\Big] \phi^{(\ell)}(\rho)=0, \label{waveequation} 
\ee
%%%%%%% 
where $\rho=\omega r$. This absorption process corresponds to quantum
mechanical tunneling through a centrifugal potential barrier in the
reduced one-dimensional system.

Klebanov solved this equation for low energy $\omega R$, by matching
solutions across two overlapping spacetime regions\footnote{ This
semi-classical approach to computing the absorption cross-section for a
field propagating in a black hole background was pioneered in the thesis
of Unruh \cite{unruh}.} \cite{kleb}. The leading term in the s-wave
absorption cross-section was
found to be
%%%%%%%%
\be
\sigma_{SUGRA}=\frac{\pi^4}{8}\omega^3 R^8.
\ee
%%%%%%%%
This result was compared to a corresponding calculation in the
Super-Yang-Mills theory, in which the dilaton couples to the operator
$\frac{T_3}{4}Tr(F^2+...)$, where $T_3$ is the D3-brane tension
\cite{kleb}. At weak-coupling, the leading-order absorption process is for
the dilaton to decay into a pair of back-to-back gluons on the
world-volume of the D3-brane. The rate for this process was found to be
%%%%%%%%%
\be
\sigma=\frac{\kappa^2 \omega^3 N^2}{32 \pi}.
\ee
%%%%%%%%%
Equating the tensions of the black 3-brane and the stack of $N$ D3-branes
yields
%%%%%%%%%
\be
R^4=\frac{\kappa}{2\pi^{5/2}}N.
\ee
%%%%%%%%%
Taking into account this relation, we find that the s-wave absorption
cross-section calculated on the supergravity side agrees with the
cross-section of the decay process in the world-volume gauge theory. This
remarkable result raises the hope of an exact relation between
Super-Yang-Mills theory and gravity.

However, the supergravity calculation is reliable in the limit of
weak curvature where $g_{YM}^2 N \rightarrow \infty$ whereas the gauge
theory calculation was performed to leading order in $g_{YM}^2 N$. Gubser
and Klebanov have resolved this discrepancy by arguing that all
higher-order corrections in the coupling vanish due to supersymmetric
non-renormalization theorems\footnote{This theorem was made explicit for 
the graviton absorption cross-section, which corresponds to the
two-point function of the stress-energy tensor of the gauge theory}
\cite{klebTASI}.

It was also found that the operator corresponding to the absorption of
higher partial waves of the dilaton should be of the form
%%%%%%%%
\be
\frac{T_3}{4\ell!} Tr(F_{ab}F^{ab}X^{(i_1}...X^{i_{\ell})})_{Traceless},
\ee
%%%%%%%%
and exact agreement between the cross-sections was found for all $\ell$.
This agreement of cross-sections provides a strong piece of evidence in
support of the exact AdS/CFT correspondence \cite{klebTASI}.

\subsection{Correlation functions and the bulk/boundary correspondence}

Absorption cross-sections are a particular example of an exact
correspondence between particle states in $AdS_5 \times S^5$ and operators
in the gauge theory. This leads to a general prescription for computing
gauge theory correlators in supergravity. We identify the generating
function of gauge theory correlators
%%%%%%%%%
\be
Z_{CFT}[\X_i (x)]=\int e^{-S_{YM}+\sum \Xi_i (x) \cal O_i}
\ee
%%%%%%%%%
with the partition function of string theory
%%%%%%%%%
\be
Z_S [\Xi_i (x)],
\ee
%%%%%%%%%
where the functions $\Xi_i (x)$ give the fixed values of the fields $\Xi_i
(x,r)$ at the AdS boundary. By differentiating the string partition
function with respect to the $\Xi$ fields and evaluate at vanishing $\Xi$,
we obtain the correlators of the operators in the Super-Yang-Mills theory 
\cite{klebTASI}.

In general, a direct comparision of the correlation functions computed in
supergravity with gauge theory computations is not possible, since the
supergravity calculations assume strong 't Hooft coupling whereas the
field theory calculations can only be performed at weak coupling. However,
it is currently believed that all two-point and three-point functions of
chiral operators are protected by supersymmetric non-renormalization
theorems, enabling one to extract weak 't Hooft coupling calculations to
the regime of strong coupling. 

The two-point function of a gauge invariant operator in the
strongly-coupled Super-Yang-Mills theory can be read off from the
absorption cross-section for the supergravity field which couples to this
operator in the worldvolume theory. The previous section outlined how
this comparision is performed. 

Consider a canonically normalized bulk scalar field which is coupled to
the D3-branes through the interaction 
%%%%%%%%%
\be
S_{int}=\int d^4 x \phi (x,0){\cal O}(x),
\ee
%%%%%%%%%
where $\phi (x,0)$ denotes the value of the field at the transverse
coordinates where the stack of D3-branes is located. Then the precise
relation is given by
%%%%%%%%%
\be
\sigma=\frac{1}{2i \omega} Disc \Pi
(p) |_{-p^2=\omega^2-i\epsilon}^{-p^2=\omega^2+i\epsilon},
\ee
%%%%%%%%%
where
%%%%%%%%%
\be
\Pi (p)=\int d^4 x e^{ip \cdot x} \langle {\cal O}(x){\cal O}(0) \rangle,
\ee
%%%%%%%%%
which depends only on $s=-p^2$. $Disc$ refers to the discontinuity of the
function $\Pi$ across the real energy axis, after extrapolating to complex
values of $s$.

Thus, as discussed in the previous section, the s-wave dilaton absorption
cross-section measures the normalized 2-point function 
$\langle {\cal O}_{\phi}(p){\cal O}_{\phi}(-p) \rangle$, where\\ ${\cal
O}_{\phi}=\frac{T_3}{4}Tr(F^2+...)$ \cite{klebTASI}.

\subsection{Exact Absorption Cross-sections}

Gubser and Hashimoto were able to solve for the dilaton absorption
cross-section on a D3-brane as an exact expansion in the
dimensionless energy parameter $\omega R$ \cite{gubserhash}. They found
that the wave
equation (\ref{waveequation}) is equivalent to Mathieu's modified
differential equation
%%%%%%%%%
\be
\big[ \partial_z^2+2q\cosh 2z-a\big] \psi(z)=0,
\ee
%%%%%%%%%
under a coordinate and wave function transformation. The exact solution of
this equation is known as a power series in $q=\omega R$, from which one
can write the complete expansion for the absorption probability of the
$\ell$th partial wave as
%%%%%%%%%
\be
P_{\ell}=\frac{4\pi^2}{[(\ell+1)!]^2[(\ell+2)!]^2}\big( \frac{\omega R}{2}
\big) ^{8+4\ell} \sum_{0 \le k \le n} b_{n,k} (\omega R)^{4n}(\ln \omega
\gamma R/2)^k, 
\ee
%%%%%%%%%
where $b_{n,k}$ are computable coefficients with $b_{0,0}=1$ and $\ln
\gamma$ is Euler's constant\footnote{Exact absorption probability has also 
been found via the Mathieu equation on the background of a six-dimensional
dyonic string \cite{clpt}. Also, it has been found that high-energy
absorption can also be expressed as an exact expansion via the Mathieu
equation as well.}.

Recently, Rastelli and Raamsdonk have shown that the first subleading
terms in the above absorption probability can be reproduced exactly in the
worldvolume calculation using a deformation of the $\cal N$ $=4$ SYM
theory by a dimension eight chiral operator \cite{KRT}. This result
provides a hint that the holographic duality may be valid beyond the
near-horizon of the stack of D3-branes.

In this chapter, we find that the absorption cross-section can be
calculated exactly for various other fields. As we shall discuss,
this is of interest from the vantage point of AdS/CFT. However, we would
like to remark that this is a useful way to probe supergravity solutions,
which are of interest in their own right. As we shall discuss, the
absorption cross-sections have a remarkable oscillatory character with
respect to the energy of the scattered fields.

\section{Massive absorption}

Scattering processes in the curved backgrounds of p-brane configurations
of M theory and string theory have been extensively studied over the past
few years \cite{dmw1,gk1,mast,cgkt,km,kleb,GKT,cl1, 
cl2,emp,ghkk,lm,gubser,taylor,ah,clpt,clpt2}. Motivation for
these 
studies is the fact that the low-energy absorption cross-sections for
different fields yield information about the two-point correlation
functions in the strongly coupled gauge theories via AdS/CFT
correspondence
\cite{malda,gkp,wit,kleb,GKT,ghkk,gubserhash,KRT,Friedman,Liu}.

Research on brane absorption has mainly concentrated on the case of
massless, minimally-coupled scalars.  Some work has been done for the
cases of the emission
of BPS particles from five- and four-dimensional black holes
\cite{gk1}.  These BPS particles can be viewed as pp-waves in a spacetime
of one higher dimension. Hence they satisfy the higher-dimensional
massless wave equations.

         We presently consider the absorption probability of
minimally-coupled massive particles by extremal $p$-branes. The
wave equation for such a scalar depends only on the metric of the
$p$-brane, which has the form
%%%%%%%
\be ds^2= \prod_{\a=1}^N H_\a^{-\ft{\td d}{D-2}}\, dx^\mu dx^\nu
\eta_{\mu\nu} + \prod_{\a=1}^N H_\a^{\ft{d}{D-2}}\, dy^mdy^m \
,\label{dmetric} \ee
%%%%%%
where $d=p+1$ is the dimension of the world volume of the
$p$-brane, $\td d =D-d-2$, and $H_\a= 1 + Q_\a/r^{\td d}$ are
harmonic functions in the transverse space $y^m$, where $r^2
=y^m\, y^m$. (Note that the ADM mass density and the physical
charges of the extreme $p$-brane solutions are proportional to
$\sum_{i=1}^NQ_i$ and $Q_i$, respectively)

 It follows that the wave equation, $\del_{\sst M} (\sqrt{g}\,
g^{\sst{MN}}\del_{\sst N} \Phi)=m^2\ \Phi$, for the massive
minimally-coupled scalar, with the ansatz, $\Phi(t, r, \theta_i)
=\phi(r)\, Y(\theta_i)\, e^{-\im \omega t}$, takes the following
form:
%%%%%%%%%
\be \fft{d^2\phi}{d\rho^2} + \fft{\td d+1}{\rho}\,
\fft{d\phi}{d\rho} + \Big[ \prod_{\a=1}^N (1 +\fft{\lambda^{\td
d}_\a}{\rho^{\td d}}) -\fft{\ell(\ell +\td d)}{\rho^2} -
\fft{m^2}{\omega^2} \prod_{\a=1}^N (1+\fft{\lambda_\a^{\td
d}}{\rho^{\td d}} )^{\ft{d}{D-2}} \Big]\, \phi =0\
,\label{genwave} \ee
%%%%%%%%%%
where $\rho=\omega\, r$ and $\lambda_\a = \omega\, Q_\a^{1/\td
d}$. Note that when $m=0$, the wave equation depends on $\td d$,
but is independent of the world-volume dimension $d$.  This
implies that the wave equation for minimally-coupled massless
scalars is not invariant under the vertical-dimensional reduction,
but is invariant under double-dimensional reduction of the
corresponding $p$-brane \cite{clpt2}.  However, for massive
scalars, the wave equation (\ref{genwave}) is not invariant under
either double or vertical reductions.

        The absorption probability of massless scalars is better
understood.  It was shown that for low frequency the
cross-section/frequency relation for a generic extremal $p$-brane
coincides with the entropy/temperature relation of the near
extremal $p$-brane \cite{clpt2}.  There are a few examples where
the wave equations can be solved exactly in terms of special
functions. Notably, the wave equations for the D3-brane
\cite{gubserhash} and the dyonic string \cite{clpt} can be cast into
modified Mathieu equations. Hence, the absorption probability can
be obtained exactly, order by order, in terms of a certain small
parameter.  There are also examples where the absorption
probabilities can be obtained in closed-form for all wave
frequencies \cite{clpt2}.

         When the mass $m$ is non-zero, we find that there are two
examples for which the wave function can be expressed in terms of
special functions and, thus, the absorption probabilities can be
obtained exactly.  One example is the wave equation in the
self-dual string background, which can be cast into a modified
Mathieu equation. Therefore, we can obtain the exact absorption
probability, order by order, in terms of a certain small
parameter. Another example
is the wave equation for the $D=4$ two-charge black hole with
equal charges.  The wave function can be expressed in terms of
Kummer's regular and irregular confluent hypergeometric functions.
It follows that we can obtain the absorption probability in
closed-form. In both of the above
examples, the massive scalar wave equation has the same form as
the massless scalar wave equation under the backgrounds where the
two charges are generically non-equal.

         However, in general, the massive scalar wave equation
(\ref{genwave}) cannot be solved analytically.  For low-frequency
absorption, the leading-order wave function can be obtained by
matching wave functions in inner and outer regions. We make use of this
technique to obtain the leading-order
absorption probability for D3-, M2- and M5-branes.

\section{Massive absorption for the self-dual string}

     For the self-dual string ($Q_1=Q_2\equiv Q$), we have $d=\td d=2$
and $\lambda_1 =\lambda_2 \equiv\lambda =\omega \sqrt{Q}$.  It
follows that the wave equation (\ref{genwave}) becomes
%%%%%%
\be \fft{d^2\phi}{d\td \rho^2} + \fft{3}{\td \rho}\,
\fft{d\phi}{d\td \rho} + \Big(1 + \fft{\td\lambda_1^2 +
\td\lambda_2^2 - \ell(\ell+2)}{\td\rho^2}
+\fft{\td\lambda_1^2\,\td\lambda_2^2}{\td \rho^4}\Big) \, \phi =0\
,\label{mselfdualstring} \ee
%%%%%%
where
%%%%
\bea &&\td \lambda_1 = \lambda\ ,\qquad \td \lambda_2 = \lambda\,
\sqrt{
1-(m/\omega)^2}\ ,\nn\\
&&\td \rho= \rho\, \sqrt{1-(m/\omega)^2}\ , \eea
%%%%%
Thus the wave equation of a minimally-coupled massive scalar on a
self-dual string has precisely the same form as that of a
minimally-coupled massless scalar on a dyonic string, where $\td
\lambda_1$ and $\td \lambda_2$ are associated with electric and
magnetic charges.  It was shown in \cite{clpt} that the wave
equation (\ref{mselfdualstring}) can be cast into the form of a
modified Mathieu equation, and hence the equation can be solved
exactly.  To do so, one makes the following definitions
%%%%%%
\be \phi(\td \rho) = \fft{1}{\rho}\, \Psi(\rho)\, \qquad \td \rho
= \sqrt{\td\lambda_1\, \td\lambda_2}\, e^{-z}\ . \ee
%%%%%
The wave equation (\ref{mselfdualstring}) then becomes the
modified Mathieu equation \cite{clpt}
%%%%%
\be \Psi'' + (8\Lambda^2\, \cosh(2z) - 4\a^2)\, \Psi =0\ , \ee
%%%%%%
where
%%%%%%
\bea
\alpha^2 &=& \ft14(\ell+1)^2 -\Lambda^2\, \Delta\ ,\nn\\
\Lambda^2 &=& \ft14 \td\lambda_1\, \td\lambda_2 = \ft14
\lambda^2\, \sqrt{1 -(m/\omega)^2}\ =\ft14
\omega\sqrt{\omega^2-m^2}Q,\label{aldpara}\\
\Delta &=& \fft{\td\lambda_1}{\td\lambda_2} +
\fft{\td\lambda_2}{\td\lambda_1} =\sqrt{1-(m/\omega)^2}
+\fft1{\sqrt{1-(m/\omega)^2}}\ . \eea
%%%%%%

       The Mathieu equation can be solved, order by order, in terms of
$\Lambda^2$.  The result was obtained in \cite{clpt}, using the
technique developed in \cite{doug}. (For an extremal D3-brane,
which also reduces to the Mathieu equation, an analogous technique
was employed in \cite{gubserhash}.)  In our case there are two
parameters, namely $\omega R$ and $m/\omega$.  We present results
for two scenarios:

\subsection{Fixed mass/frequency ratio probing}

        In this case, we have $m/\omega=\beta$ fixed.  The requirement
that $\Lambda$ is small is achieved by considering low-frequency
and, hence, small mass of the probing particles.  In this case,
$\Delta$ is fixed, and the absorption probability has the form
\cite{clpt}
%%%%%
\be P_\ell =  \fft{4\pi^2\, \Lambda^{4+4\ell}}{(\ell+1)^2 \,
\Gamma(\ell+1)^4}\, \sum_{n\ge0} \sum_{k=0}^n b_{n,k}\,
\Lambda^{2n}\,  (\log \bar\Lambda)^k\ ,\label{lamexp} \ee
%%%%%
where $\bar\lambda = e^\gamma\, \lambda$, and $\gamma$ is Euler's
constant. The prefactor is chosen so that $b_{0,0}=1$.  Our
results for the coefficients $b_{n,k}$ with $k\le n\le 3$ for the
first four partial waves, $\ell=0,1,2,3$, were explicitly given in
\cite{clpt}. In particular the result up to the order of
$\Lambda^2$ is given by \cite{clpt}
%%%%%
\be P_\ell = \fft{4\pi^2\, \Lambda^{4+4\ell}}{(\ell+1)^2\,
\Gamma(\ell+1)^4} \, \Big[ 1 -\fft{8\D}{\ell+1}\, \Lambda^2\,
\log\Lambda + \fft{4\D\, \Lambda^2}{(\ell+1)^2}\, \Big(1+
2(\ell+1)\, \psi(\ell+1)\Big) + \cdots\Big] \ , \ee
%%%%%
where $\psi(x)\equiv \Gamma'(x)/\Gamma(x)$ is the digamma
function.

\subsection{Fixed mass probing}

         Now we consider the case where the mass of the test particle
is fixed.  In this case, it is ensured that $\Lambda$ is small by
considering the limiting frequency of the probing particle, namely
$\omega \rightarrow m^+$, i.e. the particle is non-relativistic.
In this limit, the value of $\Delta$ becomes large (while at the
same time the expansion parameter $\Lambda$ can still be ensured
to remain small). Furthermore, we shall consider a special slice
of the parameter space where $\a^2$, given in (\ref{aldpara}), is
fixed.  The absorption probability for fixed $\a$ was obtained in
\cite{clpt}.  It is of particular interest to present the
absorption probability for $\a\to 0$, given by
%%%%
\be P= \fft{\pi^2}{\pi^2+ (2\log\bar\Lambda)^2}\, \Big(1
-\ft{32}{3}\Lambda^4\, (\log\bar\Lambda)^2 -\ft{16}{3}\,
(4\zeta(3)-3) \, \fft{\Lambda^4\, \log\bar\Lambda}{ \pi^2+
(2\log\bar\Lambda)^2} +{\cal O}(\Lambda^8) \Big)\ ,\label{a0} \ee
%%%%%
where $\bar\Lambda= e^\gamma\, \Lambda$.  When $\a^2<0$, we define
$\a^2 = {\rm i}\,\beta$, and find that the absorption probability
becomes oscillatory as a function of $\Lambda$, given by
\cite{clpt}
%%%%%
\be P= \fft{\sinh^2 2\pi\b}{\sinh^2 2\pi\b + \sin^2(\theta-4\b\,
\log\Lambda)} +\cdots \ , \ee
%%%%%
where
%%%%%
\be \theta={\rm arg}\, \fft{\Gamma(2\im\, \b)}{\Gamma(-2\im\,
\b)}\ . \ee
%%%%%
Note that the $\alpha\to 0$ limit is a dividing domain between the
region where the absorption probability has power dependence of
$\Lambda$ ($\a^2> 0$) and the region with oscillating behavior on
$\Lambda$ ($\a<0$).

\section{Closed-form massive absorption for the $D=4$ two-charge black
hole}

For a $D=4$ black hole, specified in general by four charges
$Q_1$, $Q_2$, $P_1$ and $P_2$~\cite{cvyoum}, we have $d=\td d=1$.
We consider the special case of two equal non-zero charges
($Q_1=Q_2 \equiv Q$ with $P_1=P_2=0$) and therefore $\lambda_1=
\lambda_2 \equiv\lambda=\omega Q$.  It follows that the wave
equation (\ref{genwave}) becomes
%%%%%%
\be \frac{d^2 \phi}{d\td \rho^2} + \frac{2}{\td \rho}\,
\frac{d\phi}{d\td \rho} + \Big[ \big( 1 + \frac{\td \lambda_1}{\td
\rho}\big) \big( 1 + \frac{\td \lambda_2}{\td \rho}\big) -
\frac{\ell (\ell +1)}{\td \rho^2} \Big]\, \phi =0\,
\label{m2chargebh} \ee
%%%%%%
where
%%%%
\bea &&\td \lambda_1 = \frac{\lambda}{\sqrt{1-(m/\omega)^2}}\
,\qquad
\td \lambda_2 = \lambda\, \sqrt{1-(m/\omega)^2}\ ,\nn\\
&&\td \rho= \rho\, \sqrt{1-(m/\omega)^2}\ , \eea
%%%%%
Thus the wave equation of a minimally-coupled massive scalar on a
$D=4$ black hole with two equal charges has precisely the same
form as that of a minimally-coupled massless scalar on a $D=4$
black hole with two different charges. The closed-form absorption
probability for the latter case was calculated in \cite{clpt2}
(see also \cite{balalar}). The absorption probability for the
former case is, therefore, given by
%%%%%%
\be P^{(\ell)}=\frac{1-e^{-2\pi
\sqrt{4\lambda^2-(2\ell+1)^2}}}{1+e^{-\pi (2\lambda  +
\sqrt{4\lambda^2-(2\ell+1)^2})} e^{-\pi \delta}}\ ,\ \ \ \ \ \
\lambda \ge \ell + \frac{1}{2}\ , \ee
%%%%%%
where
%%%%%%
\be \delta \equiv \td \lambda_1 + \td \lambda_2 - 2\lambda
=\lambda\big[(1-({m/
\omega})^2)^{1/4}-(1-(m/\omega)^2)^{-1/4}\big]^2\ge 0\ , \ee
%%%%%
with $P^{(\ell)}=0$ if $\lambda \le \ell + \frac{1}{2}$. In the
non-relativistic case ($\omega\to m^+$), the absorption
probability takes the (non-singular) form $P^{(\ell)}=1-e^{-2\pi
\sqrt{4\lambda^2-(2\ell+1)^2}}$, with $\lambda\sim mQ$.

The total absorption cross-section is given by:
%%%%%
\be \sigma^{(abs)} = \sum_{\ell\le\lambda-\ft12} \frac{\pi
\ell(\ell+1)}{\omega^2-m^2} P^{(\ell)} \ee
%%%%%
It is oscillatory with respect to the dimensionless parameter, $M
\omega\sim Q\omega=\lambda$. ($M$ is the ADM mass of the black
hole.) This feature was noted in \cite{sanchez} for Schwarzschild
black holes and conjectured to be a general property of black
holes due to wave diffraction. Probing particles feel an effective
finite potential barrier around black holes, inside of which is an
effective potential well. Such particles inhabit a quasi-bound
state once inside the barrier. Resonance in the partial-wave
absorption cross-section occurs if the energy of the particle is
equal to the effective energy of the potential barrier. Each
partial wave contributes a 'spike' to the total absorption
cross-section, which sums to yield the oscillatory pattern. As the
mass of the probing particles increases, the amplitude of the
oscillatory pattern of the total absorption cross-section
decreases.

\section{Leading-order massive absorption for D3, M2 and M5-branes}

       In the previous two sections, we considered two examples for
which the massive scalar wave equations can be solved exactly.  In
general, the wave function (\ref{genwave}) cannot be solved
analytically.  In the case of low frequency, one can adopt a
solution-matching technique to obtain approximate solutions for
the inner and outer regions of the wave equations.  In this
section, we shall use such a procedure to obtain the leading-order
absorption cross-sections for the D3, M2 and M5-branes.

       We now give a detailed discussion for the D3-brane, for which
we have $D=10$, $d=\td d=4$ and $N=1$.  We define $\lambda \equiv
\omega R$. It follows that the wave equation (\ref{genwave})
becomes
%%%%%%%%%
\be \Big( \frac{1}{\rho^5} \frac{\partial}{\partial \rho} \rho^5
\frac{\partial}{\partial \rho} + 1 + \frac{(\omega R)^4}{\rho^4} -
\sqrt{1 + \frac{(\omega R)^4}{\rho^4}} (\frac{m}{\omega})^2 -
\frac{\ell (\ell + 4)}{\rho^2} \Big) \phi (\rho) = 0.
\label{D3eqn} \ee
%%%%%%%%%
Thus, we are interested in absorption by the Coulomb potential in
6 spatial dimensions. For $\omega R \ll 1$ we can solve this
problem by matching an approximate solution in the inner region to
an approximate solution in the outer region.  To obtain an
approximate solution in the inner region, we substitute $\phi =
\rho^{- 3/2} f$ and find that
%%%%%%%%%%%
\be \Big( \frac{\partial^2}{\partial \rho^2} + \frac{2}{\rho}
\frac{\partial}{\partial \rho} - \big( \frac{15}{4} + \ell (\ell +
4) \big) \frac{1}{\rho^2} + 1 + \frac{(\omega R)^4}{\rho^4} -
\sqrt{1 + \frac{(\omega R)^4}{\rho^4}} \big( \frac{m}{\omega}
\big) ^2 \Big) f = 0. \label{inner} \ee
%%%%%%%%%%
In order to neglect $1$ in the presence of the $\frac{1}{\rho^2}$
term, we require that
%%%%%%%%%%%%
\be \rho \ll 1\ . \label{condition1} \ee
%%%%%%%%%%%
In order for the scalar mass term to be negligible in the presence
of the $\frac{1}{\rho^2}$ term, we require that
%%%%%%%%%%
\be \rho \ll \Big[ \big( \frac{\omega}{m} \big)^4 \big(
\frac{15}{4} + \ell (\ell +4)\big)^2 - (\omega R)^4\Big]^{1/4} \ .
\label{condition2} \ee
%%%%%%%%%%
Physically we must have $m \le \omega$. Imposing the low-energy
condition $\omega R \ll 1$ causes (\ref{condition1}) to be a
stronger constraint on $\rho$ than is (\ref{condition2}).  Under
the above conditions, (\ref{inner}) becomes
%%%%%%%%%
\be \Big( \frac{\partial^2}{\partial \rho^2} + \frac{2}{\rho}
\frac{\partial}{\partial \rho} - \big( \frac{15}{4} + \ell (\ell +
4) \big) \frac{1}{\rho^2} + \frac{(\omega R)^4}{\rho^4} \Big) f =
0\ , \label{approxinner} \ee
%%%%%%%%
which can be solved in terms of cylinder functions. Since we are
interested in the incoming wave for $\rho \ll 1$, the appropriate
solution is
%%%%%%%%
\be \phi_o = i \frac{(\omega R)^4}{\rho^2} \Big( J_{\ell + 2}
(\frac{(\omega R)^2}{\rho}) + i N_{\ell + 2} (\frac{(\omega
R)^2}{\rho}) \Big), \ \ \ \ \rho \ll 1, \ee
%%%%%%%%
where $J$ and $N$ are Bessel and Neumann functions.  In order to
obtain an approximate solution for the outer region, we substitute
$\phi = \rho^{-5/2} \psi$ into (\ref{D3eqn}) and obtain
%%%%%%%%%
\be \Big( \frac{\partial^2}{\partial \rho^2} - \big( \frac{15}{4}
+ \ell (\ell + 4) \big) \frac{1}{\rho^2} + 1 + \frac{(\omega
R)^4}{\rho^4} - \sqrt{1 + \frac{(\omega R)^4}{\rho^4}}
(\frac{m}{\omega})^2 \Big) \psi = 0. \label{outer} \ee
%%%%%%%%
In order to neglect $\frac{(\omega R)^4}{\rho^4}$ in the presence
of the $\frac{1}{\rho^2}$ term, we require that
%%%%%%%
\be 
(\omega R)^2\ \ll \rho. \label{condition3} 
\ee
%%%%%%%%
Within the scalar mass term, $\frac{(\omega R)^4}{\rho^4}$ can be
neglected in the presence of 1 provided that
%%%%%%%%
\be 
\omega R\ \ll \rho. \label{condition4} 
\ee
%%%%%%%%
Imposing the low-energy condition, $\omega R \ll 1$, causes
(\ref{condition4}) to be a stronger constraint on $\rho$ than
(\ref{condition3}).  Under the above conditions, (\ref{outer})
becomes
%%%%%%%%
\be 
\Big( \frac{\partial^2}{\partial \rho^2} - \big( \frac{15}{4}
+ \ell (\ell + 4) \big) \frac{1}{\rho^2} + 1 -
(\frac{m}{\omega})^2 \Big) \psi = 0. \label{approxouter} 
\ee
%%%%%%%%
Equation (\ref{approxouter}) is solved in terms of cylinder
functions:
%%%%%%%%
\be 
\phi_{\infty} = A \rho^{- 2} J_{\ell + 2} (\sqrt{1 -
(m/\omega)^2} \rho) + B \rho^{- 2} N_{\ell + 2} (\sqrt{1 -
(m/\omega)^2} \rho), \ \ \ \ \omega R \ll \rho, 
\ee
%%%%%%%%
where $A$ and $B$ are constants to be determined.

Our previously imposed low-energy condition, $\omega R \ll 1$, is
sufficient for there to be an overlapping regime of validity for
conditions (\ref{condition1}) and (\ref{condition4}), allowing the
inner and outer solutions to be matched. Within the matching
region, all cylinder functions involved have small arguments. We
use the same asymptotic forms of the cylinder functions as used by
\cite{clpt2}. We find that $B=0$ and
%%%%%%%
\be A = \frac{4^{\ell+2} \Gamma (\ell+3) \Gamma (\ell+2)}{\pi
\big( 1-(\frac{m}{\omega})^2 \big)^{\frac{\ell+2}{2}} (\omega
R)^{2\ell}} \ee
%%%%%%%
The absorption probability is most easily calculated in this
approximation scheme as the ratio of the flux at the horizon to
the incoming flux at infinity.  In general, this flux may be
defined as
%%%%%%%
\be F = i \rho^{\td d +1} \big( \bar{\phi} \frac{\partial
\phi}{\partial \rho} - \phi \frac{\partial \bar{\phi}}{\partial
\rho} \big)\ , \ee
%%%%%%%
where $\phi$ here is taken to be the in-going component of the
wave. From the approximate solutions for $\phi$ in the inner and
outer regions, where the arguments of the cylinder functions are
large, we find that the in-going fluxes at the horizon and at
infinity are given by
%%%%%%%
\be F_{horizon}=\frac{4}{\pi} \omega^4 R^8,\ \ \ \ \ \ \ \ \ \
F_{\infty}=\frac{A^2}{\pi \omega^4}\ . \ee
%%%%%%%
Thus, to leading order, the absorption probability, $P \equiv
F_{horizon}/F_{\infty}$, is
%%%%%%%
\be P^{(\ell)} = \frac{\pi^2 \big( 1-\big( \frac{m}{\omega}
\big)^2 \big)^{\ell + 2} (\omega R)^{4\ell+8}}{4^{2\ell+3}
(\ell+2)^2 [(\ell+1)!]^4} \ee
%%%%%%%
In general, the phase-space factor relating the absorption
probability to the absorption cross-section can be obtained from
the massless scalar case considered in \cite{unruh} with the
replacement $\omega \rightarrow \sqrt{\omega^2-m^2}$:
%%%%%%%%
\be \sigma^{(\ell)}=2^{n-2}\pi^{n/2-1}\Gamma(n/2-1)(\ell+n/2-1)
{\ell+n-3 \choose \ell} (\omega^2-m^2)^{(1-n)/2} P^{(\ell)} \ee
%%%%%%%%
where $n=D-d$ denotes the number of spatial dimensions. Thus, for
the D3-brane we find
%%%%%%%%
\be \sigma_{3-brane}^{(l)} = \frac{\pi^4 (\ell+3)(\ell+1)
[1-(m/\omega)^2]^{\ell-1/2}}{(3)2^{4\ell+3}[(\ell+1)!]^{4}}
\omega^{4l+3} R^{4l+8}\ . \ee
%%%%%%%%
As can be seen, within our approximation scheme, the effects of a
nonzero scalar mass amount to an overall factor in the partial
absorption cross-section. Also, the s-wave absorption
cross-section is increased by $m$ and the higher partial wave
absorption cross-sections are diminished by $m$. This is to be
expected, since the scalar mass serves to increase gravitational
attraction as well as rotational inertia.

The above approximation scheme can be applied to massive scalar
particles in all $N=1$  $p$-brane backgrounds except for the case
of $D=11$  $p$-branes with $\td d=4$ and $\td d=5$, in which cases
the scalar mass term cannot be neglected in the inner region. For
$N > 1$, we are unable to find solvable approximate equations
which give an overlapping inner and outer region.

For the M2-brane, we have $D=11$, $d=3$, $\td d =6$ and $N=1$:
%%%%%%%%%
\be \sigma_{\rm M2-brane}^{(l)} = \frac{\pi^5(\ell+5)(\ell+4)
[1-(m/\omega)^2]^{\ell-1/2}}{(15)2^{3\ell+2}\ell!(\ell+2)!
\Gamma^2(\frac{3+\ell}{2})} \omega^{3\ell+2} R^{3\ell+9} \ee
%%%%%%%%
For the M5-brane, we have $D=11$, $d=6$, $\td d =3$ and $N=1$:
%%%%%%%%
\be \sigma_{\rm M5-brane}^{(l)} = \frac{2^{2\ell+5} \pi^3
(\ell+2)(\ell+3/2)(\ell+1)[(\ell+1)!]^2
[1-(m/\omega)^2]^{\ell-1/2}}{(2\ell+3)^2[(2\ell+2)!]^4}
\omega^{6\ell+5} R^{6\ell+9} \ee
%%%%%%%%

In fact, for all $N=1$ $p$-branes, other than the two for which
the approximation scheme cannot be applied, the partial absorption
cross-sections have the same additional factor due to the scalar
mass:
%%%%%%%%
\be \sigma_{\rm massive}^{\ell}= \sigma_{\rm massless}^{\ell}
[1-(m/\omega)^2]^{\ell-1/2}, \ee
%%%%%%%%
for $m \le \omega$, and $\sigma_{\rm massless}^\ell$ has the same
form as the leading-order absorption for massless scalars.  Note
that the suppression [enhancement] of the partial cross-section
for $\ell\ge 1$ [for $\ell =0$], when the non-relativistic limit
is taken.

\section{Concluding remarks for massive absorption}

We have addressed the absorption cross-section for
minimally-coupled massive particles in extremal and near-extremal 
$p$-brane backgrounds. In particular, we found exact absorption
probabilities in the cases of the extreme self-dual dyonic string
in $D=6$ and two equal-charge extreme black hole in $D=4$. Notably
these two examples yield the same wave equations as that of the
minimally coupled massless scalar in the $D=6$ extreme dyonic
string, and two charge $D=4$ extreme black hole backgrounds,
respectively. Namely, one of the two charge parameters in the
latter (massless) case is traded for the scalar mass parameter in
the former (massive) case.  Thus, for these equal charge
backgrounds, the scattering of minimally-coupled massive particles
can be addressed explicitly, and the distinct behavior of the
absorption cross-section on the energy $\omega$ (or equivalently
momentum $p\equiv \sqrt{\omega^2-m^2}$) is studied.  In
particular, the non-relativistic limit of the particle motion
gives rise to a distinct, resonant-like absorption behavior in the
case of the self-dual dyonic string.

We have also found corrections due to the scalar mass for the
leading-order absorption cross-sections for D3-, M2- and
M5-branes. In particular, in the non-relativistic limit, there is
the expected suppression [enhancement] in the absorption
cross-section for partial waves $\ell\ge 1$ [$\ell=0$].

The results obtained for the absorption cross-section of the
minimally-coupled massive scalars, in particular those in the
extreme self-dual dyonic string background, may prove useful in
the study of AdS/CFT correspondence~\cite{malda}. Namely, the
near-horizon region of the extreme dyonic string background has
the topology of $AdS_3\times S^3$, with the $AdS_3$ cosmological
constant $\Lambda$ and the radius $R$ of the three-sphere ($S^3$)
related to the charge $Q$ of the self-dual dyonic string as
$\Lambda= R^2= \sqrt{Q}$ (see e.g., \cite{cvlar3}).  On the other
hand, the scattering of the minimally-coupled massive fields (with
mass ${\cal M}$) in the $AdS_3$ background yields
information~\cite{gkp,wit} on the correlation functions of the
operators of the boundary $SL(2,{\bf R})\times SL(2,{\bf R})$
conformal field theory~\cite{hen} with conformal dimensions
$h_{\pm}=\ft12(1\pm\sqrt{1+{\cal M}^2\Lambda^2})$~\cite{bala2}.
The scattering analyzed here corresponds to that of a
minimally-coupled massive scalar in the the full self-dual string
background, rather than in only the truncated $AdS_3$ background.
These explicit supergravity results may, in turn, shed light on
the pathologies of the conformal field theory of the dyonic string
background~\cite{seiwitt}.

%%%%%%%%%%%%%%%%%%%%%%%%%%%%%%%%%%%%%%%%%%%%%%%%%%%%%%%%%%%%%%%%%%%%%%%%%
%%%%%%%%%%%%%%%%%%%%%%%%%%%%%%%%%%%%%%%%%%%%%%%%%%%%%%%%%%%%%%%%%%%%%%%%%
%%%%%%%%%%%%%%%%%%%%%%%%%%%%%%%%%%%%%%%%%%%%%%%%%%%%%%%%%%%%%%%%%%%%%%%%%

\section{Absorption by D3-brane}

The extremal D3-brane background is of special interest, since the
correspondence there is to D=4 super Yang-Mills theory. While the
scattering processes for the massless, minimally-coupled scalar has been
extensively studied, the scattering of other fields has been
explored to a lesser extent, and that only for low-energies (see,
for example, \cite{mathur,hoso} and references therein);
nevertheless it is expected that non-renormalization theorems on
the field theory side ensure a precise agreement between the low
energy absorption cross-sections and the corresponding weak
coupling calculation of the n-point correlation functions on the
field-theory side (see, for example, \cite{Friedman,Liu,dhoker}
and references therein).

We provide an analysis of the
absorption cross-section in the extremal D3-brane background for a
broad class of massless modes and for the whole energy range. In
particular, we uncover a pattern in the energy dependence of the
absorption cross-sections for both integer and half-integer spins;
certain half-integer and integer spin pairs have identical
absorption cross-sections, thus providing an evidence on the
gravity side that such pairs couple on the dual field theory side
to the pairs of operators forming supermultiplets of strongly
coupled gauge theory.

      In the next section, we cast the wave equations of various fields
into\\ Schr\"odinger form, and obtain the effective Schr\"odinger
potentials.  We show that effective 
Schr\"odinger potentials for certain fields are identical
and hence the absorption for these fields is the same. In other
cases where Schr\"odinger potentials are not the same, we argue
that the different potentials are dual and yield the same
absorption probabilities, with numerical results supporting these
claims.  Next, we obtain numerical results for the
absorption of a large class of fields for the whole energy range
in the extremal D3-brane background.  The method for the numerical
evaluation of the absorption probabilities is given in Appendix A.1,
while the calculation of the high energy absorption cross-section
in the geometrical optics limit is given in Appendix A.2.

The  D3-brane of the type IIB supergravity is given by
%%%%%%%
\bea ds^2_{10}&=&H^{1/2} (-fdt^2 + dx_1^2 + dx_2^2 + dx_3^2) +
H^{1/2} (f^{-1} dr^2 + r^2\, d\Omega_5^2)\,,\nn\\
G_\5 &=& d^4x\wedge dH^{-1} + {*(d^4x\wedge
dH^{1}})\,.\label{d3metric} \eea where \be H=1 + {R^4\over r^4}, \
\ f=1- {{2m}\over r^4}. \ee
%%%%%
Here $R$ specifies the D3-brane charge and $m$ is the
non-extremality parameter (defined for convenience as $m\equiv \mu
R^4/\sqrt{1+\mu}$). We shall primarily concentrate on the extremal
limit $m=\mu=0$. (See, however, Appendix B for the discussion of
the high energy limit of the absorption cross-section in the
non-extreme D3-background.)

The low energy absorption probabilities for various bosonic
linearly-excited massless fields under this background were
obtained in \cite{mathur}.  The low energy absorption
probabilities for the dilatino and the two-form field are given in
\cite{hoso} and \cite{raja}, respectively (and for massive
minimally coupled modes in \cite{mass}).  In the following
subsections we study the absorption probabilities for the whole
energy range and uncover completely parallel structures.  In
particular, we shall cast the wave equation for different modes
into Schr\"odinger form and discuss the pattern of the the
Schr\"odinger potentials. We also provide a conjectured form of
the dual potentials which, in turn, yield the same absorption
probabilities. In the subsequent section we confirm the pattern
with numerical results.

\section{Effective potentials of extremal D3-brane}

The field equations are
%%%%%%%%%
\be
R_{\mu \nu}=-\frac{1}{6} F_{\mu \rho \sigma \tau \kappa} F_{\nu} ^{\rho
\sigma \tau \kappa}
\ee
%%%%%%%%%
%%%%%%%%%
\be
F_{\mu \nu \rho \sigma \tau}=\frac{1}{5!}\epsilon_{\mu \nu \rho \sigma
\tau \mu ^{\prime} \nu ^{\prime} \rho ^{\prime} \sigma ^{\prime} \tau
^{\prime}} F^{\mu ^{\prime} \nu ^{\prime} \rho ^{\prime} \sigma ^{\prime} 
\tau ^{\prime}} 
\ee
%%%%%%%%%
%%%%%%%%%
\be
D^{\mu}\partial_{[\mu}A_{\nu \rho]}=-\frac{2i}{3} \dot{F}_{\nu \rho \sigma
\tau \kappa}D^{\sigma}A^{\tau \kappa}
\ee
%%%%%%%%%
%%%%%%%%%
\be
D^{\mu}\partial_{\mu}B=0,
\ee
%%%%%%%%%
where a dot denotes the background value of a field, and the $\epsilon$
symbol is a tensor with
%%%%%%%%%
\be
\epsilon_{12r03\alpha \beta \gamma \delta \epsilon}=\sqrt{-g},
\ee
%%%%%%%%%
with $\alpha \beta \gamma \delta \epsilon$ giving a frame with positive
orientation on $S^5$. The symbol $[]$ on subscripts denotes
anti-symmetrization with strength one, i.e., $A_{[\alpha
\beta]}=\frac{1}{2}(A_{\alpha \beta}-A_{\beta \alpha})$.

In terms of the fields of ten-dimensional supergravity,
%%%%%%%%%
\be
A_{\mu \nu}=B_{\mu \nu}^{NS-NS}+iB_{\mu \nu}^{RR}.
\ee
%%%%%%%%%
The field $B$ is a complex scalar describing the dilaton and the $RR$
scalar.

In the following sections, we consider the linearized field perturbations
around their background values.

\subsection{Dilaton-axion}

The axion and dilaton of the type IIB theory are decoupled from
the D3-brane. Thus, in  the D3-brane background, they satisfy the
minimally-coupled scalar wave equation
%%%%
\be \fft{1}{\sqrt{g}} \del_\mu \sqrt{g}
g^{\mu\nu}\del_\nu\phi=0\,. \ee
%%%%%
It follows from (\ref{d3metric}) that the radial wave equation of
a dilaton-axion in the spacetime of an extremal D3-brane is given
by
%%%%%%%%
\be \Big( \frac{1}{\rho^5} \frac{\partial}{\partial \rho} \rho^5
\frac{\partial}{\partial \rho} + H - \frac{\ell (\ell +
4)}{\rho^2} \Big) \phi (\rho) = 0, \label{D3eqn} \ee
%%%%%%%%%
where
%%%%%%%%
\be H=1+\frac{e^4}{\rho^4} \ee
%%%%%%%%
and $\ell=0,1,\ldots$ corresponds to the $\ell^{th}$ partial
wave.

The quantity $e$ and $\rho$ are dimensionless energy and radial
distance parameters: $e=\omega R$ and $\rho=\omega r$.   The
leading order and sub-leading order cross-sections of the
minimally-coupled scalar by the D3-brane background were obtained
in \cite{kleb,ghkk} by matching inner and outer solutions of the
wave equations.  It was observed in \cite{gubserhash} that if one
performs the following change of variables
%%%%%
\be \rho=e\, Exp(-z)\,,\qquad \phi(r) = Exp(2z)\,\psi(r)\,, \ee
%%%%%
the wave equation (\ref{D3eqn}) becomes
%%%%%
\be \Big[\fft{\del^2}{\del z^2} + 2e\, \cosh(2z) -
(\ell+2)^2\Big]\, \psi(z)=0\,, \ee
%%%%%
which is precisely the modified Mathieu equation.  One can then
obtain analytically the absorption probability order by order in
terms of dimensionless energy $e$ \cite{gubserhash}.

In this paper, we shall express the wave equation in Schr\"odinger
form, and study the characteristics of the Schr\"odinger effective
potential.  By the substitution
%%%%%%%%
\be \phi = \rho^{-5/2} \psi, \ee
%%%%%%%%
we render (\ref{D3eqn}) into Schr\"odinger form
%%%%%%%%
\be \big(\frac{\partial^2}{\partial \rho^2}-V_{\rm eff} \big)\psi
= 0, \label{eqdil} \ee
%%%%%%%%
where
%%%%%%%%
\be V_{\rm eff}(\ell) = -H+\frac{(\ell+3/2)(\ell+5/2)}{\rho^2}
\equiv V_{\rm dilaton}(\ell). \label{V} \ee
%%%%%%%
Factors shared by the incident and outgoing parts of the wave
function cancel out when calculating the absorption probability.
Thus, the absorption probability of $\phi$ and $\psi$ are the
same.

Technically, $V_{\rm eff}$ cannot be interpreted as an effective
potential, since it depends on the particle's incoming energy. It
is straightforward to use a coordinate transformation to put the
equation in the standard Schr\"odinger form, where $V_{\rm eff}$
is independent on the energy.   However, for our purposes of
analyzing and comparing the form of the wave equations for various
fields, this is of no consequence.

Note that the first term in (\ref{V}) represents the spacetime
geometry of the extremal D3-brane, whereas the second term
represents the angular dynamics (partial modes) of the particles.
We shall see presently that some particles have effective
potentials which contain terms mixed with both ``geometrical'' and
``angular'' dynamical contributions.

\subsection{Antisymmetric tensor from 4-form}

For two free indices of the 4-form along $S^5$ and two free
indices in the remaining 5 directions, the radial wave equation
for the antisymmetric tensor derived from the 4-form is
\cite{mathur}
%%%%%%
\be \Big( \frac{1}{\rho} \frac{\partial}{\partial \rho} \rho
\frac{\partial}{\partial \rho} +H-\frac{(\ell+2)^2}{\rho^2} \Big)
\phi (\rho)=0, \label{4-form} \ee
%%%%%%
where $\ell=1,2,\ldots$.  By the substitution
%%%%%%%
\be \phi=\rho^{-1/2} \psi, \ee
%%%%%%%
we render (\ref{4-form}) into Schr\"odinger form with
%%%%%%%
\be V_{\rm 4-form}(\ell)=V_{\rm dilaton}(\ell). \ee
%%%%%%%%

\subsection{Dilatino}

The radial wave equation for the dilatino on an extremal D3-brane
was found \cite{hoso} by inserting the following spherical wave
decomposition form for the dilatino field $\lambda$ into the
covariant Dirac equation:
%%%%%%%
\be H^{\frac{1}{8}} \lambda=e^{-i \omega t} r^{-\frac{5}{2}} \Big(
F(r) \Psi_{-\ell}^{\pm} +i G(r) \big( \frac{\Gamma^{0} \Gamma^{i}
x_{i}}{r} \big) \Psi_{-\ell}^{\pm} \Big), \ee
%%%%%%%
where $\Gamma^i$ are field-independent gamma-matrices and
$i={4,\ldots,9}$ runs normal to the brane.  $\Psi_{-\ell}^{\pm}$
is the eigenspinor of the total angular momentum with $\Sigma_{ij}
L_{ij}=-\ell$, $\Gamma^{0123}=\pm i$.

The spatial momenta tangential to the branes can be made to vanish
via the Lorentz transformations and the radial wave equations are
obtained \cite{hoso}:
%%%%%%%%
\bea \omega H^{\frac{1}{2}} F+\Big(
\frac{d}{dr}+\frac{\ell+5/2}{r} \pm \frac{1}{4}(\ln H)' \Big)
G&=&0
\label{f}\\
-\omega H^{\frac{1}{2}} G+\Big( \frac{d}{dr}-\frac{\ell+5/2}{r}
\mp \frac{1}{4}(\ln H)' \Big) F&=&0 \label{ff} \eea
%%%%%%%%
Decoupling (\ref{f}) and (\ref{ff}) yields second-order
differential equations for $F$ and $G$.

Let us first consider the case of positive eigenvalue, {\it i.e.},
$\Gamma^{0123}=+i$.  In this case, the second-order wave equation
for $F$ can be cast into Schr\"odinger form by the substitution
$F=H^{1/4}\, \psi$, giving rise to the effective Schr\"odinger
potential
%%%%%%%%
\be V^F_{\rm +dilatino}(\ell)= V_{\rm
dilaton}(\ell)\,.\label{dilatino1} \ee
%%%%%%%%
Thus, the absorption probability for $F$ is identically the same
as that for the dilaton-axion.  The wave equation for $G$ can also
be cast into Schr\"odinger form by the substitution $G=H^{1/4}\,
\psi$. The corresponding Schr\"odinger potential, on the other
hand, takes a different form, given by
%%%%%%%%
\be V^G_{\rm +dilatino}(\ell)= V_{\rm dilaton}(\ell) + \frac{-3e^8
-2e^8 \ell-10e^4 \rho^4 +5\rho^8 +2\rho^8 \ell}{H^2
\rho^{10}}\,.\label{dilatino2} \ee
%%%%%%%%
Thus, we see that the effective potentials for the two components
$F$ and $G$ are quite different.  However, we expect that these
two potentials, although different, yield the same absorption
probability: they form a dual pair of potentials.  While it is not
clear to us how to present a rigorous analytical proof, our
numerical calculation (in section 4) confirm that, indeed, they
yield the same absorption probability.

Similar results are obtained for the the negative eigenvalue
solutions, {\it i.e.}, $\Gamma^{0123}=-i$:
%%%%%%%%
\bea
V^G_{\rm -dilatino}(\ell)&=& V_{\rm dilaton}(\ell +1)\,,\nn\\
V^F_{\rm -dilatino}(\ell)&=& V_{\rm dilaton}(\ell +1) +
\frac{-5\rho^8 -2\rho^8 \ell-10e^4 \rho^4 +7e^8 +2e^8 \ell}{H^2
\rho^{10}}. \eea
%%%%%%%%
Again, the numerical results in section 4 indicate that the above
two potentials yield the same absorption probabilities and hence
form a dual pair.

        In the  above discussion of the dilatino scattering equation,
we  encountered three different potentials, namely
%%%%%%
\be V^G_{\rm +dilatino}(\ell)\mapright{\rm dual} V^F_{\rm
+dilatino}(\ell)=V^G_{\rm -dilatino}(\ell-1)\mapright{\rm dual}
V^F_{\rm -dilatino}(\ell-1) \ee which all yield the same
absorption probability.

   The above structure of the dual potentials can be cast in a more
general form.  Namely, the above dual potential pairs for the
dilatino can be cast into the following form:
%%%%%%%
\be V(\ell)=-H+\frac{(2\ell+\alpha)(2\ell+\alpha \pm 2)}{4\rho^2},
\label{form} \ee
%%%%%%%
and
%%%%%%%
\be V^{\rm dual}(\ell)=V(\ell)+ \frac{\pm[(2\ell+\alpha \pm 2)e^8-
(2\ell +\alpha)\rho^8]-10e^4 \rho^4}{\rho^{10} H^2}, \label{dual}
\ee
%%%%%%%
where $\alpha=5$ and we use in the $\pm$ sign in (\ref{dual}) for
the $\mp$ eigenvalue, respectively. We have found numerically that
(\ref{form}) and (\ref{dual}) are dual potentials for integer
values of $\alpha$. In particular, for odd values of $\alpha$,
(\ref{form}) can be identified with $V_{\rm
dilaton}(\ell+(\alpha-4 \pm 1)/2)$, in which case the absorption
can be found analytically, since the wave equation is that of
Mathieu equation.

Thus, in all the subsequent examples when the potential is of the
form (\ref{dual}), we are now able to identify a dual potential
(\ref{form}) of a simpler form with a corresponding wave equation
which can be solved analytically.

\subsection{Scalar from the two-form}

For the free indices of the two-form taken to lie along the $S^5$,
the radial wave equation is \cite{mathur}
%%%%%%%%
\be \Big( \frac{H}{\rho} \frac{\partial}{\partial \rho}
\frac{\rho}{H} \frac{\partial}{\partial \rho} + H -
\frac{(\ell+2)^2}{\rho^2} \mp \frac{4e^4}{H \rho^6} (\ell+2) \Big)
\phi (\rho) = 0, \label{scalar} \ee
%%%%%%%%%
where again $\ell=1,2,\ldots$ correspond to the $\ell^{th}$
partial wave.  The sign $\pm$ corresponds to the sign in the
spherical harmonics involved in the partial wave expansion.

By the substitution
%%%%%%%%
\be \phi = \rho^{-1/2} H^{1/2} \psi, \ee
%%%%%%%%
we render (\ref{scalar}) into Schr\"odinger form.

      For the positive eigenvalue, the effective potential is of the
form given by (\ref{dual}), with $\alpha=5$ and a positive sign.
This is, in fact, the same as that for the dilatino with negative
eigenvalue. Thus, the dual potential is
%%%%%%%%
\be V_{\rm +scalar}(\ell)\mapright{\rm dual} V_{\rm dilaton}(\ell
+1). \ee
%%%%%%%%
For the negative eigenvalue, the effective potential is of the
form given by (\ref{dual}), with $\alpha=3$ and a negative sign.
Thus, the dual potential is
%%%%%%%%
\be V_{\rm -scalar}(\ell)\mapright{\rm dual} V_{\rm dilaton}(\ell
-1). \ee
%%%%%%%%

\subsection{Two-form from the antisymmetric tensor}

The equations for two-form perturbations polarized along the
D3-brane are coupled. For s-wave perturbations, they can be
decoupled \cite{mathur,raja}, and the radial wave equation is
%%%%%%%%
\be \Big( \frac{1}{\rho^5 H} \frac{\partial}{\partial \rho} \rho^5
H \frac{\partial}{\partial \rho} + H - \frac{16 e^8}{\rho^{10}
H^2} \Big) \phi (\rho) = 0, \label{2-form} \ee
%%%%%%%%%
where $\ell=0,1,\ldots.$   By the substitution
%%%%%%%%
\be \phi = \rho^{-5/2} H^{-1/2} \psi, \ee
%%%%%%%%
we render (\ref{2-form}) into Schr\"odinger form given by
(\ref{dual}), with $\alpha=5$, $\ell=0$ and the positive sign.
Thus, the dual potential is

%%%%%%%%
\be V_{\rm 2-form}(\ell=0)\mapright{\rm dual} V_{\rm
dilaton}(\ell=1). \ee
%%%%%%%%

\subsection{Vector from the two-form}

We now consider one free index of the two-form along $S^5$ and one
free index in the remaining 5 directions. For the tangential
components of this vector field, the radial wave equation is
\cite{mathur}
%%%%%%%%
\be \Big( \frac{1}{\rho^3} \frac{\partial}{\partial \rho} \rho^3
\frac{\partial}{\partial \rho} + H -
\frac{(\ell+1)(\ell+3)}{\rho^2} \Big) \phi (\rho) = 0,
\label{tang} \ee
%%%%%%%%%
where $\ell=1,2,\ldots.$   By the substitution
%%%%%%%%
\be \phi = \rho^{-3/2} \psi, \ee
%%%%%%%%
we render (\ref{tang}) into Schr\"odinger form with
%%%%%%%
\be V_{\rm tangential-vector}(\ell)=V_{\rm dilaton}(\ell). \ee
%%%%%%%%%
The radial $a_r$ and time-like  $a_0$ components of the vector
field are determined by the following coupled first order
differential equations \cite{mathur}:
%%%%%%%
\be i \frac{\partial}{\partial \rho} a_o = \big[
1-\frac{(\ell+1)(\ell+3)}{\rho^2 H} \big] a_r \label{a} \ee
%%%%%%%
and
%%%%%%%
\be i \frac{1}{\rho} \frac{\partial}{\partial \rho} \big(
\frac{\rho}{H} a_r \big) = a_o, \label{aa} \ee
%%%%%%
where $\ell=1,2,\ldots.$ Eqs. (\ref{a}) and (\ref{aa}) can be
decoupled and the wave equation for $a_r$ is
%%%%%%%%
\be \Big( H \frac{\partial}{\partial \rho} \frac{1}{\rho}
\frac{\partial}{\partial \rho} \frac{\rho}{H} + H -
\frac{(\ell+1)(\ell+3)}{\rho^2} \Big) a_r = 0. \label{ar} \ee
%%%%%%%%%
vBy the substitution
%%%%%%%%
\be a_r = \rho^{-1/2} H \psi, \ee
%%%%%%%%
we render (\ref{ar}) into Schr\"odinger form with
%%%%%%%
\be V_{\rm radial-vector}(\ell)=V_{\rm dilaton}(\ell). \ee
%%%%%%%
The wave equation for $a_0$ is
%%%%%%
\be \Big( \frac{1}{\rho} \frac{\partial}{\partial \rho}
\frac{\rho^3}{\rho^2 H-(\ell+1)(\ell+3)} \frac{\partial}{\partial
\rho} +1 \Big) a_0 =0. \label{a0} \ee
%%%%%%%
By the substitution
%%%%%%%
\be a_0=\rho^{-3/2} \sqrt{\rho^2 H-(\ell+1)(\ell+3)} \psi, \ee
%%%%%%%%
we render (\ref{a0}) into Schr\"odinger form with
%%%%%%%
\be V_{\rm 0-vector}(\ell)=V_{\rm dilaton}(\ell)+\frac{3e^8 -10e^4
\rho^4 +4\rho^6 (\ell+1) (\ell+3)-\rho^8}{\rho^{10} \big(
H-\frac{(\ell+1)(\ell+3)}{\rho^2} \big)^2}. \ee
%%%%%%%
Note that this effective potential is nonsingular only for $e^2 >
(\ell+1)(\ell+3)/2$. We have numerically confirmed that $V_{\rm
0-vector}(\ell)$ is dual to $V_{\rm dilaton}(\ell)$. Working
directly from Eqs. (\ref{a})-(\ref{aa}), we have numerically
confirmed that $a_0$ shares the same absorption probability with
the dilaton-axion, for all values of $e$.

\section{Qualitative features of absorption by the extremal D3-brane}

We have obtained numerical absorption probabilities by a method
described in Appendix A. Fig. 1 shows the s-wave absorption
probabilities for all the particles that we have studied on an
extremal D3-brane vs. the energy of incoming particles measured in
dimensionless units of $e \equiv \omega R$. %%%%%%%%
\begin{figure}
   \epsfxsize=4.0in
   \centerline{\epsffile{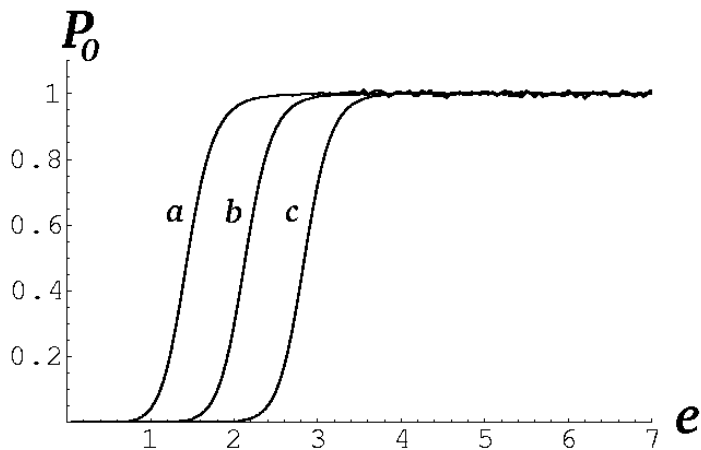}}
   \caption[FIG. \arabic{figure}.]{s-wave absorption
probabilities of various particles on an extremal D3-brane}
\end{figure}
%%%%%%%%
(a) is the s-wave absorption probability of the dilaton-axion,
dilatino with positive total angular momentum eigenvalue and
scalar from the two-form with a negative sign in the spherical
harmonic. (b) is the s-wave absorption probability of the dilatino
with negative total angular momentum eigenvalue, two-form from the
antisymmetric tensor, antisymmetric tensor from the four-form and
the longitudinal and tangential components of the vector from the
two-form. (c) is the s-wave absorption probability of the scalar
from the two-form with a positive sign in the spherical harmonic.
These absorption probabilities are simply related by $\ell
\rightarrow \ell \pm 1$. This structure is suggestive of a
particle supermultiplet structure-- namely, different multiplets.

This demonstrates the similar and surprisingly simple structure of
absorption probabilities between the various particles, which is
not apparent from previous analytical low-energy absorption
probabilities. These numerical results also support the idea of
dual potentials.

Fig. 2 shows the partial absorption probabilities for a
dilaton-axion on an extremal D3-brane vs. the energy of incoming
particles.
%%%%%%%%
\begin{figure}
   \epsfxsize=4.0in
   \centerline{\epsffile{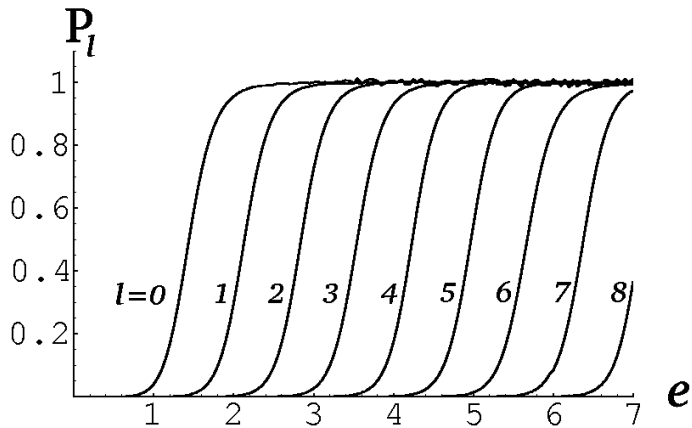}}
   \caption[FIG. \arabic{figure}.]{Partial absorption probabilities for a
dilaton-axion on an extremal D3-brane}
\end{figure}
%%%%%%%%
There is no absorption at zero energy and total absorption is
approached at high energy. Thus, the high energy absorption
cross-section is found by setting $P=1$ in (\ref{prob}):
%%%%%%
\be \sigma^{(\ell)}=\frac{8 \pi^2 (\ell+3)(\ell+2)^2(\ell+1)}{3
(\omega R)^2} \ee
%%%%%%%%
In fact, for all types of waves absorbed by all branes and black
holes studied thus far \cite{mynotes}, total partial wave
absorption occurs at high energy. It is conjectured that this is a
general property of absorption for all objects that have an event
horizon. Results for objects other than D3-branes will be
published shortly by the present authors.

The partial absorption probabilities of a massive
minimally-coupled scalar on an extremal D3-brane vs. the energy of
the incoming particles is shown in Fig. 3.
%%%%%%%%
\begin{figure}
   \epsfxsize=4.0in
   \centerline{\epsffile{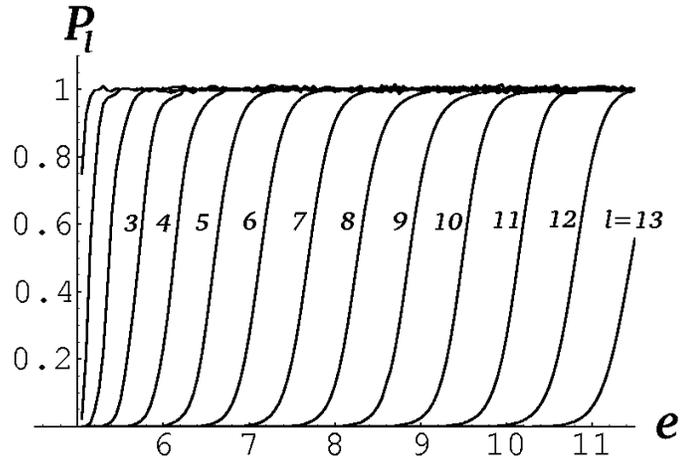}}
   \caption[FIG. \arabic{figure}.]{Partial absorption probabilities for
a massive minimally-coupled scalar on an extremal D3-brane}
\end{figure}
%%%%%%%%
As the mass of the scalar is increased, the partial absorption
probabilities occur at lower energies. Physically, there is
greater absorption at low energy due to the additional
gravitational attraction that is present from the nonzero mass.
The low energy absorption probability for this case were obtained
in \cite{mass}.

Fig. 4 shows the numerical s-wave absorption cross-section for a
dilaton-axion on an extremal D3-brane (continuous line),
super-imposed with previously obtained low energy semi-analytical
results (short dashes) \cite{gubserhash}, as well as high energy total
absorption (long dashes) vs. the energy of incoming particles.
Throughout the rest of this paper, absorption cross-sections are
plotted in units of $R^5$.
%%%%%%%%
\begin{figure}
   \epsfxsize=4.0in
   \centerline{\epsffile{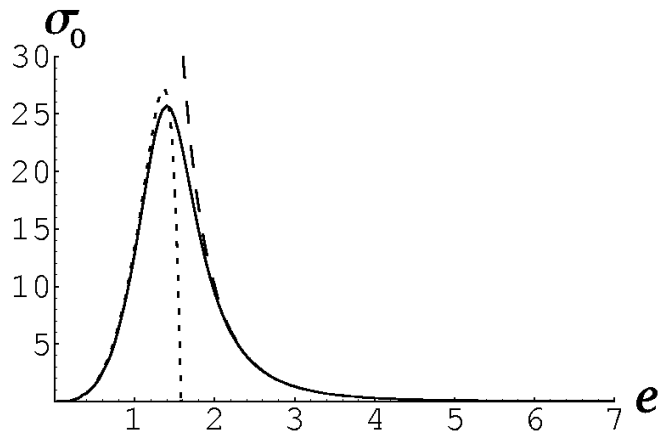}}
   \caption[FIG. \arabic{figure}.]{s-wave absorption
cross-section for a dilaton-axion on an extremal D3-brane}
\end{figure}
%%%%%%%%
The resonance roughly corresponds to the region in which there is
a transition from zero absorption to total absorption.

As evident from Fig. 5, the energies at which there is a resonance
in the partial absorption cross-section are proportional to the
partial-wave number.
%%%%%%%%
\begin{figure}
   \epsfxsize=4.0in
   \centerline{\epsffile{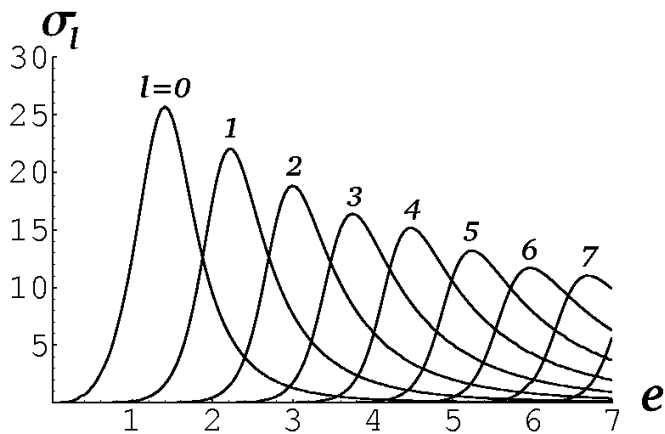}}
   \caption[FIG. \arabic{figure}.]{Partial absorption cross-sections for a
dilaton-axion on an extremal D3-brane}
\end{figure}
%%%%%%%%
Also, the magnitude of the peak of each partial absorption
cross-section decreases with the partial-wave number. These
characteristics seem reasonable if one considers particle
dynamics; in terms of radial motion, rotational kinetic energy
counteracts gravitational attraction.

In Fig. 6, we plot the partial-wave effective potentials vs.
radial distance (in dimensionless units) for a dilaton-axion on an
extremal D3-brane. These are plotted at the energies of the maxima
of the corresponding partial-wave absorption cross-sections. For
the s-wave absorption cross-section, the maximum is at $e=1.4$.
The maxima of higher partial-wave absorption cross-sections are
separated by energy gaps of approximately $\Delta e=.75$, with
increasing $\ell$.

%%%%%%%%
\begin{figure}
   \epsfxsize=4.0in
   \centerline{\epsffile{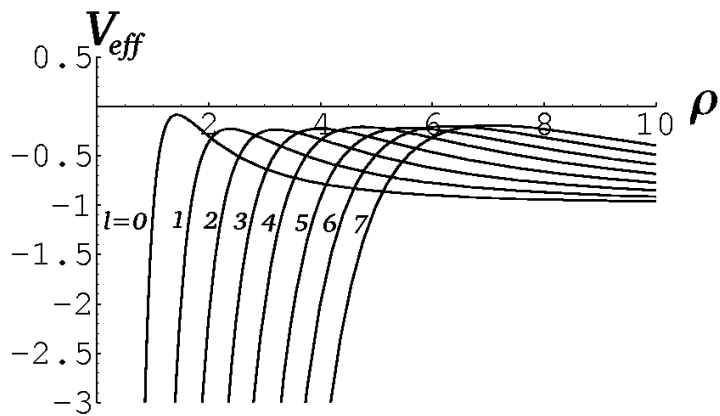}}
   \caption[FIG. \arabic{figure}.]{Partial effective potentials absorption
for a dilaton-axion on an extremal D3-brane}
\end{figure}
%%%%%%%%
As can be seen, the incoming particle must penetrate an effective
barrier in order to be absorbed by the D3-brane. Once absorbed,
the waves inhabit quasi-bound states until they quantum tunnel to
asymptotically flat spacetime. As is shown, the heights of the
partial-wave effective potential barriers at the energies of the
maxima of the partial-wave absorption cross-sections are roughly
equal, which partly explains the similar structure of the partial
absorption probabilities of different partial-wave numbers. Thus,
the decreasing maximum values of the partial-wave absorption
cross-sections with increasing $\ell$ arises purely as a result of
the phase-factors.

The superposition of maxima of partial-wave absorption
cross-section leads to the oscillatory character of the total
absorption cross-section with respect to the energy of the
incoming particles. This is shown in Fig. 7 for the case of the
dilaton-axion, dilatino and scalar from the two-form for
comparison.
%%%%%%%%
\begin{figure}
   \epsfxsize=4.0in
   \centerline{\epsffile{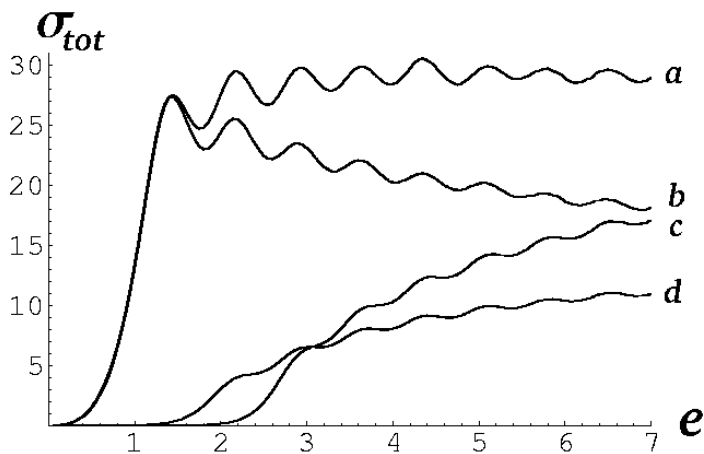}}
   \caption[FIG. \arabic{figure}.]{Total absorption cross-sections for
the dilaton-axion, dilatino and scalar from two-form on an
extremal D3-brane}
\end{figure}
%%%%%%%%
(a) is the total absorption cross-section of the dilaton-axion on
an extremal D3-brane, (b) is that of the dilatino with positive
total angular momentum eigenvalue, (c) is that of the scalar from
the two-form with positive sign in the spherical harmonic, and (d)
is that of the dilatino with negative total angular momentum
eigenvalue.

The amplitude of oscillation decreases exponentially with energy.
For the dilaton-axion, the total absorption cross-section
converges to the geometrical optics limit at high energy, which we
have calculated in a previous section.  The oscillatory behavior
is shared by all total absorption cross-sections that have been
studied as of this time \cite{mynotes}, for various particles in
various spacetime backgrounds. The oscillatory structure of the
absorption cross-section of a scalar on a Schwarzschild black hole
has been noted by Sanchez \cite{sanchez}.

It is interesting to note that extinction cross-sections which
arise in the field of optics have similar oscillatory properties
\cite{hulst}.

Fig. 8 shows the total absorption cross-section of the scalar from
the two-form with negative total angular momentum eigenvalue.
%%%%%%%%
\begin{figure}
   \epsfxsize=4.0in
   \centerline{\epsffile{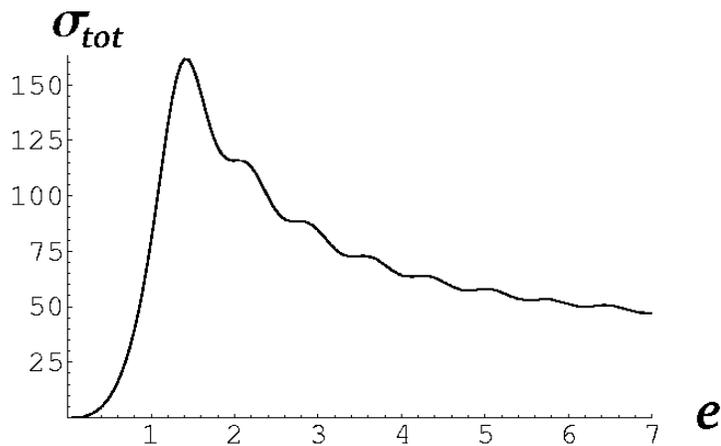}}
   \caption[FIG. \arabic{figure}.]{Total absorption cross-section for
the scalar from two-form with negative total angular momentum
eigenvalue}
\end{figure}
%%%%%%%%

As already noted and shown in Fig. 3, as the mass of the
minimally-coupled scalar is increased, the partial-wave absorption
probabilities occur at lower energies. This causes the drastic
qualitative difference between the massless and massive scalar
cases at low energy, i.e., the divergent total absorption
cross-section at zero energy for the massive scalar, as is shown
in Fig. 9.
%%%%%%%%
\begin{figure}
   \epsfxsize=4.0in
   \centerline{\epsffile{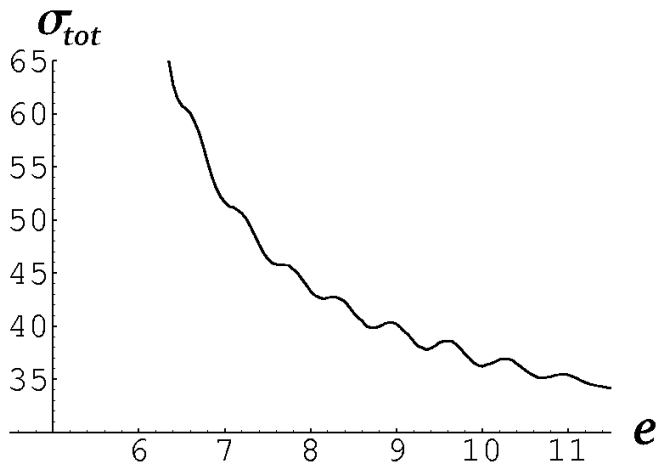}}
   \caption[FIG. \arabic{figure}.]{Total absorption cross-section for
a massive minimally-coupled scalar on an extremal D3-brane}
\end{figure}
%%%%%%%%

%%%%%%%%%%%%%%%%%%%%%%%%%%%%%%%%%%%%%%%%%%%%%%%%%%%%%%%%%%%%%%%%%%%%%%
%%%%%%%%%%%%%%%%%%%%%%%%%%%%%%%%%%%%%%%%%%%%%%%%%%%%%%%%%%%%%%%%%%%%%%
%%%%%%%%%%%%%%%%%%%%%%%%%%%%%%%%%%%%%%%%%%%%%%%%%%%%%%%%%%%%%%%%%%%%%%

\section{Absorption by near-extremal D3-branes}

It has been shown, through
the analysis of wave equations in Schr\"odinger form, that certain
half-integer and integer spin massless modes of the extremal
D3-brane have identical absorption probabilities \cite{poritz},
which implies that such fields couple on the dual field theory to
operators forming supermultiplets of strongly coupled gauge
theory. In an effort to enhance our understanding of non-BPS
states, we continue this study for near-extremal
D3-branes, which have already been probed by minimally-coupled massless 
scalars \cite{siopsis,siopsis2}.

In the next section, we present
the wave equations for various linear field perturbations in a
non-extremal D3-brane background, for which the ten-dimensional
metric is left unperturbed. Then we calculate the
low-energy absorption probabilities for these field perturbations
in the background of near-extremal D3-branes, using a method that
was initially applied to a minimally-coupled scalar probe
\cite{siopsis,siopsis2}. 

\section{Effective potentials of non-extremal D3-branes}

The D3-brane of type IIB supergravity is given by
%%%%%%%
\begin{eqnarray}
ds^2_{10}&=&H^{-1/2} (-f dt^2+ dx_1^2 +dx_2^2+dx_3^2) + H^{1/2}
(f^{-1} dr^2 + r^2 d\Omega_{5}^2), \\ \nonumber G_{(5)}&=&d^4 x
\wedge dH^{-1}+\ast (d^4 x \wedge dH^{-1}). \label{d3metric}
\end{eqnarray}
%%%%%%%
where $H=1+\frac{R^4}{r^4}$ and $f=1 - \frac{r_o^4}{r^4}$. $R$
specifies the D3-brane charge and $r_o$ is the non-extremality
parameter. The field equations are
%%%%%%%
\begin{eqnarray}
R_{\mu \nu} &=& -\frac{1}{6}F_{\mu \rho \sigma \tau \kappa}F_{\nu}
^{\rho \sigma \tau \kappa}, \\ \nonumber F_{\mu \nu \rho \sigma
\tau}&=&\frac{1}{5!}\epsilon_{\mu \nu \rho \sigma \tau \mu^{'}
\nu^{'} \rho^{'} \sigma^{'} \tau^{'}}F^{\mu^{'} \nu^{'} \rho^{'}
\sigma^{'} \tau^{'}}, \\ \nonumber D^{\mu} \partial_{[\mu}A_{\nu
\rho ]} &=& -\frac{2i}{3} \dot{F}_{\nu \rho \sigma \tau \kappa}
D^{\sigma} A^{\tau \kappa}, \\ \nonumber D^{\mu} \partial_{\mu} B
&=& 0,
\end{eqnarray}
%%%%%%%%
where a dot above a symbol for a field denotes its background
value. We study the wave equations for linear perturbations that
leave the ten-dimensional background metric unperturbed. Deriving
the radial wave equations in the background of non-extremal
D3-branes is a straightforward generalization of what is done in
\cite{mathur} for the extremal case.

\subsection{Dilaton-axion}

The dilaton and axion are decoupled from the D3-brane in type IIB
theory, satisfying the minimally-coupled scalar wave equation
%%%%%%%%
\be \frac{1}{\sqrt{-g}} \partial_{\mu} \sqrt{-g} g^{\mu \nu}
\partial_{\nu} \phi=0. \label{dilaton} \ee
%%%%%%%%
Thus, the radial wave equation of a dilaton-axion in the spacetime
of a non-extremal D3-brane is given by
%%%%%%%%
\be \Big( \frac{1}{r^5} \frac{\partial}{\partial r} f r^5
\frac{\partial}{\partial r} + \omega^2 \frac{H}{f} - \frac{\ell
(\ell + 4)}{r^2} \Big) \phi = 0, \label{D3eqn} \ee
%%%%%%%%%
where $\ell=0,1,..$

We shall express the wave equation in Schr\"odinger form, and
study the characteristics of the Schr\"odinger effective
potential.  By the substitution
%%%%%%%%
\be \phi = r^{-5/2} f^{-1/2} \psi, \ee
%%%%%%%%
we render (\ref{D3eqn}) in the Schr\"odinger form
%%%%%%%%
\be \big(\frac{\partial^2}{\partial r^2}-V_{\rm eff} \big)\psi =
0, \label{eqdil} \ee
%%%%%%%%
where
%%%%%%%%
\be V_{\rm eff}(\ell)
=-\frac{\omega^2}{f^2}H+\frac{(\ell+3/2)(\ell+5/2)}{f
r^2}+\frac{(f-1)(-f+16)}{4f^2r^2}. \label{V} \ee
%%%%%%%
Factors that are shared by the incident and outgoing parts of the
wave function cancel out when calculating the absorption
probability. Thus, the absorption probability of $\phi$ and $\psi$
are the same.

Technically, $V_{\rm eff}$ cannot be interpreted as an effective
potential, since it is dependent on the particle's incoming
energy. However, this is of no consequence for our analysis of the
form of the wave equations for various particles.

\subsection{Scalar from the two-form}

For the free indices of the two-form taken to lie along the $S^5$,
the radial wave equation is
%%%%%%%%
\be \Big( \frac{H}{r} \frac{\partial}{\partial r} \frac{f r}{H}
\frac{\partial}{\partial r} + \frac{H}{f} \omega^2 -
\frac{(\ell+2)^2}{r^2} \mp \frac{4R^4}{H r^6} (\ell+2) \Big)
a_{(\alpha \beta)} = 0, \label{scalar} \ee
%%%%%%%%%
where $\ell=1,2,..$ The sign $\pm$ corresponds to the sign in the
spherical harmonic involved in the partial wave expansion.

By the substitution
%%%%%%%%
\be a_{(\alpha \beta)} = \big( \frac{H}{f r} \big)^{1/2}
\psi_{(\alpha \beta)}, \ee
%%%%%%%%
we render (\ref{scalar}) into the Schr\"odinger form:
%%%%%%%%
\begin{eqnarray}
V_{\rm eff}(\ell) = -\frac{\omega^2
H}{f^2}+\frac{(\ell+3/2)(\ell+5/2)}{f
r^2}+\frac{(f-1)(15f+16)}{4r^2 f^2}+ \nn\\
\frac{(H-1)[-f^2(15H+49)+f (\pm 16(\ell+2)H+30H+2)+H-1]}{4r^2 f^2
H^2}. \label{Vscalar}
\end{eqnarray}
%%%%%%%%
\subsection{Vector from the two-form}

We now consider one free index of the two-form along $S^5$ and one
free index in the remaining 5 directions. For the tangential
components of the vector, the radial wave equation is
%%%%%%%%
\be \Big( \frac{1}{r^3} \frac{\partial}{\partial r} r^3 f
\frac{\partial}{\partial r} + \frac{\omega^2 H}{f} -
\frac{(\ell+1)(\ell+3)}{r^2} \Big) a_1 = 0, \label{tang} \ee
%%%%%%%%%
where $\ell=1,2,..$

By the substitution
%%%%%%%%
\be a_1  = (r^3 f)^{-1/2} \psi, \ee
%%%%%%%%
we render (\ref{tang}) in the Schr\"odinger form with
%%%%%%%
\be V_{\rm eff}(\ell) = -\frac{\omega^2
H}{f^2}+\frac{\ell+3/2)(\ell+5/2)}{f r^2}
+\frac{(f-1)(3f+16)}{4f^2r^2}. \ee
%%%%%%%%
The radial and time-like components of the vector can be
determined from each other by
%%%%%%%
\be \frac{\partial}{\partial r} a_o = \frac{1}{i \omega} \big[
\omega^2-\frac{(\ell+1)(\ell+3)}{r^2 H}f \big] a_r \label{a} \ee
%%%%%%%
and
%%%%%%%
\be \frac{1}{r^3}\frac{\partial}{\partial r} \big[ r^3 \big(
\frac{\partial}{\partial r} a_0+i \omega a_r \big)
\big]-\frac{(\ell+1) (\ell+3)}{fr^2} a_0=0, \label{aa} \ee
%%%%%%
where $\ell=1,2,..$

(\ref{a}) and (\ref{aa}) can be decoupled. The wave equation for
$a_r$ is
%%%%%%%%
\be \Big( H \frac{\partial}{\partial r} \frac{f}{r}
\frac{\partial}{\partial r} \frac{r f}{H} + \omega^2 H -
\frac{(\ell+1)(\ell+3)}{r^2}f \Big) a_r = 0. \label{ar} \ee
%%%%%%%%%
By the substitution
%%%%%%%%
\be a_r = (r f^3)^{-1/2} H \psi, \label{aar} \ee
%%%%%%%%
we render (\ref{ar}) into the Schr\"odinger form with
%%%%%%%
\be V_{\rm eff}(\ell) = -\frac{\omega^2
H}{f^2}+\frac{\ell+3/2)(\ell+5/2)}{f r^2}
+\frac{(f-1)(35f+16)}{4f^2r^2}. \ee
%%%%%%%
The wave equation for $a_0$ is
%%%%%%
\be \Big( \frac{f}{r} \frac{\partial}{\partial r} \frac{r^3
f}{\omega^2 r^2 H-(\ell+1)(\ell+3)f} \frac{\partial}{\partial r}
+1 \Big) a_0=0. \label{a0} \ee
%%%%%%%
By the substitution
%%%%%%%
\be a_0=(\frac{r^3 f}{\omega^2 r^2 H-(\ell+1)(\ell+3)f})^{-1/2}
\psi, \label{ao} \ee
%%%%%%%%
(\ref{a0}) can be rendered into the Schr\"odinger form. However,
the corresponding potential is singular. Instead, we determine
$a_0$ directly from $a_r$ via
%%%%%%%%%
\be a_0=\frac{i f}{\omega r} \frac{\partial}{\partial r} \Big(
\frac{rf}{H} a_r \Big). \label{aoar} \ee
%%%%%%%%%

\subsection{Antisymmetric tensor from 4-form}

For two free indices of the 4-form along $S^5$ and two free
indices in the remaining 5 directions, the coupled radial wave
equations for the components of the antisymmetric tensor derived
from the 4-form are
%%%%%%%
\be b_{3r}=\mp \frac{\omega rH}{\ell+2} \frac{1}{f} b_{12}, \ee
%%%%%%%
%%%%%%%
\be \frac{\partial}{\partial r}b_{03}-i \omega b_{3r}=\pm
\frac{i}{r} (\ell+1) b_{12} \ee
%%%%%%%
and
%%%%%%%
\be \frac{\partial}{\partial r} b_{12}=\mp \frac{i}{r f}
(\ell+2)b_{03}, \ee
%%%%%%%
where $\ell=1,2,..$ Eliminating $b_{03}$ and $b_{3r}$ we get
%%%%%%
\be \Big( \frac{1}{r} \frac{\partial}{\partial r} r f
\frac{\partial}{\partial r} +\frac{\omega^2
H}{f}-\frac{(\ell+2)^2}{r^2} \Big) b_{12}=0, \label{4-form} \ee
%%%%%%
By the substitution
%%%%%%%
\be b_{12}=(r f)^{-1/2} \psi, \ee
%%%%%%%
we render (\ref{4-form}) into Schr\"odinger form with
%%%%%%%
\be V_{\rm eff}(\ell) = -\frac{\omega^2
H}{f^2}+\frac{(\ell+3/2)(\ell+5/2)}{f r^2}
+\frac{(f-1)(15f+16)}{4f^2r^2}. \ee
%%%%%%%

\subsection{Two-form from the antisymmetric tensor}

The equations for two-form perturbations polarized along the
D3-brane are coupled. For s-wave perturbations, they can be
decoupled \cite{raja,mathur}, and the radial wave equation is
%%%%%%%%
\be \Big( \frac{1}{r^5 H} \frac{\partial}{\partial r} r^5 f H
\frac{\partial}{\partial r} + \frac{\omega^2 H}{f} - \frac{16
R^8}{r^{10} H^2} \Big) \phi = 0, \label{2-form} \ee
%%%%%%%%%
where $\ell=0,1,..$

By the substitution
%%%%%%%%
\be \phi = (r^5 f H)^{-1/2} \psi, \ee
%%%%%%%%
we render (\ref{2-form}) into Schr\"odinger form with
%%%%%%%
\be V_{\rm eff}(\ell) = -\frac{\omega^2 H}{f^2}+\frac{15}{4f r^2}
+\frac{12(H-1)^2}{r^2 f H^2}+\frac{(f-1)(15f+16)}{4r^2 f^2}.
\label{2form} \ee
%%%%%%%

\subsection{General form of effective potential}

For an extremal D3-brane, the effective potentials for the
dilaton-axion, vector from the two-form and the antisymmetric
tensor from the 4-form all reduce to
%%%%%%%
\be V_{\rm eff}(\ell) = -\omega^2
H+\frac{(\ell+3/2)(\ell+5/2)}{r^2}. \ee
%%%%%%%
Thus, the above fields have identical absorption probabilities.
The effective potentials for the scalar from the two-form and the
two-form from the antisymmetric tensor include additional terms,
which merely have the effect of changing the partial wave number
by $\ell \rightarrow \ell \pm 1$ \cite{poritz}.

For a non-extremal D3-brane, the effective potentials for the
dilaton-axion, vector from the two-form and the antisymmetric
tensor from the 4-form are of the form
%%%%%%%
\be V_{\rm eff}(\ell) = -\frac{\omega^2 H}{f^2}
+\frac{(\ell+3/2)(\ell+5/2)}{f r^2}+
\frac{\big(f-1\big)\big((a-1)(a+1) f+16\big)}{4f^2 r^2},
\label{general} \ee
%%%%%%%
where $a=0$ for the dilaton-axion, $a=2$ for the tangential
components of the vector from the two-form, $a=4$ for the
antisymmetric tensor from the 4-form, and $a=6$ for the radial
component of the vector from the two-form. For the scalar from the
two-form and the two-form from the antisymmetric tensor, the
effective potential has additional terms that are proportional to
the charge of the D3-brane. However, for a chargeless D3-brane,
the scalar and two-form fields fit into the above scheme with
$a=4$. It is interesting to note the change of grouping of
absorption probabilities between the cases of extremal and
chargeless D3-branes. The effective potential for the
time-component of the vector from the two-form does not appear to
fit into this scheme, though certain potential terms may merely
have the effect of changing the partial wave number.

We conjecture that $a$ is a parameter that depends on the
polarization of the angular momentum. Note that this parameter
only plays a role in absorption away from extremality.

\section{Low-energy absorption probabilities for a near-extremal
D3-brane}

\subsection{Dilaton-axion, vector from two-form, and antisymmetric tensor
from 4-form}

We will solve the wave equations in the approximation that
%%%%%%%%
\be r_o \ll R \ll 1/\omega. \ee
%%%%%%%%
We work in the limit in which the frequency is large compared to
the temperature, which means that $r_o \leq R^2 \omega$. We render
the wave function in a form that will be convenient for isolating
the singularity at $r=r_o$:
%%%%%%
\be \psi=(r^5 f)^{+1/2} \tilde{\phi}. \label{tilde} \ee
%%%%%%
Substituting (\ref{tilde}) into (\ref{eqdil}) together with
(\ref{general}) yields
%%%%%%%
\be \Big( \frac{1}{r^5} \partial_r fr^5 \partial_r +\omega^2
\frac{H}{f} -\frac{\ell(\ell+4)}{r^2}+\frac{a^2}{4r^2} (1-f) \Big)
\tilde{\phi}=0. \label{general2} \ee
%%%%%%%%
Note that $\phi=F(r) \tilde{\phi}$, where $F(r)$ is a nonsingular
function of $r$ for $r>r_o$.

We will now use the same approximation and procedure as that by
Siopsis in the case of a minimally-coupled scalar
\cite{siopsis,siopsis2}. We use three matching regions. In the
outer region defined by $R^2 \omega \ll r$, (\ref{general2})
becomes
%%%%%%%
\be \Big( r^2 \partial_r^2 +5r\partial_r +\omega^2
r^2-\ell(\ell+4) \Big) \tilde{\phi}=0. \label{outerV} \ee
%%%%%%%
The solution is
%%%%%%%
\be \tilde{\phi}=\frac{1}{r^2}J_{(\ell+2)}(\omega r).
\label{outersol} \ee
%%%%%%%
In the intermediate region defined by $r_o \ll r-r_o$ and $r \ll
R$, expressed in terms of the dimensionless quantity $z \equiv
(r-r_o)/r_o$,
%%%%%%
\be \Big( \frac{1}{z^3} \partial_z z^5 \partial_z +\frac{16
\kappa^2}{z^2}-\ell(\ell+4) \Big) \tilde{\phi}=0, \ee
%%%%%%%
where
%%%%%%%
\be \kappa=\frac{\omega r_o}{4} \big( 1+\frac{R^4}{r_o^4} \big)
^{1/2} \approx \frac{R^2 \omega}{4 r_o}=\frac{\omega}{4 \pi T_H},
\ee
%%%%%%%
and $T_H=\frac{r_o}{\pi R^2}$ is the Hawking temperature. The
corresponding wave function solution is
%%%%%%%%
\be \tilde{\phi}= \frac{A}{r_o^2 z^2} H_{\ell+2}^{(1)} \big
(\frac{4\kappa} {z}\big), \label{intermsol} \ee
%%%%%%%%
where we consider the purely incoming solution. Matching
asympototic form of the intermediate solution  (\ref{intermsol})
for large $z$ with the asymptotic form of the outer solution
(\ref{outersol}) for small $\omega r$ yields the amplitude
%%%%%%%
\be A=\frac{i \pi}{(\ell+2)!(\ell+1)!} \big( \frac{\omega
R}{2}\big) ^{2\ell+4}. \label{A} \ee
%%%%%%%
For $z \ll \kappa$, we expand the intermediate wave function
solution (\ref{intermsol}):
%%%%%%%%
\be \tilde{\phi}= \frac{(-i)^{\ell+5/2} A}{r_o^2 \sqrt{2\pi
\kappa}} z^{-3/2} e^{4i\kappa /z} \Big(
1+\frac{i(\ell+3/2)(\ell+5/2)z}{8\kappa}+O(\kappa^{-2})\Big).
\label{small} \ee
%%%%%%%%
The inner region is defined by $r-r_o \ll R^2 \omega$. The inner
and intermediate regions overlap, since $r_o \ll R^2 \omega$.
There is a singularity in the wave function at $r=r_o$, which can
be isolated by taking the wave function to be
\cite{siopsis,siopsis2}
%%%%%%
\be \tilde{\phi}= f^{i \kappa} \varphi. \label{innersol} \ee
%%%%%%
Isolating this singularity in the wavefunction enables us to
calculate the dominant term in the near-extremal absorption
probability. For the inner region, we substitute (\ref{innersol})
into (\ref{general2}) and express the result in terms of the
dimensionless parameter $x\equiv r/r_o$:
%%%%%%%
\be x^2 f(x) \partial_x^2 \varphi+x\big[ 5-(1-8i\kappa) x^{-4}
\big] \partial_x \varphi +\Big(16 \kappa^2 h(x)- \ell(\ell+4)
+\frac{a^2}{4}(1-f(x)) \Big) \varphi=0, \label{innerequation} \ee
%%%%%%
where
%%%%%%
\be f(x)=1-x^{-4} \label{f} \ee
%%%%%%
and
%%%%%
\be h(x) \equiv \frac{1+x^2+x^4}{(1+x^2)x^4}.\label{h} \ee
%%%%%%%%%
Since $x \ll \kappa$, we shall expand the inner wave function
solution in $\kappa^{-1}$: \be \varphi = A B e^{i\kappa \alpha
(x)} \beta(x) \big( 1+\frac{i}{\kappa} \gamma(x) + O(\kappa^{-2})
\big). \label{solform} \ee
%%%%%%
We will solve for B such that $\varphi(1)=B$, which implies that
$\alpha(1)=\gamma(1)=0$ and $\beta(1)=1$. We solve for the
functions $\alpha$,$\beta$ and $\gamma$ by plugging
(\ref{solform}) into (\ref{innerequation}):
%%%%%%%%
\be \alpha (x) = -4 \int_{1}^{x} dy
\frac{1+\sqrt{1+h(y)y^4(y^4-1)}}{y(y^4-1)}. \ee
%%%%%%%%%
%%%%%%%%%
\be \beta (x) = \exp \Big( -\frac{1}{2} \int_{1}^{x} dy
\frac{y(y^4-1)\alpha^{''}
+(5y^4-1)\alpha^{'}}{4+y(y^4-1)\alpha^{'}} \Big). \ee
%%%%%%%%%%
%%%%%%%%%%
\be \gamma (x)=-\frac{1}{2} \int_{1}^{x} dy \frac{y^4
\ell(\ell+4)-y(y^4-1)
\frac{\beta^{''}}{\beta}-(5y^4-1)\frac{\beta^{'}}{\beta}-\frac{a^2}{4y}}
{y(y^4-1)\alpha^{'}+4}. \ee
%%%%%%%%
We match the expressions for $\alpha$, $\beta$ and $\gamma$ in the
large $x$ limit with the asymptotic form of the intermediate
solution (\ref{small}) and solve for $B$:
%%%%%%%%%
\be B=\frac{(-i)^{\ell+5/2}}{r_o^2 \sqrt{2\pi \kappa}} \big(
1-\frac{i(\ell+3/2)(\ell+5/2)}{8\kappa} \big). \ee
%%%%%%%%%
The absorption probability is
%%%%%%%%
\be P=8\pi \kappa r_o^4 |A|^2 |B|^2. \label{prob} \ee
%%%%%%%
Plugging in the amplitudes $A$ and $B$, we find that
%%%%%%%%
\be P^{near-extremal}= \Big( 1+\big( \frac{4(\ell+2)^2-1}{32
\kappa} \big)^2 \Big) P^{extremal}, \label{nearprob} \ee
%%%%%%%%%
where
%%%%%%%%
\be P^{extremal}=\frac{4 \pi^2}{\big( (\ell+1)!(\ell+2)! \big)^2}
\big( \frac{\omega R}{2} \big)^{4\ell+8}. \label{extremal} \ee
%%%%%%
Our result agrees with \cite{siopsis,siopsis2} for the
minimally-coupled scalar. As we have shown, for near-extremal
D3-branes, the dilaton-axion, tangential and radial components of
the vector from the two-form and the antisymmetric tensor from the
4-form have identical absorption probabilities, as in the extremal
case. As can be seen, as $\kappa \rightarrow \infty$, we recover
the result previously obtained for an extremal D3-brane.

\subsection{Time-like component of vector from two-form}

Using (\ref{aar}),(\ref{aoar}) and (\ref{tilde}) we find the wave
function for the longitudinal component of the vector from the
two-form directly in terms of the radial component wave function:
%%%%%%%
\be a_o=\frac{if}{\omega r} \partial_r (r^3 \tilde{\phi}). \ee
%%%%%%%%
We can find the wave function solutions for $a_o$ in the three
regions directly from the dilaton wave functions (\ref{outersol}),
(\ref{intermsol}) and (\ref{solform}). We must be careful to
expand to $O(\kappa^{-3})$ in the dilaton wave functions in order
for $a_o$ to be of order $O(\kappa^{-2})$. Matching $a_o$ in the
three regions yields the same absorption probability as for the
dilaton given by (\ref{nearprob}). Thus, for near-extremal
D3-branes, the longitudinal component of the vector from the
two-form and dilaton-axion have identical absorption
probabilities, as in the extremal case.

\subsection{Two-form from the antisymmetric tensor}

Substituting (\ref{tilde}) into (\ref{eqdil}) together with
(\ref{2form}) yields
%%%%%%%
\be \Big( \frac{1}{r^5} \partial_r fr^5 \partial_r +\omega^2
\frac{H}{f} +\frac{4}{r^2}(1-f)-\frac{12(H-1)^2}{r^2 H^2} \Big)
\tilde{\phi}=0. \label{gen2form} \ee
%%%%%%%%
We require four regions for matching. Consider $r_1$ and $r_2$
such that $r_2 \ll R \ll r_1$. The outer region is given by $r >
r_1$, where $\omega r_1 \ll 1$. In the outer region, the wave
equation and solution are given by (\ref{outerV}) and
(\ref{outersol}), respectively, with $\ell=0$.

The outer intermediate region is given by $r_1 > r > r_2$ and
$\omega R^2 \ll r$, so that we can ignore the $\omega^2$ term in
the wave equation:
%%%%%%%
\be \Big( \frac{1}{r^5} \partial_r r^5 \partial_r
-\frac{12(H-1)^2}{r^2 H^2} \Big) \tilde{\phi}=0. \ee
%%%%%%%
The corresponding wave function solution is
%%%%%%%
\be \tilde{\phi}= C H^{3/2} +D H^{-1/2}. \ee
%%%%%%%
Matching the outer and outer intermediate regions yields
%%%%%
\be C+D=\frac{\omega^2}{8}. \ee
%%%%%%
The inner intermediate region is defined by $r_o \ll r-r_o$ and $r
\ll R$. Expressed in terms of the dimensionless quantity $z \equiv
(r-r_o)/r_o$, the wave equation in this region is
%%%%%%%
\be \big( \frac{1}{z^3} \partial_z z^5 \partial_z
+\frac{16\kappa^2}{z^2}-12 \big) \tilde{\phi}. \ee
%%%%%%%
The corresponding wave function solution is
%%%%%%%%
\be \tilde{\phi}= A \frac{1}{r_o^2 z^2} H_4^{(1)} \big(
\frac{4\kappa}{z} \big),\label{interm2form} \ee
%%%%%%
where we take the purely incoming solution. Matching the wave
function across the two intermediate regions yields
%%%%%%
\be A=\frac{i \pi}{12} \big( \frac{\omega R}{2}\big)^{6},\ \ \ \ \
C=0. \ee
%%%%%%
For $z \ll \kappa$, we expand the inner intermediate wave function
solution (\ref{interm2form}):
%%%%%%%%
\be \tilde{\phi}= \frac{(-i)^{1/2} A}{r_o^2 \sqrt{2\pi \kappa}}
z^{-3/2} e^{4i\kappa /z} \big(
1+\frac{i(7/2)(9/2)z}{8\kappa}+O(\kappa^{-2})\big).\label{small2form}
\ee
%%%%%%%%
For the inner region, we substitute (\ref{innersol}) into
(\ref{gen2form}) and express the result in terms of the
dimensionless parameter $x$:
%%%%%%%
\be x^2 f(x) \partial_x^2 \varphi+x\big[ 5-(1-8i\kappa) x^{-4}
\big] \partial_x \varphi +\Big(16 \kappa^2 h(x) +4\big(
1-f(x)\big)-12\Big) \varphi=0. \label{innerequation2form} \ee
%%%%%%
where $f(x)$ and $h(x)$ are given by (\ref{f}) and (\ref{h}),
respectively.

Following the case for the dilaton, we plug the ansatz for the
inner wave function solution, given by (\ref{solform}), into
(\ref{innerequation2form}) and solve for the amplitude $B$ by
matching the inner solution with the inner intermediate solution.
The result is that
%%%%%%%%%
\be B=\frac{(-i)^{1/2}}{r_o^2 \sqrt{2\pi \kappa}} \big(
1-\frac{i(7/2)(9/2)}{8\kappa} \big). \ee
%%%%%%%%%
Plugging amplitudes $A$ and $B$ into (\ref{prob}), we find that
the absorption probability is
%%%%%%%%
\be P^{near-extremal}= \Big( 1+\big( \frac{63}{32 \kappa} \big)^2
\Big) P^{extremal}, \ee
%%%%%%%%%
where $P^{extremal}$ is the same as that given for the dilaton in
(\ref{extremal}) with $\ell \rightarrow \ell + 1$. As with
previous cases, as $\kappa \rightarrow \infty$, we recover the
result previously obtained for an extremal D3-brane. We have shown
that, for near-extremal D3-branes, the two-form from the
antisymmetric tensor does not have the same absorption probability
as the dilaton-axion with $\ell \rightarrow \ell+1$, while it does
in the extremal case.

\subsection{Scalar from two-form}

Substituting (\ref{tilde}) into (\ref{eqdil}) together with
(\ref{scalar}) yields
%%%%%%%
\begin{eqnarray}
\Big( \frac{1}{r^5} \partial_r fr^5 \partial_r +\omega^2
\frac{H}{f} -\frac{\ell(\ell+4)}{r^2}+\frac{4}{r^2} (1-f) - \\
\nonumber \frac{(H-1)[-f^2(15H+49)+f (\pm
16(\ell+2)H+30H+2)+H-1]}{4r^2 f H^2} \Big) \tilde{\phi}=0.
\label{genscalar}
\end{eqnarray}
%%%%%%%%
We require four regions for matching, as defined in the previous
section for the two-form. In the outer region, the wave equation
and solution are given by (\ref{outerV}) and (\ref{outersol}),
respectively. In the outer intermediate region, we may ignore the
$\omega^2$ term in the wave equation:
%%%%%%%
\be \Big( \frac{1}{r^5} \partial_r r^5 \partial_r
-\frac{\ell(\ell+4)}{r^2}-\frac{(H-1)[4\big( 1\pm (\ell+2) \big)
H-12]}{r^2 H^2} \Big) \tilde{\phi}=0. \ee
%%%%%%%
The corresponding wave function solution is
%%%%%%%
\be \tilde{\phi}=C r^{\pm(\ell+2)-2} +D r^{\mp(\ell+2)-2} \big(1+
\frac{\ell+2}{\ell+2\pm 2} \frac{R^4}{r^4} \big). \ee
%%%%%%%
Matching the outer and outer intermediate regions yields
%%%%%
\be C=\frac{1}{(\ell+2)!}\big( \frac{\omega}{2}\big) ^{\ell+2},\ \
\ \ D=0, \ee
%%%%%%
for the positive sign and
%%%%%
\be D=\frac{1}{(\ell+2)!}\big( \frac{\omega}{2}\big) ^{\ell+2},\ \
\ \ C=0, \ee
%%%%%%
for the negative sign.

Expressed in terms of the dimensionless quantity $z \equiv
(r-r_o)/r_o$, the wave equation in the inner intermediate region
is
%%%%%%%
\be \big( \frac{1}{z^3} \partial_z z^5 \partial_z
+\frac{16\kappa^2}{z^2} -\ell(\ell+4) \mp 4(\ell+2)-4 \big)
\tilde{\phi}. \ee
%%%%%%%
The corresponding wave function solution is
%%%%%%%%
\be \tilde{\phi}= A \frac{1}{r_o^2 z^2} H_{\ell+2\pm 2}^{(1)}
\big( \frac{4\kappa}{z} \big),\label{intermscalar} \ee
%%%%%%
where we take the purely incoming solution.

Matching the wave function across the two intermediate regions
yields
%%%%%%
\be A_{\pm}=\frac{i \pi}{(\ell+2\pm 1)!(\ell+1\pm 1)!} \big(
\frac{\omega R}{2}\big)^{2\ell+4\pm 2}. \ee
%%%%%%
For $z \ll \kappa$, we expand the inner intermediate wave function
solution (\ref{intermscalar}):
%%%%%%%%
\be \tilde{\phi}= \frac{(-i)^{\ell+1/2} A}{r_o^2 \sqrt{2\pi
\kappa}} z^{-3/2} e^{4i\kappa /z} \big( 1+\frac{i(\ell+3/2\pm
2)(\ell+5/2\pm 2)z}{8\kappa}+O(\kappa^{-2})\big).
\label{smallscalar} \ee
%%%%%%%%
For the inner region, we substitute (\ref{innersol}) into
(\ref{genscalar}) and express the result in terms of the
dimensionless parameter $x$:
%%%%%%%
\begin{eqnarray}
x^2 f(x) \partial_x^2 \varphi+x\big[ 5-(1-8i\kappa) x^{-4} \big]
\partial_x \varphi +\Big(16 \kappa^2 h(x) + \\ \nonumber 4\big(
1-f(x)\big)-\ell(\ell+4) \mp 4(\ell+2)+\frac{15}{4}
f(x)-\frac{15}{2}-\frac{1}{f(x)} \Big) \varphi=0.
\label{innerequationscalar}
\end{eqnarray}
%%%%%%
where $f(x)$ and $h(x)$ are given by (\ref{f}) and (\ref{h}),
respectively.

Following the case for the dilaton, we plug the ansatz for the
inner wave function solution, given by (\ref{solform}), into
(\ref{innerequationscalar}) and solve for the amplitude $B$ by
matching the inner solution with the inner intermediate solution.
The result is that
%%%%%%%%%
\be B=\frac{(-i)^{\ell+1/2}}{r_o^2 \sqrt{2\pi \kappa}} \big(
1-\frac{i(\ell+3/2\pm 2)(\ell+5/2\pm 2)}{8\kappa} \big). \ee
%%%%%%%%%
Plugging amplitudes $A$ and $B$ into (\ref{prob}), we find that
the absorption probability is
%%%%%%%%
\be P^{near-extremal}= \Big( 1+\big( \frac{(\ell+3/2\pm
2)(\ell+5/2\pm 2)}{8 \kappa} \big)^2 \Big) P^{extremal}, \ee
%%%%%%%%%
where $P^{extremal}$ is the same as that given for the dilaton in
(\ref{extremal}) with $\ell \rightarrow \ell \pm 1$. As with
previous cases, as $\kappa \rightarrow \infty$, we recover the
result previously obtained for an extremal D3-brane.

We have shown that, for near-extremal D3-branes, the scalar from
the two-form does not have the same absorption probability as the
dilaton-axion with $\ell \rightarrow \ell \pm 1$, while it does in
the extremal case.

\section{Discussion of absorption by near-extremal D3-branes}

We have obtained a simple relation between the extremal and
near-extremal absorption probabilities of a D3-brane:
%%%%%%%%
\be P^{near-extremal}= \Big( 1+\big( \frac{4\nu^2-1}{32 \kappa}
\big)^2 \Big) P^{extremal}, \label{final} \ee
%%%%%%%%%
where $\nu=\ell+2$ for the dilaton-axion, the vector from the
two-form and the antisymmetric tensor from the 4-form, $\nu=4$ for
the two-form from the antisymmetric tensor, and $\nu=\ell+4$ for
the scalar from the two-form. Note that (\ref{final}) has the same
form as that of various fields scattered by a four-dimensional
$N=4$ supergravity equal-charge black hole \cite{gubser2}.

As we would expect, a perturbation from extremality increases the
absorption probability. At near-extremality, the absorption
probability of the dilaton-axion, vector from the two-form and
antisymmetric tensor from the 4-form are identical, just as they
are at extremality \cite{poritz}. At extremality, the absorption
probability of the scalar from the two-form and the two-form from
the antisymmetric tensor are simply related to that of the
dilaton-axion by a change in the partial wave number $\ell
\rightarrow \ell \pm 1$. For near-extremality, this relation
breaks down, and the scalar and two-form fields are absorbed more
than if this were the case.

We have found numerically that, further away from extremality, the
dilaton-axion, vector from the two-form and antisymmetric tensor
from the 4-form no longer share the same absorption probability
\cite{unpublished}. As can be seen from the effective potential
(\ref{general}), this is due to a parameter $a$ which we
conjecture depends on the angular momentum polarization. $a=0$ for
the dilaton-axion, $a=2$ for the tangential components of the
vector from the two-form, $a=4$ for the antisymmetric tensor from
the 4-form, and $a=6$ for the radial component of the vector from
the two-form. There are additional potential terms for the scalar
from the two-form and the two-form from the antisymmetric tensor.
However, these terms vanish for a chargeless D3-brane, enabling
the scalar and two-form fields to fit into the above scheme with
$a=4$. It is rather curious how the grouping of absorption
probabilities changes between the cases of extremal and chargeless
D3-branes. At first look, it appears that the time-component of
the vector from the two-form does not fit into this scheme, though
certain potential terms may merely be equivalent to a change in
the partial wave number. We conjecture that the parameter $a$,
which only plays a role away from extremality, depends on the
polarization of the angular momentum.

The absorption probability can be expressed in terms of Gamma
functions:
%%%%%%
\be P \approx \frac{8 \pi^3 \kappa}{\big( \nu!(\nu-1)! \big)^2}
\frac{| \Gamma (\nu /2+1/4+2i \kappa )\Gamma (\nu/2+3/4+2i
\kappa)|^2}{| \Gamma (1+4i \kappa)|^2} \big( \frac{\omega r_o}{2}
\big)^{2\nu}, \label{form} \ee
%%%%%%
so long as one keeps in mind that the formula holds only to order
$\kappa^{-2}$. For $\nu=\ell+2$, this absorption probability is
the same as that derived in \cite{siopsis,siopsis2} for a
minimally-coupled scalar. Siopsis expressed the absorption
probability in the form (\ref{form}) in order to interpret factors
in terms of left and right-moving temperatures $T_L$ and $T_R$. In
particular, the general form of a black hole grey-body factor is
%%%%%%%%%
\be P_{b.h.} \sim \Big| \frac{\Gamma \big(
\frac{\ell+2}{2}+i\frac{\omega}{4\pi T_L}\big) \Gamma \big(
\frac{\ell+2}{2}+i\frac{\omega}{4\pi T_R}\big) }{\Gamma \big(
1+\frac{\omega}{2\pi T_H}\big) } \Big|^2. \label{bh} \ee
%%%%%%%%%
Thus, by comparing (\ref{form}) with (\ref{bh}), we are led to the
conclusion that both left and right-moving modes contribute to the
absorption probability at temperatures
%%%%%%%%
\be T_L=T_R=T_H \ee

%% file: thesisworld.tex
\chapter{Brane Worlds in B Fields}

\section{Introduction}

\subsection{Another Look at Higher-Dimensional Spacetime}

String theories and M-theory require that our universe has more
than three spatial dimensions. Studies along more phenomenological
lines have recently led to new insights on how extra dimensions
may manifest themselves, and how they may help solve long-standing
problems such as hierarchy or the cosmological constant problem.

M-theory determines the dimensionality of spacetime to be eleven.
However, it remains a matter of speculation as to the mechanism by
which extra dimensions are hidden, so that spacetime is
effectively four-dimensional in a low-energy regime. Until
recently, the extra dimensions were usually assumed to be a size
of roughly the Planck length $l_{Pl} \sim 10^{-33}$cm. With a
corresponding energy scale of $M_{Pl} \sim 10^{19}$ GeV, probing
extra dimensions directly appeared to be a hopeless pursuit.

In order to explain why there are three large spatial dimensions,
Brandenberger and Vafa have proposed that all dimensions were
initially compact and three spatial dimensions grew in size as a
result of string-string annihilations-- a process that is roughly
analogous to breaking a rubber band that had been keeping a paper
rolled up tight. On the other hand, extra dimensions may have once 
been large, compactified by a dynamical mechanism-- such as the
requirement that the effective four-dimensional mass density remains
nonzero.

\subsection{Sub-millimeter Extra Dimensions}

Recently there has been renewed interest in the notion that our
observable world is a three-dimensional brane embedded in a
higher-dimensional space, whose extra dimensions may be large or
even infinite. Gauge fields are naturally trapped on the brane by
way of open strings whose ends are confined to the worldvolume of
a D-brane. Indeed, the distances at which non-gravitational
interactions cease to be four-dimensional are determined by the
dynamics on the brane and may be much smaller than $R$, the size
of extra dimensions (for simplicity, we assume that all extra
dimensions are of the same size).

Gravity, on the other hand, becomes multi-dimensional at scales
just below $R$, at which point the gravitational force is
%%%%%%%
\be F_{grav}=\frac{G_D m_1 m_2}{r^{D-2}}, \ee
%%%%%%%
where $D$ is the dimensionality of spacetime. The four-dimensional
gravitational law has been verified experimentally down to
distances of about 0.2 mm, which means that $R$ can actually be as
large as 0.1 mm!

This possibility provides a novel way of addressing the hierarchy
problem, i.e., why the electroweak scale ($M_{EW} \sim 1$ TeV) is
drastically different from the Planck scale ($M_{Pl} \sim
10^{16}$ TeV). In higher-dimensional theories, the four-dimensional Planck
scale is not a fundamental parameter. Instead, the mass scale $M$ of
higher-dimensional gravity is fundamental. The four-dimensional Planck
mass goes as
%%%%%%%%%%%
\be
M_{Pl} \sim M(M R)^{d/2}, \label{M} 
\ee
%%%%%%%%%%
where $d$ is the number of extra
dimensions. Thus, if the size of the extra dimensions is large relative to
the fundamental length scale $M^{-1}$, then the Planck mass is much
greater than the fundamental gravity scale.

If, motivated by unification, one supposes that the fundamental gravity
scale is of the same order as the electroweak scale, $M \sim 1$ TeV, then
the hierarchy between $M_{Pl}$ and $M_{EW}$ is entirely due to the large
size of extra dimensions. Thus, the hierarchy problem now takes on a
geometrical setting, i.e., why is $R$ large?

Assuming that $M \sim 1$ TeV, then from equation (\ref{M}) we find that
%%%%%%%
\be
R \sim M^{-1} \big( \frac{M_{Pl}}{M} \big) ^{2/d} \sim
10^{\frac{32}{d}-17} cm.
\ee
%%%%%%%%
For one extra dimension $R$ is unacceptably large. However, for
increased number of extra dimensions, $R$ decreases. For $d=2$, $R
\sim 1$ mm, which has been the motivation for the recent experimental
search for deviations from Newton's gravitational law at sub-millimeter
distances. However, cosmology excludes a mass scale as low as $M \sim 1$
TeV for $d=2$\footnote{Kaluz-Klein graviton modes may be produced at
high temperatures, such as during the Big Bang
nucleosynthesis. Thus, having an upper bound on the mass density of KK 
gravitons yields an lower bound on $M$ in terms of $d$ and the maximum
temperature that ever ocurred in the universe.}. A more realistic value
$M \sim 30$ TeV implies $R \sim 1-10 \mu$m. While an experimental search
for deviations from four-dimensional gravity is difficult in the
micro-meter range, it is not impossible.

M-theory requires 7 unobserved dimensions. If these are all of the size
$R$, then with the above assumptions an experimental search for violations
in four-dimensional gravity appear to be hopeless. However, the
compactification scales of extra dimensions are not guaranteed to be of
the same order, which means that some large extra dimensions may be
observable in this manner.

\subsection{Warped Geometry}

We saw above how the accuracy in our observations of a
four-dimensional law of gravity impose an upper limit on the size
of an extra dimension. However, we shall soon discuss how Randall
and Sundrum have found that, in the case of a certain warped
higher-dimensional spacetimes, an extra dimension can be very
large, or even infinite, without destroying the four-dimensional
gravitation at low energy.

When considering distance scales much larger than the brane thickness, we
can model the brane as a delta-function source of the gravitational
field. The brane is characterized by the energy density per unit
three-volume $\sigma$, which is also known as the brane tension. We shall
focus on the case of one extra dimension. The five-dimensional
gravitational action in the presence of the brane is
%%%%%%%%
\be
S_{grav}=-\frac{1}{16\pi G^{(5)}} \int d^4 x dz \sqrt{g^{(5)}}
R^{(5)}-\Lambda \int d^4 x dz \sqrt{g^{(5)}}-\sigma \int d^4 x
\sqrt{g^{(4)}}, 
\ee
%%%%%%%%
where $\Lambda$ is the five-dimensional cosmological constant, and the
superscripts denote the dimensionality of the space for which Newton's
gravitational constant, the metric and the curvature applies. The
existence of a four-dimensionally flat solution requires fine-tuning
between $\Lambda$ and $\sigma$:
%%%%%%%%
\be
\Lambda=-\frac{4\pi}{3} G^{(5)} \sigma^2. \label{fine}
\ee
%%%%%%%%
This is similar to fine-tuning the four-dimensional consmological constant
to zero. If equation (\ref{fine}) does not hold, then the intrinsic
geometry on the brane is AdS, a case which will be discussed shortly. With
(\ref{fine}) satisfied, the metric solution is
%%%%%%%%%%%
\be
ds^2=e^{-2k|z|}(-dt^2+dx_1^2+dx_2^2+dx_3^2)+dz^2.
\ee
%%%%%%%%%%%
This spacetime is Anti-de Sitter with $Z_2$ symmetry and of radius
$k^{-1}$.

The graviton obeys a the wave equation for a massless,
minimally-coupled scalar propagating in this spacetime:
%%%%%%
\be
\partial_{\mu}\sqrt{-g}g^{\mu \nu}\partial_{\nu}\Phi(x_{\mu},z)=0.
\ee
%%%%%%
Consider a wave function of the form
%%%%%%
\be \Phi(x_{\mu},z)=e^{x_{\mu}p^{\mu}}h(z). \ee
%%%%%%
The mass $m$ of the graviton is $m^2=-p^2$. The massless graviton
wave function is
%%%%%%%
\be h_0(z)=e^{-kz}, \ee
%%%%%%%
and is, therefore, normalizable with its maximum at $z=0$. That
is, the massless graviton is localized at $z=0$. The higher modes
of the graviton are given by
%%%%%%%
\be h_m(z)=J_2 \big( \frac{m}{k}e^{kz}\big) +A N_2 \big(
\frac{m}{k}e^{kz}\big), \ee
%%%%%%%
where $J_2$ and $N_2$ are Bessel functions. We solve for $A$ by
satisfying the boundary condition imposed by the delta-function
source:
%%%%%%%
\be h_m (z)=N_1 \big( \frac{m}{k}\big) J_2 \big( \frac{m}{k}
e^{kz} \big) -J_1 \big( \frac{m}{k} \big) N_2 \big(
\frac{m}{k}e^{kz} \big). \ee
%%%%%%%
Thus, these massive graviton modes propagate in the extra
dimension, providing a high-energy correction to the
four-dimensional gravitational law, which we will see in more
detail for a specific model.

\subsection{A Higher-Dimensional Solution to the Hierarchy Problem}

One approach which uses a warped higher-dimensional spacetime
compactifies the extra dimension over $S^1/Z_2$ by imposing a
brane at each end of the dimension. The branes have tensions of
opposite sign. Randall and Sundrum initially considered a brane
world model in which our observable universe resides on the
negative tension brane at $z=0$ (RS1 scenario) \cite{randall2}.

Newton's gravitational constant in four and five dimensions are
related by
%%%%%%%
\be G_{(4)}=G_{(5)} \frac{k}{e^{2kz_c}-1}, \label{newton} \ee
%%%%%%%%
where $z_c$ is the length of the extra dimension. Thus, for
relatively large $z_c$, the gravitational interactions of matter
living on the negative tension brane are weak.

If one takes the five-dimensional gravity scale and the Anti-de
Sitter radius $k$ to be of the order of the weak scale, $M_{EW}
\sim 1$ TeV, then (\ref{newton}) implies that the effective
four-dimensional Planck mass is of the order
%%%%%%%
\be M_{Pl} \sim e^{k z_c}M_{EW}. \ee
%%%%%%%
Thus, for $z_c$ about $37$ times larger than the Anti-de Sitter
radius $k^{-1}$, the value of $M_{Pl}/M_{EW}$ is of the right
order of magnitude to solve the hierarchy problem.

\subsection{Solving the Cosmological Constant Problem?}

In the context of the brane world, the cosmological constant problem
amounts to the question of why the vacuum energy density has almost no
effect on the curvature induced on our brane. It may be plausible that
perhaps the vacuum energy density induces a non-trivial warp factor on the
higher-dimensional geometry, while the four-dimensional Poincar\'{e}
invariance is maintained. The suggestion has been put forth that a
hypothetical bulk scalar field conformally coupled to brane world matter
may play an important role. That is, a nonzero vacuum energy may be
compensated by a shift in this scalar field.

Unfortunately, this solution involves a naked singularity at finite proper
distance to the brane, and it has been argued that a possible resolution
of this singularity reintroduces the cosmological constant problem. For
example, a second brane can be introduced, so that the space between the
branes is completely non-singular. However, in order to be a solution
to the Einstein equations, the tension of the second brane must be
fine-tuned, and this fine-tuning is no more desireable than that of the
cosmological constant in four-dimensional theories. Nevertheless, this
idea may point the way to a new solution. 

\subsection{An Alternative to Compactification}

The massless graviton wave function remains normalizable as $z_c
\rightarrow \infty$. Thus, even if there exists a single brane of
positive tension in an infinite extra dimension, gravity as well
as gauge fields are localized on the brane (RS2 scenario). For the
author, this is the most intriguing aspect of brane worlds, and why
simple models that may not be readily embedded in fundamental
string/M-theories are still of interest. For the first time in 70 years,
this offers an alternative to the Kaluza-Klein compactified extra
dimensions. 

The wave functions of the Kaluza-Klein gravitons are normalized by
%%%%%%%%
\be \int dz e^{2k|z|} h_m (z)h_{m^{\prime}}(z)=\delta
(m-m^{\prime}). \ee
%%%%%%%%
The measure $dz e^{2k|z|}$ is due to the warping factor in the
geometry. Using the asymptotics of the Bessel functions, the normalized
KK graviton modes are found to be
%%%%%%%%%%
\be
h_m (z)=\sqrt{\frac{m}{k}} \frac{J_1 \big( \frac{m}{k}\big) N_2 \big(
\frac{m}{k}e^{kz}\big) -N_1 \big( \frac{m}{k}\big) J_2 \big(
\frac{m}{k}e^{kz}\big) }{\sqrt{\big[ J_1 \big( \frac{m}{k}\big) \big] ^2 
+\big[ N_1 \big( \frac{m}{k}\big) \big] ^2}}. 
\ee 
%%%%%%%%%%
At large $z$, these wave functions oscillate:
%%%%%%%%%%
\be
h_m (z) \sim \sin \big( \frac{m}{k}e^{kz} \big),
\ee
%%%%%%%%%%%
and they are suppressed at $z=0$:
%%%%%%%%%
\be
h_m (0) \sim \sqrt{\frac{m}{k}}.
\ee
%%%%%%%%%%
These KK modes correspond to gravitons escaping into the extra dimension
or coming in towards the brane. The coupling of matter living on the brane
to these modes is fairly weak at small $m$, and so they have a low
production rate at low energy.

Consider the Yukawa-type contribution of KK graviton exchange to the
gravitational potential between two unit point masses living on the braneL
%%%%%%%%
\be
V_{KK} (r)=-G_{(5)} \int_0^{\infty} dm |h_m(0)|^2
\frac{e^{-mr}}{r}=-\frac{G_{(4)}}{r}\big( 1+\frac{const}{k^2r^2}\big).
\ee
%%%%%%%%%
Thus, the four-dimensional gravitational potential, including the graviton
zero mode, is given by
%%%%%%%%
\be
V(r)=-\frac{G_{(4)}}{r} \big( 1+\frac{const}{k^2r^2} \big). \label{Npot}
\ee
%%%%%%%%
In contrast to compact extra dimensions, for which corrections are
suppressed exponentially at large distances, the correction to Newton's
law has power law behavior at large $r$. However, at distances greater
than the Anti-de Sitter radius $k^{-1}$, this correction is negligible. 

Duff showed, in his PhD thesis, that one-loop quantum corrections to the
graviton propogator lead to a Newtonian potential of the form
(\ref{Npot}). Since one path to this potential is the result of a
five-dimensional classical calculation and the other results from a
four-dimensional quantum calculation, they seem to be completely
unrelated-- at first sight. However, the AdS/CFT correspondence can be
invoked to show that these are two dual ways of describing the same
physics.

In the simple models discussed here, the hierarchy between the Planck and
weak scales is not explained by the physics of infinite dimensions, since
in this limit $G_{(4)}=G_{(5)}k$. However, a modest extension of this
model, which includes a "probe" brane placed some distance from our brane
world, has been shown to lead to exponential hierarchy even if the extra
dimension is infinite.

\subsection{Lifting Brane Worlds to Higher Dimensions}

The first step in embedding brane world scenarios such as RS2 into
higher dimensional, fundamental theories such as string/M-theory is to
demonstrate how the warped geometry arises from the near-horizon limit of
branes. Consider the standard isotropic $p$-brane in D dimensions
with the metric
%%%%%%%%%
\be
ds_D^2=H^{\frac{\tilde{d}N}{D-2}} \eta_{\mu
\nu}dx^{\mu}dx^{\nu}+H^{\frac{dN}{D-2}}(dr^2+r^2d\Omega_{\tilde{d}+1}^2),
\ee
%%%%%%%%%
where the harmonic function $H$ is
%%%%%%%%%
\be
H=1+\frac{R^{\tilde{d}}}{r^{\tilde{d}}}
\ee
%%%%%%%%%%
and $d=p+1$ and $\tilde{d}=D-d-2$. We consider the near-horizon limit
by dropping the "1" in $H$ and make the coordinate transformation
%%%%%%%%%
\be
r=(1+k|z|)^{-\frac{2}{\tilde{d}N-2}},
\ee
%%%%%%%%%
for $\tilde{d}N \ne 2$ and
%%%%%%%%%
\be
r=e^{-\frac{k}{\tilde{d}}|z|},
\ee
%%%%%%%%
for $\tilde{d}N =2$. We then perform a dimensional reduction on the 
$\tilde{d}+1$-sphere to recover the lower-dimensional brane world
scenario. Our brane world is located at $z=0$, some distance
away from the fundamental brane at $z \rightarrow \infty$. The absolute
value imposes $Z_2$ symmetry about $z=0$, and corresponds to a
delta-function source that is manually added to the
Lagrangian\footnote{The singular sources in the energy-momentum
tensor are composed of two elements. One is a fundamental brane source
with tension opposite in sign to the brane world tension. The other
is the manually added $\delta$-function. This has been found to
resolve an apparent discrepancy between supersymmetry and the sign
and magnitude of the brane world tension.}. The requirement of this
additional source makes an embedding of this simple brane world scenario
in M-theory questionable. Nevertheless, such brane world scenarios make
for interesting toy models which, as previously discussed, show potential
higher-dimensional solutions to long-standing problems in particle physics
and cosmology, as well as offer an alternative to Kaluza-Klein
dimensional compactification.

Fundamental brane configurations whose near-horizon regions yield a
five-\\ dimensional brane world scenario after dimensional reduction
include:

A D3-brane reduced on $S^5$

A D4-brane reduced on $S^1 \times S^4$, and wrapped around $S^1$. The
M-theoretic equivalent of this is an M5-brane reduced on $T^2 \times S^4$,
and wrapped around $T^2$.

A D5-brane reduced on $T^2 \times S^3$, and wrapped around $T^2$.

Two M5-branes intersecting on $3+1$ dimensions, reduced on $T^4 \times
S^2$ (or $K3 \times S^2$), and wrapped around $T^4$ (or $K3$).

\subsection{Branes in B Fields}

In this chapter, we will mainly focus on the effects of a background NS B
field on the four-dimensional graviton. Our motivation stems from duality
between the near-horizon region sof a $Dp$-branes in background B fields
and the large $N$ limit of non-commutative gauge theories, and how
non-commutative effects may be manifest in the localization of gravity on
the brane world in the dual gravity theory.

We show, for an example, how a D3-brane in a constant NS B field
background can be obtained. Beginning with a D3-brane in type IIB theory,
we T-dualize the solution and end up with a D2-brane in type IIA
theory, smeared along one direction. A spatial rotation in one
world-volume and one transverse direction yields a smeared
D2-brane on a tilted torus. A second T-duality transformation leads to a
D3-brane with a background B field:
%%%%%%
\begin{eqnarray}
ds_{10}^2&=&H^{-1/2}\Big( h^{-1/4} (-dt^2+dx_3^2) +h^{3/4}(dx_1^2+dx_2^2)
\Big)\\
\nonumber & &+H^{1/2} h^{-1/4} (dr^2+r^2
d\Omega_5^2),\\ \nonumber
H&=&1+\frac{R^4}{r^4},\\ \nonumber
h^{-1}&=&H^{-1} \sin^2 \theta +\cos^2 \theta,
\end{eqnarray}
%%%%%%
where the B field is given by
%%%%%%%%
\be
B_{23}=\frac{\sin \theta}{\cos \theta} H^{-1}h,
\ee
%%%%%%%%%
the dilaton is
%%%%%%%%%%
\be
e^{2\phi}=h,
\ee
%%%%%%%%%%%%
and the form fields are
%%%%%%%%%%
\be
F_{01r}=\sin \theta \partial_2 H^{-1}
\ee
%%%%%%%%%%
and
%%%%%%%%%
\be
F_{0123r}=\cos \theta h \partial_r H^{-1}.
\ee
%%%%%%%%%

Note that one may alternatively perform a Lorentz boost rather
than a spatial rotation and end up with a background field of space-time
components,i.e., an E field. We discuss this, as well as more complicated
brane configurations and intersections, in appendix B.

The asymptotic value of the B field is $B \rightarrow \tan \theta$. The
parameter $R$ is defined by $\cos \theta R^4=4\pi gN$, where $N$ is the
number of D3-branes and $g$ is the asymptotic value of the coupling
constant\footnote{While it is possible to gauge away the B field, this
introduces a constant flux for the worldvolume gauge field, corresponding
to a D1-brane charge.}. 

In a certain decoupling limit which we will explore in more detail, this
supergravity solution is dual to a non-commutative gauge theory, $\theta$
is the non-commutativity parameter, and is related to the coordinate
commutators by
%%%%%%%%%
\be
[x^{\mu},x^{\nu}]=i \theta^{\mu \nu}.
\ee
%%%%%%%%%

\subsection{Massive Gravity}

Randall and Sundrum \cite{randall2} have shown that, with
fine tuned brane tension, a flat 3-brane embedded in $AdS_5$ can
have a single, massless bound state. Four-dimensional gravity is
recovered at low-energy scales. It has also been proposed that
part or all of gravitational interactions are the result of
massive gravitons. For example, in one model, gravitational
interactions are due to the net effect of the massless graviton
and ultra-light Kaluza-Klein state(s)
\cite{kogan1,mous,kogan2,kogan3}. In another proposal, there is no
normalizable massless graviton and four-dimensional gravity is
reproduced at intermediate scales from a resonance-like behavior
of the Kaluza-Klein wave functions
\cite{kogan2,kogan3,gregory,csaki2,dvali}.

Recently, it has been shown that an $AdS_4$ brane in $AdS_5$ does
not have a normalizable massless graviton. Instead, there is a
very light, but massive bound state graviton mode, which
reproduces four-dimensional gravity
\cite{karch,kogan,kogan4,porrati1}. The bound state mass as a
function of brane tension, as well as the modified law of gravity,
were explored in \cite{schwartz,miemiec}.

A brane world in which two extra infinite spatial dimensions do
not commute has recently been used in order to localize scalar
fields as well as fermionic and gauge fields \cite{pilo}. Also,
the cosmological evolution of the four-dimensional universe on the
probe D$3$-brane in geodesic motion in the curved background of
the source D$p$-brane with nonzero NS $B$ field has been explored
\cite{youm}.

We consider brane world models which arise from a sphere reduction
in M-theory or string theory, as the near-horizon of $p$-branes
\cite{cvetic} with a constant, background $B$-field on the
world-volume. The dual field theory is non-commutative Yang-Mills
\cite{itzhaki,maldacena}.

We consider string theoretic $p$-brane solutions for $p=3,4,5$
with $0,1,2$ world-volume dimensions wrapped around a compact
manifold, respectively. For $p>5$, the space of the extra large
dimension has finite-volume, with a naked singularity at the end,
for which case the localization of gravity is trivial.

For all of the cases we study, we find that there is no
normalizable massless graviton, but there is a massive bound state
graviton which plays the role of the four-dimensional graviton.
This yields another class of examples for which gravitational
interactions are entirely the result of massive gravitons. As we
will see, the bound mass increases continuously from zero as a
constant, background $B$ or $E$-field is turned on, except for the
case $p=5$, for which there is a mass gap.

Thus, from the vantage point of four-dimensional gravity, a background B
field has the same effect as a negative four-dimensional cosmological
constant. Namely, both yield a massive four-dimensional graviton.

\section{The graviton wave equation}
\subsection{With background B field}

Consider the metric of a $p$-brane (expressed in the string frame)
in a constant NS $B$ field in the $2$,$3$ directions \cite{maldacena,oz}:
%%%%%
\begin{eqnarray}
ds_{10}^2 &=& H^{-1/2}\big(
-dt^2+dx_1^2+h(dx_2^2+dx_3^2)+dy_i^2\big) +l_s^4 H^{1/2}(du^2+u^2
d\Omega_5^2),\\ \nonumber H &=& 1+\frac{R^{7-p}}{l_s^4 u^{7-p}},\
\ \ \ \ \ h^{-1}=H^{-1} \sin^2\theta+\cos^2\theta,\label{metric}
\end{eqnarray}
%%%%%%%%
where $u=r/l_s^2$ is a dimensionless radial parameter and
$i=1,..,p-3$. The $B$ field and dilaton are given by
%%%%%%%
\be
B_{23}=\tan \theta H^{-1}h,\ \ \ \ \ e^{2\varphi}=g^2
H^{(3-p)/2}h,
\ee
%%%%%%
%%%%%%
\be
\cos \theta R^{7-p}=(4\pi)^{(5-p)/2}\Gamma{\Big( \frac{7-p}{2}
\Big) } g_s l_s^{p-3} N,
\ee
%%%%%%
where $N$ is the number of $p$-branes and $g_s$ is the asymptotic
value of the coupling constant. $\theta$ is known as the non-commutativity
parameter, and is related to the asymptotic value of the
$B$-field: $B_{23}^{\infty}=\tan \theta$.

Consider the following decoupling limit, in which we take the $B$ field to
infinity \cite{maldacena}:
%%%%%%
\begin{eqnarray}
l_s \rightarrow 0,\ \ \ \ \ \ \ l_s^2 \tan \theta =b,
\\ \nonumber \tilde{x}_{2,3}=\frac{b}{l_s^2} x_{2,3},
\end{eqnarray}
%%%%%
where $b,u,\tilde{x}_{\mu}$ and $g_s l_s^{p-5}$ stay
fixed. Thus,
%%%%%%
\be
l_s^{-2} ds^2=\big( \frac{u}{R}\big)^{(7-p)/2}
\big(-dt^2+dx_1^2+\tilde{h}(d\tilde{x}_2^2+d\tilde{x}_3^2)+dy_i^2\big)+\big(
\frac{R}{u}\big)^{(7-p)/2} (du^2+u^2 d\Omega_{8-p}^2),
\ee
%%%%%%
where $\tilde{h}^{-1}=1+b^2 (u/R)^{7-p}$.
The dual field theory is Yang-Mills with noncommuting $2,3$
coordinates \cite{maldacena}. For small $u$, the above solution reduces to
$AdS_5 \times S^5$, corresponding to ordinary Yang-Mills living in the
IR region of the dual field theory, which is also the case if we had
taken the decoupling limit with finite $B$-field. From this point
on, we drop the $\tilde{}$ on the coordinates. For $p \ne 5$, we change
coordinates to $u/R \equiv (1+k|z|)^{2/(p-5)}$ \cite{cvetic}:
%%%%%%%%
\be
l_s^{-2} ds^2=(1+k|z|)^{\frac{7-p}{p-5}}\big(-dt^2+dx_1^2+
\tilde{h}(dx_2^2+dx_3^2)+dy_i^2 +dz^2\big)+(1+k|z|)^{\frac{p-3}{p-5}}
d\Omega_{8-p}^2,
\ee
%%%%%%
For $p=5$, we change coordinates to $u/R \equiv e^{-k|z|/2}$
\cite{cvetic}:
%%%%%%%
\be
l_s^{-2} ds^2=e^{-k|z|/2}\big(-dt^2+dx_1^2+\tilde{h}(dx_2^2+dx_3^2)+
dy_i^2 +dz^2+d\Omega_3^2\big).
\ee
%%%%%%%
After dimensional-reduction over $S^{8-p}$, the above metric corresponds
to a $p+2$-dimensional domain wall at $z=0$. We will consider the case
where $y_i$ are wrapped around a compact manifold, so that there is a
$1+3$-dimensional dual field theory on the remaining world-volume
coordinates. The equation of motion for the graviton fluctuation
$\Phi=g^{00}h_{01}$ is
%%%%%%
\be
\partial_M \sqrt{-g}e^{-2\varphi}g^{MN}\partial_N \Phi=0,
\label{eom}
\ee
%%%%%%
where $h_{01}$ is associated with the energy-momentum tensor
component $T_{01}$ of the Yang-Mills theory. For $p \ne 5$, consider
the ansatz
%%%%%%%
\be
\Phi=\phi(z)e^{ip \cdot x}=(1+k|z|)^{\frac{9-p}{2(5-p)}}\psi(z)e^{ip
\cdot x}. \label{ansatz}
\ee
%%%%%%%%
Thus,
%%%%%
\be
-\psi^{''}+U\psi=m^2 \psi, \label{wave}
\ee
%%%%%%
%%%%%%
\be
U=\frac{(p-9)(3p-19)k^2}{4(5-p)^2(1+k|z|)^2}+\frac{p-9}{5-p}k\delta(z)+
\frac{\alpha^2}{(1+k|z|)^{2\frac{(7-p)}{5-p}}}, \label{U} \ee
%%%%%%
where
%%%%%%
\be
\alpha \equiv b \sqrt{(p_2^2+p_3^2)},
\ee
%%%%%%
and $m^2=-p^2$.

For $p=5$, consider the ansatz
%%%%%%%
\be
\Phi=\phi(z)e^{ip \cdot x}=e^{k|z|/2}\psi(z)e^{ip\cdot x}. \label{a5}
\ee
%%%%%%%%
Thus, $\psi$ satisfies (\ref{wave}) with
%%%%%%
\be
U=\frac{1}{4}k^2-k\delta(z)+\alpha^2 e^{-k|z|}.
\label{u5}
\ee
%%%%%%
For more general directions of the world-volume $B$ field, $\alpha
\equiv b \sqrt{p_i^2}$, where $p_i$ are the momenta along the large,
non-commuting directions in the world-volume. Thus, if only the wrapped
$y_i$ coordinates are  non-commuting, then $\alpha=0$.

\subsection{With background B and E fields}

Consider the metric of a $p$-brane in a constant two-component
NS $B$ field background \cite{maldacena,oz}:
%%%%%
\begin{eqnarray}
ds^2 &=& H^{-1/2}\big(
h_e(-dt^2+dx_1^2)+h_m(dx_2^2+dx_3^2)+dy_i^2\big)
+l_s^4 H^{1/2}(du^2+u^2 d\Omega_{8-p}^2),\\ \nonumber
H &=& 1+\frac{R^{7-p}}{l_s^4 u^{7-p}},\ \ \ \ h_m^{-1}=H^{-1}
\sin^2\theta_m+\cos^2\theta_m,\ \ \ \ h_e^{-1}=-H^{-1}\sinh^2
\theta_e+\cosh^2 \theta_e.
\end{eqnarray}
%%%%%%%%
The $B$ and $E$ field components and dilaton are given by
%%%%%%%
\be
B_{23}=\tan \theta_m H^{-1}h_m,\ \ \ \ \ E_{01}=-\tanh \theta_e
H^{-1} h_e,\ \ \ \ \ e^{2\varphi}=g^2 H^{\frac{3-p}{2}}h_e h_m.
\ee
%%%%%%
Consider the following decoupling limit \cite{maldacena}:
%%%%%%
\begin{eqnarray}
l_s \rightarrow 0,\ \ \ \ \ \cosh \theta_e=
\frac{b^{'}}{l_s},\ \ \ \ \ \cos \theta_m=\rm{fixed},\\
\nonumber x_{0,1}=\frac{b^{'}}{l_s^2}\tilde{x}_{0,1},
\ \ \ x_{2,3}=l_s \cos \theta_m \tilde{x}_{2,3},
\end{eqnarray}
%%%%%
where $b^{'},u,\tilde{x}_{\mu}$ and $g_s l_s^{p-7}$ remain fixed. Thus,
%%%%%%
\be
l_s^{-2} ds^2=H^{1/2}\Big( \big(\frac{u}{R}\big)^{7-p}
(-d\tilde{t}^2+d\tilde{x}_1^2)+\big(\frac{u}{R}\big)^{7-p}
\tilde{h}_m(d\tilde{x}_2^2+d\tilde{x}_3^2)+H^{-1}dy_i^2+du^2+
u^2 d\Omega_{8-p}^2\Big),
\ee
%%%%%%%%
where we define
%%%%%%%%
\be
\tilde{h}_m^{-1}=1+\big(\frac{u}{R}\big)^{7-p}\cos^{-2} \theta_m.
\ee
%%%%%%%%
The dual field theory is Yang-Mills with noncommuting $0,1$ and
$2,3$ coordinates \cite{maldacena}. Once again, we drop the $\tilde{}$ on
the coordinates. For $p \ne 5$, we change coordinates to $u/R \equiv
(1+k|z|)^{2/(p-5)}$:
%%%%%%%%
\bea
l_s^{-2}ds^2
&=& H^{1/2}\Big((1+k|z|)^{\frac{2(p-7)}{5-p}}\big(-dt^2+dx_1^2+
\tilde{h}_m(dx_2^2+dx_3^2)+dz^2\big)+\\ & & H^{-1}dy_i^2+
(1+k|z|)^{\frac{4}{p-5}}d\Omega_{8-p}^2 \Big).
\eea
%%%%%%
For $p=5$, we change coordinates to $u/R \equiv e^{-k|z|/2}$:
%%%%%%%%
\be
l_s^{-2}ds^2=H^{1/2} \big( e^{-k|z|} (-dt^2+dx_1^2+
\tilde{h}_m(dx_2^2+dx_3^2)+dz^2+d\Omega_3^2)+H^{-1}dy_i^2\big).
\ee
%%%%%%
As in the case of a one-component $B$-field, we consider a
dimensional-reduction over $S^{8-p}$ and $y_i$ wrapped around a
compact manifold. One effect of the $E$-field, as opposed to the
$B$-field, is to add a breathing-mode to $S^{8-p}$. For $p \ne 5$, we
insert this metric into the graviton equation of motion (\ref{eom}) with
the ansatz (\ref{ansatz}) and obtain (\ref{wave}) with $U$ given by
(\ref{U}) and with
%%%%%%
\be
\alpha \equiv \frac{\sqrt{p_2^2+p_3^2}}{\cos \theta_m}.
\ee
%%%%%%
For $p=5$, we use the wave function ansatz (\ref{a5}) and obtain
(\ref{wave}) with $U$ given by (\ref{u5}). Again, in general $\alpha
\equiv \sqrt{p_i^2}/\cos \theta_m$, where $p_i$ are the momenta along the
large, non-commuting directions.

We will focus on the $p=3$ case, for which the volcano
potential $U$ is plotted in Figure 1, for $\alpha=0$ (solid line) and
$\alpha=2$ (dotted line).
%%%%%%%%
\begin{figure}
   \epsfxsize=4.0in
   \centerline{\epsffile{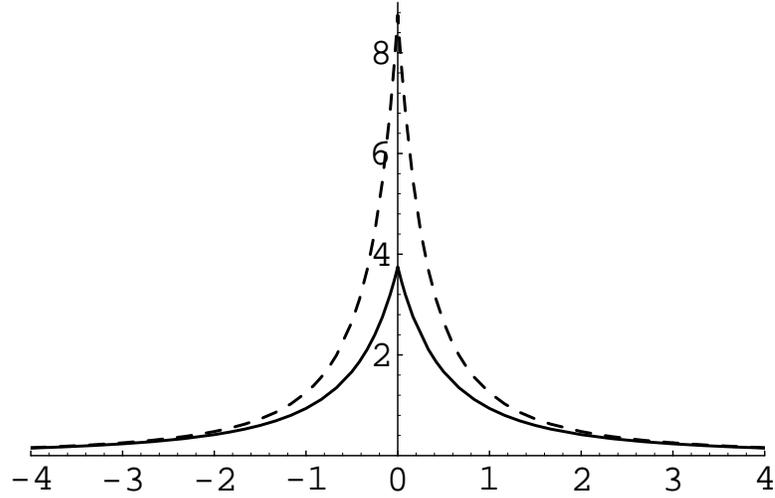}}
   \caption[FIG. \arabic{figure}.]{$U(z)$ for $\alpha=0$ (solid line) and
$\alpha=2$ (dotted line) for $p=3$.}
\end{figure}
%%%%%%%%
As can be seen, the effect of the B and E fields is the raise
the peak of the potential. This is the case in general for $p \le 5$.

\section{The massless graviton}

Consider the coordinate transformation
%%%%%%%
\be
u/R \equiv (1+k|w|)^{-1}\label{ztrans}
\ee
%%%%%%%
and the wave function transformation
%%%%%%%
\be
\Theta(w)=(1+k|w|)^{\frac{p-3}{4}}\psi(z).\label{wavetrans}
\label{aa}
\ee
%%%%%%%%
applied to the graviton wave equation (\ref{wave}).
Thus,
%%%%%
\be
-\Theta^{''}+V\Theta=\frac{m^2}{(1+k|w|)^{p-3}}\Theta,\label{Wave}
\ee
%%%%%%
where
%%%%%%
\be
V=\frac{(6-p)(8-p)k^2}{4(1+k|w|)^2}-(6-p)k\delta(w)+
\frac{\alpha^2}{(1+k|w|)^4}.
\label{V}
\ee
%%%%%%
For the massless case, this is a Schr\"{o}dinger-type equation. The
wave function solution can be written in terms of Bessel functions as
%%%%%%
\be
\Theta(w)=N_0 (1+k|w|)^{1/2} \big[ J_{\nu} \Big(\frac{i\alpha}{1+k|w|}
\Big) +A(\alpha) Y_{\nu} \Big( \frac{i \alpha}{1+k|w|}\Big) \big],
\label{solution}
\ee
%%%%%%
where $\nu=(7-p)/2$ and $N_0$ is the normalization constant, in the
event that the above solution is normalizable. $A(\alpha)$ is solved by
considering the $\delta (w)$ boundary condition:
%%%%%
\be
A(\alpha)=\frac{2\nu J_{\nu}(i \alpha)+i\alpha[J_{\nu+1}(i\alpha)-
J_{\nu-1}(i\alpha)}{-2\nu Y_{\nu}(i\alpha)+i\alpha[Y_{\nu-1}(i\alpha)-
Y_{\nu+1}(i\alpha)]},\label{A}
\ee
%%%%%
where we used $2J_n^{\prime}(x)=J_{n-1}(x)-J_{n+1}(x),$ and likewise for
$Y_n^{\prime}(x)$. In order for the massless graviton to be localized on
the brane, the corresponding wave function must be normalizable, i.e.,
$\int_0^{\infty} |\psi(w,m=0)|^2 dw < \infty$. (\ref{solution}) can be
written in the form
%%%%%%%
\be
\Theta(w)=(normalizable\ part)+A(non-renormalizable\ part),
\ee
%%%%%%%%
as can be seen from the asymptotic forms of the Bessel functions. Thus,
the massless graviton is only localized on the brane if $A=0$. In this
limit, there are no background B or E fields and we recover the
Randall-Sundrum model.

As shown in Figure 2, for the case of the D$3$-brane, as $\alpha$
increases, the modulus of $A$ approaches unity.
%%%%%%%%
\begin{figure}
   \epsfxsize=4.0in
   \centerline{\epsffile{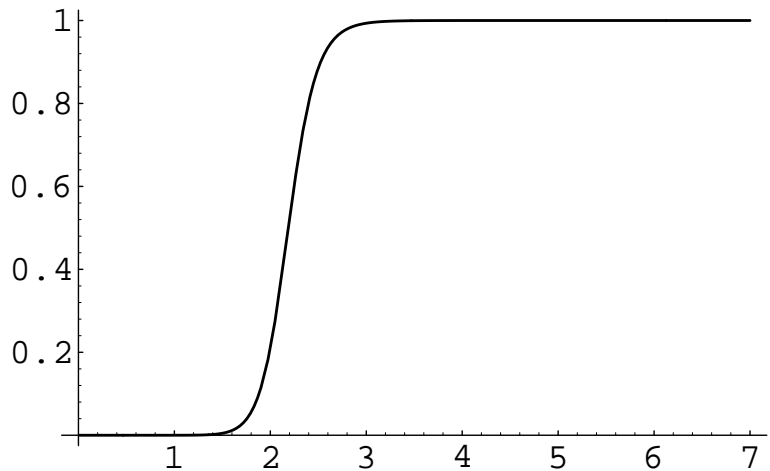}}
   \caption[FIG. \arabic{figure}.]{$|A|^2$ versus $\alpha$ for
$p=3$.}
\end{figure}
%%%%%%%%
$A$ has the same characteristics for $p \le 5$ in general. The issue
arises as to whether there is a massive graviton state bound to the
brane for nonzero $A$.

\section{Localization of the massive graviton}

%%%%%%%%
\begin{figure}
   \epsfxsize=4.0in
   \centerline{\epsffile{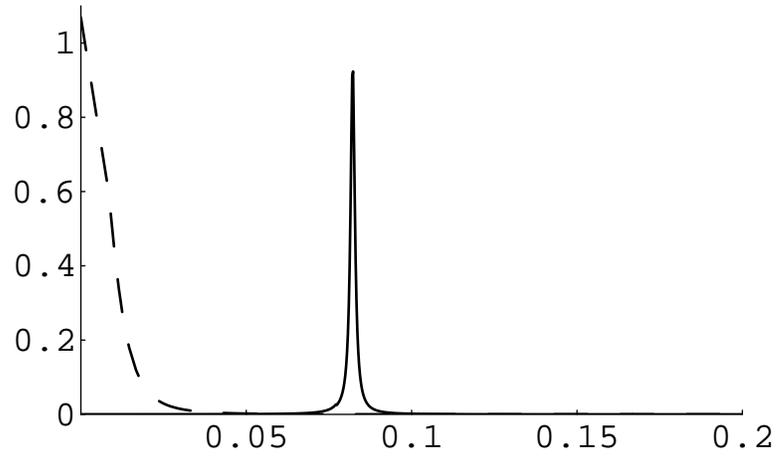}}
   \caption[FIG. \arabic{figure}.]{$|\psi (z=0)|^2$ versus $m$ for
$\alpha=0$ (dotted line) and $\alpha=.02$ (solid line) for $p=3$.}
\end{figure}
%%%%%%%%
We will first concentrate on the case of the brane world as the
dimensionally-reduced near-horizon of the D$3$-brane. For this case, the
massive wave equation  (\ref{wave}) with $U$ given by (\ref{U}) is a
modified Mathieu's equation \cite{hashimoto}, supplemented by the $\delta
(w)$ boundary condition. The exact solution of the modified Mathieu's
equation is known, and has been applied to absorption by $D3$-branes and
six-dimensional dyonic strings
\cite{gubser,park}. However, it is
rather laborious to use the exact solution in this context and so we
content ourselves with plotting numerical solutions-- except for the case
$p=5$, for which the exact solution is a sum of Bessel functions.

In order to solve for the unnormalized wave function, we input the
$\delta(z)$ boundary condition and solve the Schr\"{o}dinger-type equation
outwards. We then numerically integrate in order to find the correct
normalization factor.

%%%%%%%%
\begin{figure}
   \epsfxsize=4.0in
   \centerline{\epsffile{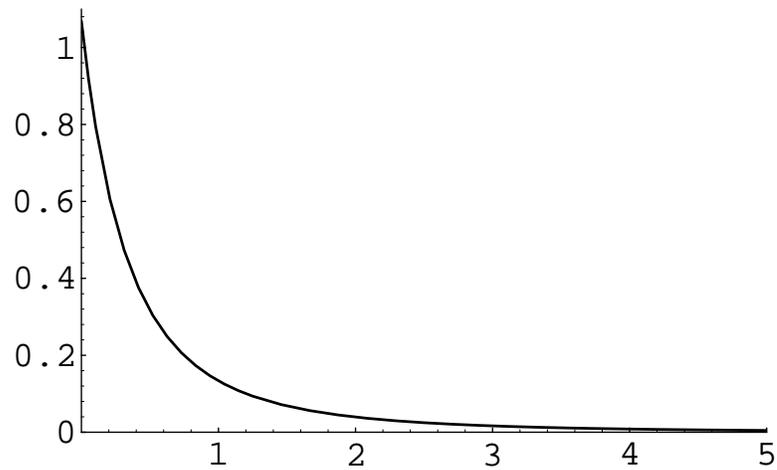}}
   \caption[FIG. \arabic{figure}.]{$|\psi(z)|^2$ for $m/k=0$ for
$p=3$.}
\end{figure}
%%%%%%%%
%%%%%%%%
\begin{figure}
   \epsfxsize=4.0in
   \centerline{\epsffile{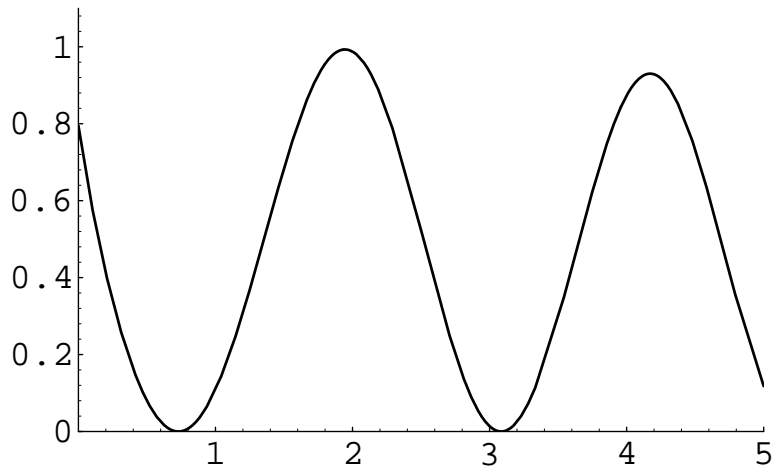}}
   \caption[FIG. \arabic{figure}.]{$|\psi(z)|^2$ for $m/k=1.5$ for
$p=3$.}
\end{figure}
%%%%%%%%
As can be seen from the dotted line in Figure 3, for the case of the
D$3$-brane with no background B and E fields, there is a resonance in the
modulus of the wave function on the brane for $m=0$, which implies that the
massless graviton is localized on the brane. This behavior can also be seen
from Figure 2, by the fact that $A=0$ and the massless solution is
normalizable.

We may arrive at the same conclusion via Figures 4 and 5, where the
massless wavefunction has a peak on the brane at $z=0$ whereas the
wavefunction of mass $m=1.5 k$ oscillates without feeling the presence of
the brane \cite{karch}.

In the case of nonzero B or E fields, the resonance in the modulus of the
wave function on the brane is at a nonzero mass. Thus, for a nonzero,
constant background B or E field, a massive graviton is localized on the
brane. This behavior is shown by the solid line in Figure 3, where the
$|\psi(z=0)|^2$ is plotted for $\alpha=.02$, and there is a large
resonance in the wave function on the brane for $m/k=.0822$. Again, this
behavior may also be ascertained from plots of $|\psi(z)|^2$ for various
values of $m/k$. For a mass significantly less than the resonance
mass, the wave function has a peak at $z=0$ that is not relatively high,
while a wave function with a mass significantly greater than the resonance
mass oscillates without feeling the brane.

It is straight-forward to repeat the above analysis for $p=4,5$. Note that
a D$4$-brane reduced on $S^1 \times S^4$ yields the same graviton wave
equation as a M$5$-brane reduced on $T^2 \times S^4$. Likewise, the
D$5$-brane reduced on $T^4 \times S^3$ and a M$5$-brane intersection
reduced on $T^4 \times S^2$ or K3 $\times S^2$ yield the same graviton
wave equation. This was initially observed in \cite{cvetic} in the case of
zero $B$-field.
%%%%%%%%
\begin{figure}
   \epsfxsize=4.0in
   \centerline{\epsffile{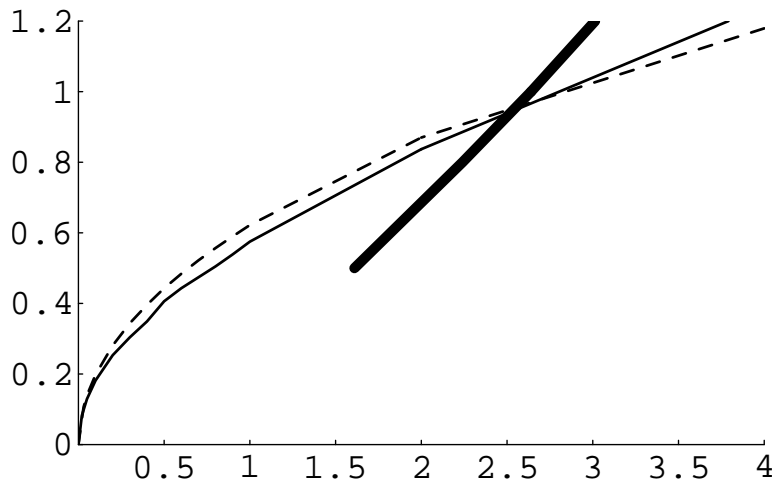}}
   \caption[FIG. \arabic{figure}.]{Resonance mass versus $\alpha$ for
$p=3$ (solid line), $4$ (dotted line) and $5$ (bold line).}
\end{figure}
%%%%%%%%
Figure 6 shows the resonance mass versus $\alpha$ for $p=3,4,5$. We classify
them by the corresponding near-horizon brane configurations in ten or
eleven dimensions. The solid line corresponds to the D$3$-brane. The
dotted line corresponds to the D$4$ or M$5$-brane. The bold line
corresponds to the D$5$-brane or an M$5$-brane intersection. It is
rather curious that the three lines appear to intersect at the same point,
at approximately $\alpha=2.688$ and $m/k=.976$.

%% file: thesiscon.tex
\chapter{Conclusions}

\section{Top Ten Unsolved Problems}

During the Strings 2000 Conference, a list of the ten most important
unsolved problems in fundamental physics was compiled by a selection panel
consisting of David Gross, Edward Witten and Michael Duff. Conference
participants were invited to submit one well-defined and clearly
stated candidate problem for consideration. 

These questions are meaningful to me as a guiding light for future
research-- that of my own generation as well as future generations whom I
may someday have the pleasure to teach.\\ 
\\
(1) {\it What are the fundamental degrees of freedom of M-theory (the
theory whose low-energy limit is eleven-dimensional supergravity and which
subsumes the five consistent superstring theories) and does the theory
describe Nature?}\\
Louise Dolan, University of North Carolina, Chapel Hill; Annamaria
Sinkovics,\\ Spinoza Institute; Billy and Linda Rose, San Antonio
College\\
\\
(2) {\it What is the resolution of the black hole information paradox?}\\
Tibra Ali, Department of Applied Mathematics and Theoretical Physics,
Cambridge; Samir Mathur, Ohio State University\\
\\
(3) {\it Are all the (measurable) dimensionless parameters that
characterize the physical universe calculable in principle or are some
merely determined by historical or quantum mechanical accident and
uncalculable?}\\
David Gross, Institute for Theoretical Physics, University of California,
Santa Barbara\\
\\
(4) {\it What physics explains the enormous disparity between the
gravitational scale and the typical mass scale of the elementary
particles?}\\
Matt Strassler, University of Pennsylvania\\
\\
(5) {\it Why does the universe appear to have one time and three space
dimensions?}\\
Shamit Kachru, University of California, Berkeley; Sunil Mukhi, Tata
Institute of Fundamental Research; Hiroshi Ooguri, California Institute of
Technology\\
\\
(6) {\it What is the lifetime of the proton and how do we understand
it?}\\
Steve Gubser, Princeton University and California Institute of
Technology\\
\\
(7) {\it Can we quantitatively understand quark and gluon confinement in
Quantum Chromodynamics and the existence of a mass gap?}\\
Igor Klebanov, Princeton University; Oyvind Tafjord, McGill University\\
\\
(8) {\it Is Nature supersymmetric, and if so, how is supersymmetry
broken?}\\
Sergio Ferrara, CERN European Laboratory of Particle Physics; Gordon Kane,
University of Michigan\\
\\
(9) {\it Why does the cosmological constant have the value that it has,
is it really zero and is it really constant?}\\
Andrew Chamblin, Massachusetts Institute of Technology; Renata Kallosh,
Stanford University\\
\\
(10) {\it How can quantum gravity help explain the origin of the
universe?}\\
Edward Witten, California Institute of Technology and Institute for
Advanced Study, Princeton\\
\\

Perhaps the most dramatic consequence of Grand unification is that
nucleons are unstable. A rough calculation, making use of the resemblance
of the muon decay process to the Feynman diagrams relevant to proton
decay, it is estimated that the lifetime of a proton is approximately
%%%%%%%%
\be
\tau_p \sim \tau_{\mu} \big( \frac{m_{\mu}}{m_p}\big) ^5 \big(
\frac{u}{M_W}\big) ^4,
\ee
%%%%%%%%
where all coupling constants have been assumed to be equal and the
unification scale $u$ has been taken to represent the mass of the
leptoquark bosons. Using the observed muon lifetime $\tau_{\mu}=7 \times
10^{-14}$ years and a plausible but by no means well-established value of
the unification scale $u \approx 10^{15} GeV/c^2$, the proton lifetime is
estimated to be about $\tau_p \sim 10^{34}$ years. More detailed
calculation show that the proton lifetime is likely to be in the range
$\tau_p=3 \times 10^{30\pm 2}$ years. The sector of M-theory in which the
AdS/CFT correspondence lies may yield information on quark interactions
which may in order to understand the process of proton decay.

Why are the magnitudes of the electron and proton charges the same? Why is
the electron charge three times that of the Down quark charge? Why
do the coupling constants have the values that they have? The Standard
Model offers no answers to these questions. Instead, measurements must
be taken in order to determine the values of about twenty different
parameters. Many physicists hope that these parameters are ultimately
derivable from pure theory. String/M-theory might be this ultimate
theory; it is believed that there are no free parameters in
string/M-theory, other than possibly the string tension.

Guided by an abundance of fundamental symmetries in Nature, it is rather
curious why the coupling constants are such that there is a dramatic
difference between the gravitational scale and that of the forces of the
Standard Model. There have been several attempts to explain the presence
of large numbers in terms of a higher-dimensional geometry. For example,
Randall and Sundrum used a brane world scenario to explain the Hierarchy
Problem in terms of the size of the extra dimension. However, this leads
one to ask what determines the size of the extra dimension.

Furthermore, if some spacetime dimensions are compactified to a
sub-millimeter size or
less, then what determines which dimensions are large and which are
small? Is there a dynamical process controlling the evolving size of
dimensions? Were all of the dimensions intially the same size? The idea of
brane worlds makes the notion more viable that there may be hidden
dimensions of the same size as the ones we see. Otherwise, what breaks the
symmetry between the sizes of dimensions?

Limited regions of M-theory can be probed using perturbative methods 
for string theoretic calculations. However, it is evident with the
existence of $p$-branes that there is a wealth of non-perturbative
phenomena. This has paved the way to understanding black hole microscopics
via intersections of wrapped $p$-branes.

While black hole microscopics has increased our understanding of black
hole thermodynamics, as well as scattering and decay processes, there
remains an unresolved issue. Namely, on the level of quantum field theory
in curved spacetime, it would appear that information pertaining to all
energy entering a black hole is lost to the outside universe. The
Holographic Principle offers tantilizing hints that this information may
actually be living on the surface of the event horizon, rather than
trapped within the central singularity. The portion of M-theory
corresponding to AdS/CFT-type examples of the Holographic Principle may 
offer the answer to this puzzle.

While quantizing gravity in the framework of particle physics
poses a major problem, quantum gravity arises naturally from
string/M-theory. I have discussed the importance of understanding a full
theory of quantum gravity in cases where there are strong gravitational
fields, such as near the central singularity of black holes or during early
cosmological epochs. M-theory may supply the full quantum theory of
gravity which will allow us to probe the first moment of the Big Bang,
when the universe itself may have been formed.

%% file: appendixA.tex
\chapter{Absorption by branes: calculations in detail}

\section{Outline of Numerical Method}

We will now outline the numerical method for finding the
absorption cross-section. We will use the well-studied case of the
dilaton-axion on an extremal D3-brane. In this case, the radial
wave equation is given by (\ref{eqdil}) and (\ref{V}). We take the
wave close to the horizon, at $\rho = .01$, to be purely incoming:
%%%%%%%%%
\be \psi (\rho) = \rho \exp{\big(\frac{i (\omega R)^2}{\rho}\big)}
\label{bndry} \ee
%%%%%%%%
The solution in the far region is: \be \psi (\rho) = A_{in}
\exp{(i \rho)} + A_{out} \exp{(- i \rho)}. \label{far} \ee
%%%%%%%

We use Mathematica to numerically integrate (\ref{eqdil}) with the
boundary condition given by (\ref{bndry}). At $\rho=45$, we match
the result with (\ref{far}) to find $A_{in}$ and $A_{out}$. The
absorption probability is given by
%%%%%%%
\be P = 1 - |\frac{A_{out}}{A_{in}}|^2. \ee The absorption
cross-section for a scalar is
%%%%%%%
\be \sigma^{(\ell)}=\frac{8 \pi^2 (\ell+3)(\ell+2)^2(\ell+1)}{3
(\omega R)^2} P^{(\ell)} \label{prob} \ee
%%%%%%%%
Numerical integration for other types of particles incident on
other objects with event-horizons is straight-forward.

\section{High-energy absorption cross-section for a
dilaton-axion on a nonextremal D3-brane}

In this appendix, we obtain the analytical high energy absorption
cross-section for a dilaton-axion in a {\it nonextremal} D3-brane
background, by employing the geometrical optics limit for the
classical motion of a particle \cite{teuk}.

For a nonextremal D3-brane, the metric takes the form
(\ref{d3metric}). The classical Lagrangian for a particle  is of
the form:
%%%%%%%%%
\be L=\frac{1}{2} g_{\alpha \beta} \dot{x}^{\alpha}
\dot{x}^{\beta}. \ee
%%%%%%%%%
$\dot{x}^{\alpha} \equiv dx^{\alpha}/d\lambda$, where $\lambda$ is
an affine parameter.  The Euler-Lagrange equations are
%%%%%%%%%%
\be \frac{d}{d\lambda} \big( \frac{\partial L}{\partial
\dot{x}^{\alpha}} \big) - \frac{\partial L}{\partial x^{\alpha}} =
0. \ee
%%%%%%%%%
The equation of motion for $\theta$ is
%%%%%%%%%
\be \frac{d}{d\lambda} (H^{1/2} r^2 \dot{\theta}) = H^{1/2} r^2
\sin \theta \cos \theta (\dot{\phi_3}^2 - \dot{\phi}^2 - \cos^2
\phi \dot{\phi_1}^2 - \sin^2 \phi \dot{\phi_2}^2). \label{theta}
\ee
%%%%%%%%%
The solution of (\ref{theta}) is $\theta = \pi /2$ and
$\dot{\theta} = 0$. The equations of motion for $\phi_3$ and $t$
are of the form
%%%%%%%%
\be \frac{d}{d\lambda} (H^{1/2} r^2 \dot{\phi_3} )=0 \label{phi}
\ee
%%%%%%%%
and
%%%%%%%%
\be \frac{d}{d\lambda} (H^{-1/2} f \dot{t} )=0. \label{t} \ee
%%%%%%%%
(\ref{phi}) and (\ref{t}) each imply a constant of motion:
%%%%%%%
\be H^{1/2} r^2 \dot{\phi_3} = {\rm constant} \equiv \ell \ee
%%%%%%
and
%%%%%%%
\be H^{-1/2} f \dot{t} = {\rm constant} \equiv E, \ee
%%%%%%%
where $\ell$ and $E$ are interpreted as the angular momentum and
energy of the particle, respectively.

Also, since the particle  scatters only in a direction transverse
to the D3-brane ({\it i.e.}, it does not  travel along the
D3-brane), $\dot{x_i} = 0$, for $i = 1,2,3$.  Substituting our
solutions for $\theta$ and $\dot{x_i}$ into the Lagrangian yields
%%%%%%%%%
\be 2L = - H^{-1/2} f \dot{t}^2 + H^{1/2} f^{-1} \dot{r}^2 +
H^{1/2} r^2 \dot{\phi_3}^2. \ee
%%%%%%%%
Instead of finding a rather complicated equation of motion for
$r$, we use the fact that, for a massless particle,
%%%%%%%%
\be 2L = g_{\alpha \beta} p^{\alpha} p^{\beta} = 0, \ee
%%%%%%%
where $p^{\alpha} = \dot{x}^{\alpha}$. Substituting our results
for $\dot{\phi_3}$ and $\dot{t}$ into the previous equation yields
%%%%%%%%
\be \dot{r}^2 = E^2 - \frac{\ell^2}{r^2} \frac{f}{H}. \ee
%%%%%%%%
We introduce a new parameter, $\lambda^{\prime} \equiv \ell
\lambda$, so that
%%%%%%%
\be \big( \frac{dr}{d\lambda^{\prime}} \big) ^2 = \frac{1}{b^2} -
V_{\rm effective}, \ee
%%%%%%%
where $b \equiv \ell /E$ is the impact parameter and
%%%%%%%
\be V_{\rm effective} = \frac{1}{r^2} \frac{f}{H}. \ee
%%%%%%%
The absorption cross-section for particles  at high energy  can be
obtained by determining  the classical trajectory  of the
scattered particle and  using the optical limit result:
%%%%%%%%%
\be
\sigma_{\rm abs}=\frac{8}{15} \pi^2 b_{\rm crit}^5 ,\nn\\
\ee
%%%%%%%%%
where the  critical impact parameter separating absorption from
scattering orbits is given by $1/b_{\rm crit}^2 = V_{\rm
maximum}$. Thus,
%%%%%%%%%
\bea \sigma_{\rm abs} &=&\frac{8}{15} \pi^2
\Big(\frac{1}{2}R^4+3m+\frac{1}{2} \sqrt{(R^4+6m)^2+8mR^4}
\Big)^{5/4} \times \nn\\
& &\Big( \frac{\frac{3}{2}R^4+3m+\frac{1}{2} \sqrt{(R^4+6m)^2
+8mR^4}}{\frac{1}{2}R^4+m+\frac{1}{2} \sqrt{(R^4+6m)^2+8mR^4}}
\Big)^{5/2}, \eea
%%%%%%%%%
where $m$ is the nonextremality parameter. In the extremal limit,
$m=0$, this result reduces to:
%%%%%%%%%
\be \sigma_{\rm abs} = \frac{32 \sqrt{2}}{15} \pi^2
R^5(1+5\frac{m}{R^4}). \ee
%%%%%%%%%%

%% file: appendixB.tex
\chapter{Brane Intersections with B field}
%\label{ch:Appendices}

\section{Intersection of two 5-branes, a string and a pp-wave
with B field}

We begin with the intersection of two D5-branes, a D1-brane and a pp-wave
\cite{lima}. We perform a T-duality transformation
along the common spatial direction $x$, which yields the intersection of
two smeared D4-branes, a smeared D0-brane and an NS-string. We apply the
Lorentz boost
%%%%%%%%
\be
t= t^{\prime} \cosh \theta + x^{\prime} \sinh \theta,\ \ \
t= t^{\prime} \sinh \theta + x^{\prime} \cosh \theta.
\ee
%%%%%%%
Performing a T-duality transformation along $x^{\prime}$, and restoring
the labels $x$ and $t$, we find
%%%%%%%
\begin{eqnarray}
ds_{10}^2 &=&
K^{-3/4}H^{-1/4}\tilde{H}^{-1/4}h^{3/4}\Big(-W^{-1}dt^2+W\big( dx+
(W^{-1}-1)dt\big)^2\Big)\\ \nonumber
& &+K^{1/4}H^{3/4}\tilde{H}^{-1/4}h^{-1/4}(dy_1^2+..+dy_4^2)\\ \nonumber
& &+K^{1/4}H^{-1/4}\tilde{H}^{3/4}h^{-1/4}(dz_1^2+..+dz_4^2),\\ \nonumber
H&=&1+\frac{R_1^2}{r_1^2}\ \ \
\tilde{H}=1+\frac{R_2^2}{r_2^2},\ \ \ K=W=H_1 H_2,\\ \nonumber
\phi&=&-\frac{1}{2} \log \big[ \frac{K h}{H \tilde{H}} \big],\ \ \
h^{-1}=-(H \tilde{H} K)^{-1} \sinh^2 \theta +\cosh^2 \theta,\\ \nonumber
F_{(1)}&=&d\chi =\sinh \theta dK^{-1},\ \ \
F_{(5)}=\sinh \theta [\tilde{H} d^4z\wedge dH^{-1}+H d^4y\wedge
d\tilde{H}^{-1}],\\ \nonumber
F_{(3)}&=&h \cosh \theta [e^{\phi} \ast \tilde{H} dt\wedge dx\wedge
d^4z\wedge dH^{-1}+e^{\phi} \ast H dt\wedge dx\wedge d^4y\wedge
d\tilde{H}^{-1}\\ \nonumber & &+dt\wedge dx\wedge dK^{-1}],\ \ \
B_{(2)}=-(H_1 H_2 H_3)^{-1} h \tanh \theta dt \wedge dx.
\end{eqnarray}
%%%%%%%
where
%%%%%%
\be
dy_1^2+..+dy_4^2=dr_1^2+r_1^2 d\Omega_3^{(1)2}
\ee
%%%%%
and likewise for $dz_1^2+..+dz_4^2$.
$H$,$\tilde{H}$,$K$ and $W$ are harmonic functions associated with the two
D5-branes, the D1-brane and the pp-wave, respectively. Note that this is a
non-standard intersection, since there is no overlap in the coordinate
dependence of $H$ and $\tilde{H}$ and a multi-charge $p$-brane
cannot be obtained via dimensional reduction \cite{lima}.

\section{Intersection of three M5-branes and a pp-wave with
B-field}

We begin with three non-extremal M5-branes, each pair intersecting at a
3-brane with a boost along a string common to all 3-branes. We
dimensionally reduce to ten dimensions, ending up with an NS-NS 5-brane
intersecting with two smeared D4-branes and a boost along the common
string. Performing a T-duality transformation along the common spatial
direction yields an NS5-brane intersecting with two doubly smeared
D3-branes and an NS-string. Applying T-duality along a Lorentz-boosted
common spatial direction and oxidizing back to eleven dimensions gives the
boosted M5-brane intersection with a background B-field:
%%%%%%%%
\begin{eqnarray}
ds_{11}^2&=&(F_1 F_2 F_3)^{-2/3} h^{-1/3} \Big( h\big( F_1 F_2 F_3
(-K^{-1} f dt^2+K (dy_1+(K^{\prime -1}-1)dt)^2)\\ \nonumber & & +F_2 F_3
dy_2^2\big)+F_2 F_3 dy_3^2+F_1 F_3(dy_4^2+dy_5^2)+F_1
F_2(dy_6^2+dy_7^2)+f^{-1}dr^2+r^2d\omega_2^2\Big),\\ \nonumber
h^{-1}&=&-F_2 F_3 f\sinh^2 \theta +\cosh^2 \theta,\\ \nonumber
F_{(4)} &=& 3\sinh \theta (dy_3\wedge dy_6\wedge dy_7\wedge dF_2^{\prime}+
dy_3\wedge dy_4\wedge dy_5\wedge dF_3^{\prime})\\ \nonumber
& &+3h\cosh \theta \ast
(dt\wedge dy_1\wedge dy_2\wedge dy_3\wedge dy_6\wedge dy_7\wedge
dF_2^{\prime}\\ \nonumber & & +dt\wedge dy_1\wedge dy_2\wedge dy_3\wedge
dy_4\wedge dy_5\wedge dF_3^{\prime})\\ \nonumber & & +3\ast (dt\wedge
dy_1\wedge dy_4\wedge dy_5\wedge dy_6\wedge dy_7\wedge dF_1^{\prime})\\
\nonumber & & -\tanh \theta d(hF_2 F_3 f)\wedge dt\wedge dy_1\wedge dy_2.
\end{eqnarray}
%%%%%%%%
where
%%%%%%%%
\begin{eqnarray}
K&=&1+\frac{Q}{r},\ \ \ K^{\prime -1}=1-\frac{Q^{\prime}}{r}K,\ \ \
F_i^{-1}=1+\frac{P_i}{r},\ \ \ F_i^{\prime -1}=1+\frac{P_i^{\prime}}{r},\\
\nonumber Q&=&\mu \sinh^2 \beta,\ \ \ Q^{\prime}=\mu \sinh \beta \cosh
\beta,\ \ \ P_i=\mu \sinh^2 \gamma_i,\ \ \ P_i^{\prime}=\mu \sinh \gamma_i
\cosh \gamma_i.
\end{eqnarray}
%%%%%%%%

\section{D4-D8 intersection with B field}

We perform a T-duality transformation along a spatial direction $x_4$
in the D4-D8 intersection to obtain a D3-D7 intersection. A spatial
rotation $(x_1,x_4) \rightarrow (x_1^{\prime},x_4^{\prime})$ followed by a
T-duality transformation along $x_4^{\prime}$ yields a D4-D8 intersection in a
B-field:
%%%%%%%%
\begin{eqnarray}
ds_{10}^2&=&(gz)^{1/12} \Big( H^{-3/8}\big(-h^{-1/4}
dt^2+h^{3/4}(dx_1^2+dx_4^2)\ \ \+h^{-1/4}(dx_2^2+dx_3^2) \big)\\
\nonumber & &+H^{5/8}h^{-1/4}(dy_1^2+..dy_4^2+dz^2),\ \ \
\phi=-\frac{1}{2} \log \big[ (gz)^{-5/3} H^{-1/2}h \big],\\
\nonumber H&=&1+\frac{R^{10/3}}{(y_1^2+..+ y_4^2+dz^2)^{5/3}},\ \ \
h^{-1}=(gz)^{-2/3} H^{-1}\sin^2 \theta +\cos^2 \theta,\\ \nonumber
F_{(4)}&=& e^{-\frac{1}{2}\phi} \ast (h \cos \theta d^5 x \wedge
dH^{-1})+\sin \theta dt \wedge dx_2 \wedge dx_3 \wedge dH^{-1},\\
\nonumber F_{(2)}&=& -g (gz)^{-2/3} \sin \theta h H^{-1} dx_1 \wedge
dx_4,\ \ \ B_{2}=(gz)^{-2/3} H^{-1} h \tan \theta dx_1 \wedge dx_4.
\end{eqnarray}
%%%%%%%%
where $g=\frac{3}{\sqrt{2}} \frac{m}{\cos \theta}$.

If instead of the spatial rotation we do a Lorentz boost with $t$
and $x_4$, then we end up with a D4-D8 intersection in an E-field:
%%%%%%%%
\begin{eqnarray}
ds_{10}^2&=&(gz)^{1/12}
\Big(H^{-3/8}\big(\tilde{h}^{3/4}(-dt^2+dx_4^2)+\tilde{h}^{-1/4}
(dx_1^2+dx_2^2+dx_3^2)\big)\\ \nonumber & &
+H^{5/8}\tilde{h}^{-1/4}(dy_1^2+..dy_4^2+dz^2),\\ \nonumber
H&=&1+\frac{R^{10/3}}{(y_1^2+..+y_4^2+dz^2)^{5/3}},\ \ \
\tilde{h}^{-1}=-(gz)^{-2/3} H^{-1}\sinh^2 \tilde{\theta} +\cosh^2
\tilde{\theta},\\ \nonumber \phi&=&-\frac{1}{2} \log \big[ (gz)^{-5/3}
H^{-1/2}\tilde{h} \big],\\ \nonumber & &
F_{(4)}= e^{-\frac{1}{2}\phi} \ast (\tilde{h} \cosh \tilde{\theta} d^5 x
\wedge dH^{-1})+\sinh \tilde{\theta} d^3 x \wedge dH^{-1},\\ \nonumber
F_{(2)}&=& g (gz)^{-2/3} \sinh \tilde{\theta} \tilde{h} H^{-1} dt \wedge
dx_4,\ \ \
B_{2}=-(gz)^{-2/3} H^{-1} \tilde{h} \tanh \tilde{\theta} dt \wedge dx_4.
\end{eqnarray}
%%%%%%%%
where $g=\frac{3}{\sqrt{2}} \frac{m}{\cosh \tilde{\theta}}$.

\section{D3-brane and pp-wave with B field}

We begin with a D2-brane and a pp-wave propagating along a direction
in the world-volume. A spatial rotation in one
world-volume and one transverse direction followed by a
T-duality transformation leads to a D3-brane and pp-wave with a
background B field:
%%%%%%
\begin{eqnarray}
ds_{10}^2&=&H^{-1/2}\Big( h^{-1/4} \big( -K^{-1}
dt^2+K(dx_3+(K^{-1}-1)dt)^2\big)\\ \nonumber & & 
+h^{3/4}(dx_1^2+dx_2^2)\Big)+H^{1/2} h^{-1/4} (dr^2+r^2
d\Omega_5^2),\\ \nonumber
H&=&1+\frac{R^4}{r^4},\\ \nonumber
h^{-1}&=&H^{-1} \sin^2 \theta +\cos^2 \theta.
\end{eqnarray}
%%%%%%
Alternatively, one may perform a Lorentz boost rather than a spatial
rotation and end up with a background field of a different component:
%%%%%%
\begin{eqnarray}
ds_{10}^2&=&H^{-1/2}\Big( \tilde{h}^{3/4} \big( -K^{-1}
dt^2+K(dx_3+(K^{-1}-1)dt)^2\big)\\ \nonumber
& &+\tilde{h}^{-1/4}(dx_1^2+dx_2^2)\Big)+H^{1/2} \tilde{h}^{-1/4}
(dr^2+r^2 d\Omega_5^2),\\ \nonumber H&=&1+\frac{R^4}{r^4},\\ \nonumber
\tilde{h}^{-1}&=&-H^{-1} \sinh^2 \tilde{\theta} +\cosh^2 \tilde{\theta}.
\end{eqnarray}
%%%%%%

%% file: appendixC.tex
\chapter{Wrapped D5-brane as Brane World}
%\label{ch:Appendices}

\section{$p=5$ as an exactly solvable model}

Consider the equations (\ref{Wave}) and (\ref{V}). Although this
is not in Schr\"{o}dinger form, for $p=5$ this equation is easily
solvable for the massive case. The solution for $\Theta$ is
%%%%%%
\be
\Theta=N(1+k|w|)^{1/2}\big[J_{-i\gamma}\Big(\frac{i\alpha}{1+k|w|}\Big)
+A(\alpha)Y_{-i\gamma}\Big(\frac{i\alpha}{1+k|w|}\Big)], \ee
%%%%%%
where
%%%%%
\be A(\alpha)=\frac{2 J_{-i\gamma}(i
\alpha)+i\alpha[J_{-i\gamma+1}(i\alpha)-
J_{-i\gamma-1}(i\alpha)}{-2Y_{-i\gamma}(i\alpha)+i\alpha[Y_{-i\gamma-1}
(i\alpha)-Y_{-i\gamma+1}(i\alpha)]},\label{A5} \ee
%%%%%
and $\gamma \equiv \sqrt{4(m/k)^2-1}$. Transforming back to the
$z$ coordinate and $\psi$ wave function using (\ref{ztrans}) and
(\ref{wavetrans}) we obtain the wave function solution
%%%%%%%
\be \psi(z)=N_0 \big[ J_{-i\gamma} (i\alpha\ e^{-k|z|/2}) +
A(\alpha) Y_{-i\gamma}(i \alpha\ e^{-k|z|/2}) \big]. \ee
%%%%%%%
For the massless graviton
%%%%%%%
\be \psi(z)\sim e^{-k|z|/2}+A(\alpha) e^{k|z|/2}, \ee
%%%%%%%%
and so the massless graviton wave function is not normalizable for
nonzero $B$ or $E$ fields.

As in the zero $B$-field case, there is a mass gap and so we must
have $m^2 \ge \frac{1}{4}k^2$ for the massive wave functions.
Expanding the Bessel functions for large $k|z|$ yields
%%%%%%%
\begin{eqnarray}
\psi(z) &=& N\frac{e^{\pi/2}}{\Gamma(1-i\gamma)} \Big(
[1+iA(\alpha)\rm{csch}(\pi\gamma)\cosh(\pi\gamma)]e^{iq|z|-i\gamma
\ln(\alpha/2)}-\\ \nonumber & &
iA(\alpha)\rm{csch}(\pi\gamma)e^{-\pi}\frac{\Gamma(1-i\gamma)}
{\Gamma(1+i\gamma)}e^{-iq|z|+i\gamma \ln(\alpha/2)}\Big),
\end{eqnarray}
%%%%%%%
where $q \equiv \sqrt{m^2-k^2/4}$. We numerically normalize the
wave function, and find the resonant mass by solving for the
maximum of $|\psi(z=0)|^2$ for a given $\alpha$. The resonant mass
versus $\alpha$ is plotted for $p=5$ as the bold line in Figure 6.
With this exact solution to the wave function, one can then find
an analytic expression for how the gravitational potential is
modified by the presence of the constant, background $B$ and $E$
fields.

\section{Nonextremality and the delocalization of gravity}

Since the present entropy density of our universe is small, but
nonzero, it is of pertinence to consider brane world models
arising from nonextremal $p$-branes. We consider the case of the
D$5$-brane in type IIB supergravity, which tends to lead to a
simpler graviton wave equation than that of other $p$-branes. The
metric is
%%%%%%%%%%%
\be
ds^2=H^{-1/4}(-f dt^2+dx_i^2+dy_j^2)+H^{3/4}(f^{-1}dr^2+r^2 d\Omega_3^2),
\ee
%%%%%%%%%%%
where
%%%%%%%%%%
\be
H=1+\frac{R^2}{r^2},\ \ \ \ \ \ \ f=1-\frac{r_o^2}{r^2},
\ee
%%%%%%%%%%%
and $i=1,2,3$ and $j=1,2$. Consider the near-horizon limit $r<<R$:
%%%%%%%%%
\be
ds^2=(\frac{r}{R})^{1/2}(-f dt^2+dx_i^2+dy_j^2+R^2
d\Omega_3^2)+(\frac{r}{R})^{-3/2} f^{-1}dr^2.
\ee
%%%%%%%%%
After applying the coordinate transformation
%%%%%%%
\be
\frac{r}{R}=(e^{-k|z|}+r_o^2/R^2)^{1/2},
\ee
%%%%%%
we obtain
%%%%%%
\be
ds^2=(e^{-k|z|}+c)^{1/4}\big( f(-dt^2+dz^2)+dx_i^2+dy_j^2+R^2 d\Omega_3^2
\big).
\ee
%%%%%%%%%
Note that in the extremal limit this corresponds to the conformally-flat
frame. After dimensional reduction over $S^3$, the above metric
corresponds to a 7-dimensional domain wall at $z=0$. We will consider the
case where $y_j$ are wrapped around a compact manifold, so that the
brane world has 1+3 large dimensions. Note that our coordinate system
ensures that the domain-wall solution is $Z_2$ symmetric. The graviton
wave equation is $\partial_{\mu}(\sqrt{-g}g^{\mu \nu}\partial_{\nu}
\Phi=0$. Consider the Ansatz
%%%%%%%%
\be
\Phi=\phi(z)e^{ip\cdot x}=(e^{-k|z|}+c)^{-1/2}\psi(z)e^{ip\cdot x},
\ee
%%%%%%%
which satisfies the Schr\"{o}dinger-type equation,
%%%%%%%
\be
-\partial_z^2 \psi+V\psi=m^2\psi,
\ee
%%%%%%%
where $m^2$ is the eigenvalue of the 4-dimensional wave equation.
The schr\"{o}dinger potential is given by
%%%%%%%%%
\be
V=k^2\Big( \frac{1}{2(1+r_o^2/R^2 e^{k|z|})}+\frac{1}{4(1+r_o^2/R^2
e^{k|z|})^2}\Big) -\frac{k}{1+r_o^2/R^2} \delta(z).
\ee
%%%%%%%%
The massless wave-function is given by
%%%%%%%%%
\be
\psi=N(e^{-k|z|}+r_o^2/R^2)^{1/2}, \label{psi}
\ee
%%%%%%%%%
which is the solution that satisfies the $\delta(z)$ boundary
condition. The trapping of gravity requires that the wave-function be
normalizable, i.e., $\int |\psi|^2 dz < \infty$, in order to have a finite
leading-order correction to gravitational field on the brane. However, the
wave-function given by (\ref{psi}) is only normalizable in the extremal
limit where $r_o$ strictly vanishes. This is due to the definite, nonzero
temperature of the nonextremal brane solution, which is evident if we
represent the nonextremality parameter in the form
%%%%%%%%
\be \frac{r_o^2}{R^2}=\frac{1}{(2\pi RT_H)^2}-(2\pi RT_H)^2, \ee
%%%%%%%
where $T_H$ is the Hawking temperature of the D$5$-brane. Thus,
gravitons are radiated away from the nonextremal brane.
Considering the nonextremality of the D$5$-brane as a perturbation
and we shall express the (modulus of) the graviton wave function
in terms of the wave functions which arise in the extremal case.
For a brane world arising from an extremal D$5$-brane, the bound
graviton wave function is given by
%%%%%%%
\be \Theta_0=N_0 e^{-k|z|/2}, \ee
%%%%%%%%
and the wavefunctions of the Kaluza-Klein modes of mass $m$ are
given by
%%%%%%%
\be \Theta_m=N_m (k\sin q|z|-2q\cos q|z|), \ee
%%%%%%%
where $q=\sqrt{m^2-\frac{1}{4}k^2}$ and $N_0$, $N_m$ are
normalization constants. There is a mass gap between the localized
massless state and the massive Kaluza-Klein modes, where $m \ge
k/2$.

The modulus of the wavefunction (\ref{psi}) can be expressed as
%%%%%%%
\be |\psi|^2=|\Theta_0|^2 +A\ee
%%%%%%%